\pdfoutput=1
\documentclass[prb,twocolumn]{revtex4-2}
\usepackage{graphicx}
\usepackage{dcolumn}
\usepackage{pgffor}
\usepackage{bm}
\usepackage[colorlinks=true,
            linkcolor=blue,
            citecolor=blue,
            urlcolor=blue]{hyperref}
\usepackage{pdfpages}

\makeatletter
\AtBeginDocument{\let\LS@rot\@undefined}
\makeatother

\begin{document}

\title{Orbital Hybridization Induces Giant Cubic Rashba Effect at Cu/WO$_{3}$ Interface}

\author{Md Aktar Hossain}
 \email{aktarbinmotiur@kgpian.iitkgp.ac.in}
 \affiliation{Department of Physics, Indian Institute of Technology Kharagpur, Kharagpur-721302, India}
\author{Saikat Das}
 \email{saikat@phy.iitkgp.ac.in}
 \affiliation{Department of Physics, Indian Institute of Technology Kharagpur, Kharagpur-721302, India}

\begin{abstract}
Harnessing Rashba spin-orbit interaction and related spintronic functionalities has traditionally relied on metallic surfaces or interfaces containing elemental heavy metals. Here, using first-principles calculations and Cu(001)/WO$_3$(001) as a model heterostructure, we show that interfacing a light metal, Cu, with a band insulator, WO$_3$, yields an interface state, which exhibits a robust Rashba spin-splitting, arising from the interplay between linear and cubic Rashba effects. The spin-splitting is driven by the strong spin-orbit coupling of W atoms and enabled by W-Cu orbital hybridization at the interface. The cubic Rashba contribution asymptotically grows with Cu thickness and can be explained in terms of cross‑coupling between the vacuum/Cu and Cu/WO$_3$ interfaces. This interfacial cross-coupling, however, diminishes at larger Cu thicknesses, allowing us to extract the intrinsic cubic Rashba parameter, which amounts to a giant value of $\approx$ -1.93 eV$\cdot$\AA$^3$. In contrast, the linear Rashba parameter is only weakly affected by this cross‑coupling and varies from $\approx$ +0.3 to +0.49 eV$\cdot$\AA~. We further show that sizable linear and cubic Rashba effects persist across several interface geometries and Cu surface orientations including (110) and (111). Our work identifies the Cu/WO$_{3}$ interface as a novel light metal/heavy-element-based oxide platform for exploring the rich spectrum of Rashba physics, including linear and non-linear spin-orbit phenomena.

\end{abstract}

\maketitle

\section{Introduction}
The Rashba spin-orbit interaction, characterized by a momentum-dependent spin-splitting, allows for a facile interconversion between charge and spin currents--a key enabler for spintronics applications, including non-volatile spin-orbit torque magnetic random access memory, magneto-electric spin-orbit logic, and terahertz emitters~\citep{RevModPhys.91.035004, manipatruni2019scalable, Seifert2016}. This phenomenon arises from a spin-split band structure driven by broken inversion symmetry, with the strength of the splitting directly correlating to the efficiency of charge-to-spin current interconversion~\citep{Rashba, rashba1960properties}. The ability to modulate this splitting via an external electric field offers a pathway to ultra-low-power, voltage-controlled spintronic devices~\citep{PhysRevApplied.15.064015}. Besides, Rashba spin-orbit interaction is at the heart of a range of exotic phenomena such as non-linear spin-charge conversion and quantum metric effects~\citep{Quantum_metric}. Consequently, several materials including LaAlO$_3$/SrTiO$_3$~\citep{Ohtomo2004, LaAlO3_STO}, KTaO$_3$/SrTiO$_3$~\citep{KTaO3}, van der Waals heterostructures~\citep{two-dimensional}, Si-terminated TbRh$_{2}$Si$_{2}$~\citep{si-terminated} have been extensively investigated.

While any conducting surface or interface lacking the inversion symmetry is allowed to exhibit Rashba spin-orbit interaction, surfaces/interfaces involving light metals, with an intrinsically weak atomic spin-orbit coupling, have remained underexplored. Surprisingly, recent experimental works seem to challenge this perception. For example, Rashba spin-splitting and efficient spin-to-charge current conversion are reported at the interface between light metal, such as Cu and amorphous Bi$_2$O$_3$~\citep{cu-bi2o3, metal_Bi}. A robust charge-to-spin current conversion with spin-orbit torque efficiencies comparable to conventional spintronic materials is realized at the interface between Ni$_{0.8}$Fe$_{0.2}$ (Py) and epitaxial (001)-oriented Bi$_2$WO$_6$ thin films~\citep{das2023observation}. Furthermore, this spin current is found to enhance transverse thermoelectric conversion in Ni/(001)-Bi$_2$WO$_6$ heterostructures~\citep{itoh2024enhancement}. Bi$_2$O$_3$ and Bi$_2$WO$_6$ are wide-bandgap insulators ($\sim$2.1--2.8 eV) that cannot support electronic transport. However, they contain heavy elements like Bi and W that can interact with the light atoms from the metallic layer. Specific to the case of Bi$_2$WO$_6$, a W-rich termination layer forms the interface. Thus, it is proposed that the Rashba spin splitting emerges at the interface and is possibly amplified by the hybridization of W orbitals with those of the constituting atoms of Py ~\citep{das2023observation}.

Inspired by these findings, this study aims to theoretically investigate the emergent Rashba effect that could arise due to proximity coupling at the interface between a light metal and a heavy-element based oxide insulator. As a model system, we use Cu/WO$_3$ heterointerface. We choose the paraelectric WO$_3$ instead of the ferroelectric Bi$_2$WO$_6$ to avoid complexities that might arise from the polar displacement of W atoms and the resulting Rashba spin-splitting of bulk electronic structure ~\citep{Djani2019, acs_applied}. Crucially, it can replicate the interfacial W-rich environment like that in the Py/Bi$_2$WO$_6$ system. The Cu layer also mimics the interfacial non-magnetic metallic phase observed in the Py layer of the Py/Bi$_2$WO$_6$ \citep{das2023observation}, and simplifies the analysis of Rashba spin-splitting by eliminating the effect of exchange interaction. Additionally, while Cu/amorphous Bi$_2$O$_3$ interfaces have been studied using density functional theory (DFT)~\citep{cu-bi2o3}, similar studies on insulating tungsten oxide compounds like WO$_3$ are missing.
\begin{figure*}[htbp]
    \centering    
    \includegraphics[scale=0.9]{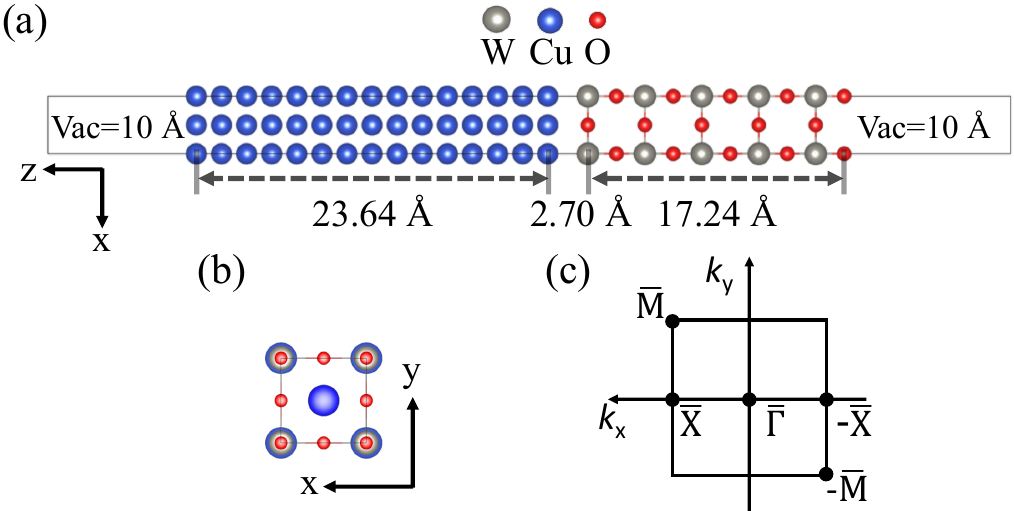}
    \caption{Structural model of the WO$_{2}$-terminated  Cu(001)/WO$_3$(001) heterostructure with a Type-I interface configuration. (a) Side view of the slab model showing the interfacial atomic arrangement, (b) top view of the  slab model; for visual clarity, the Cu atoms are shown enlarged, and a deep blue shade is used for the face-centered Cu atom. (c) Corresponding surface Brillouin zone with high-symmetry points labeled.}
    \label{fig:structure_001}
\end{figure*}

In this work, we present a systematic density functional theory (DFT) study of Cu/WO$_3$(001) heterostructures with different Cu surface orientations: (001), (110), and (111), which are found to promote strong orbital hybridization between Cu and W atoms. This hybridization leads to the emergence of robust Rashba-split interface states with robust linear and cubic Rashba effects. For the Cu(001)/WO$_3$(001) heterostructure we find the intrinsic linear Rashba parameter, $\alpha_R \approx +0.49 $~eV$\cdot$\AA~, and remarkably a giant cubic Rashba parameter, $\tilde{\alpha}_R \approx -1.93 $~eV$\cdot$\AA$^{3}$.
\section{Methodology}
We employed the OpenMX code for first-principles density functional theory (DFT)-based electronic structure calculations \citep{ozaki2003variationally,ozaki2004numerical,Norm}. We used the GGA-PBE exchange-correlation functional \citep{GGA} and applied an energy cutoff of 220~Ry for all calculations. We set the self-consistent field (SCF) convergence criterion to 2.72~$\times$~10$^{-5}$~eV. The W $5d$ states are treated within the DFT+U approach using an optimized Hubbard parameter of $U = 5.65$~eV~\citep{DFT+U-1,DFT+U-2,DFT+U-3}.

We first optimized the bulk lattice parameters of Cu and WO$_3$, starting from a face-centered cubic phase (space group $Fm\overline{3}m$) and a cubic phase (space group $Pm\overline{3}m$), respectively. Bulk structural relaxations were performed until the force on each atom was less than 1~meV/\AA. The optimized lattice parameters are 3.67~\AA\ and 3.83~\AA\ for cubic Cu and WO$_3$, respectively. These values compare fairly well with the experimental ones \citep{cu_exp,wo3_exp,wo3_tetragonal}. Using the optimized lattice parameters, we constructed slab models of the Cu/WO$_3$(001) heterostructures. We then calculated interface and formation energies for different Cu orientations: (001), (110), and (111), while maintaining the (001) orientation of the WO$_3$ layer. In all cases, the WO$_3$ surface is terminated by a WO$_2$ layer. For each orientation, we considered three interface configurations: Cu beneath W atoms (Type-I), Cu beneath O atoms (Type-II), and Cu at bridge sites between two O atoms (Type-III) (see Supplementary Fig. S1 and Table S1 in the Supplementary Information \citep{supplementary})~\nocite{suppl-1}. For Cu(001)/WO$_3$(001), Type-II is lowest in energy, whereas Type-I is favored for Cu(110)/WO$_3$(001) and Cu(111)/WO$_3$(001). All three heterostructures have negative formation energies, indicating potential experimental stability. Since Type-I is preferred for two of the three orientations and can be kinetically stabilized by non-equilibrium growth techniques, we use this configuration as the canonical model in the main text. Detailed results for the Type-II and Type-III configurations are provided in the Supplementary Information~\citep{supplementary}.

\begin{figure*}[htbp]
\centering
\includegraphics[scale=0.75]{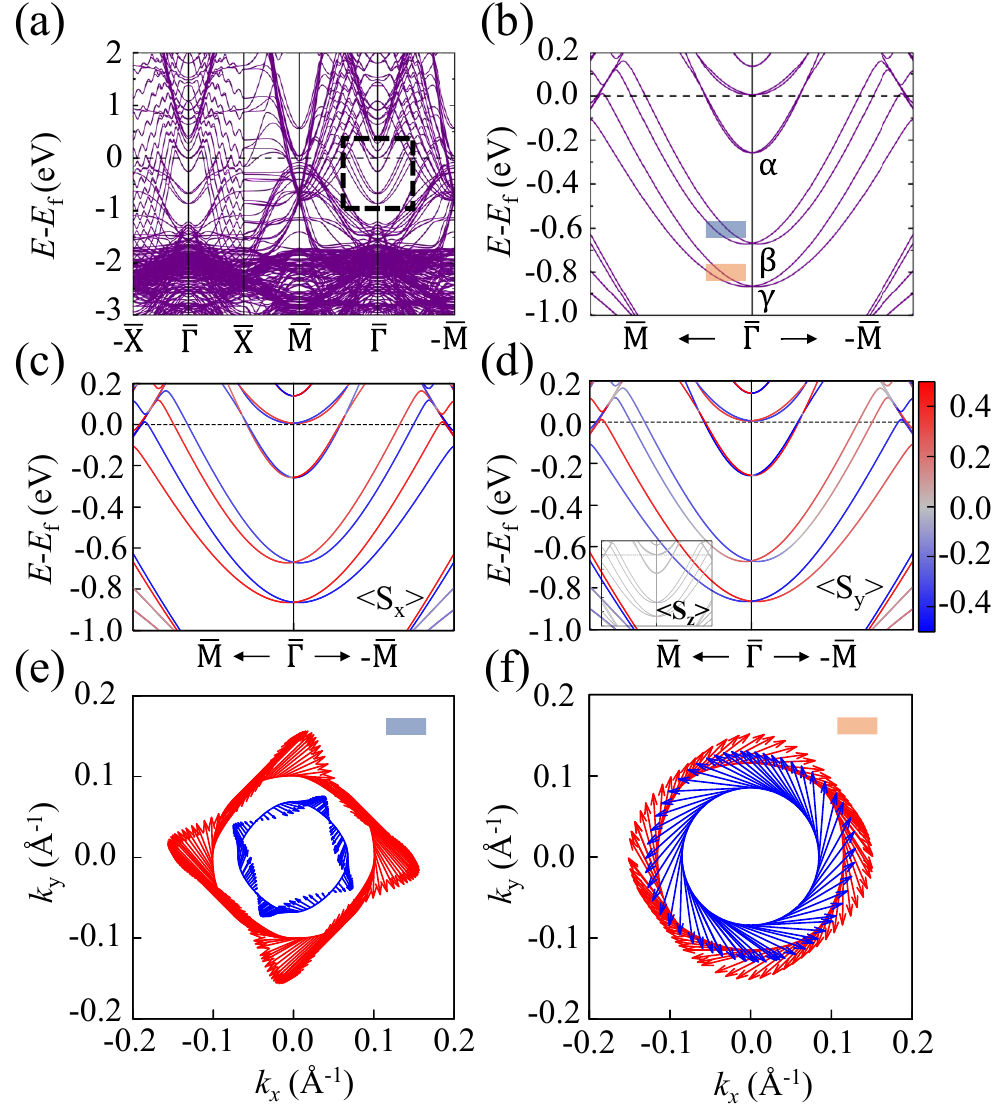}
\caption{
Electronic band structure and spin textures of the WO$_{2}$-interface terminated Cu(001)/WO$_3$(001) heterostructure. (a) Band structure. (b) Magnified view of the $\alpha$, $\beta$, and $\gamma$ bands near the $\overline{\Gamma}$ point, corresponding to the dashed box in (a). (c), (d) Spin-projected band structures showing the in-plane $\langle S_x \rangle$ and $\langle S_y \rangle$ components, respectively; the inset in (d) shows the negligible $\langle S_z \rangle$ contribution. (e), (f) Constant-energy spin textures for the $\beta$-band  and $\gamma$ band, corresponding to the energies marked by the colored rectangles in (b). The length of arrows is proportional to the magnitude of spin expectation value.}
\label{fig:band_spin_w-ter}
\end{figure*}

Figure \ref{fig:structure_001} shows the Type-I slab model consisting of 15 monolayers (ML) thick Cu(001) and a 4 unit-cells-thick WO$_3$(001) layer. The interfacial separation is optimized to 2.70 \AA~(see Supplementary Fig. S2~\citep{supplementary}). We passivated the WO$_3$/vacuum interface with hydrogen to avoid unwanted electronic states. A vacuum region of 20~\AA\ (10~\AA\ on each side) was introduced along the $c$-axis to eliminate spurious interactions between periodic images. The in-plane lattice parameters are set to $a = b = 3.83$~\AA for both Cu and WO$_3$ layers. Consequently, the WO$_3$ layer imposes an in-plane tensile strain of +4.3 $\%$ on the Cu. For the band structure calculations, spin–orbit coupling (SOC)~\citep{relativistic,relativistic-1} is included using $j$-dependent pseudopotentials. Self-consistent-field (SCF) calculation is performed using a $k$-point mesh of 16~$\times$~16~$\times$~1 in the first Brillouin zone. 

Notably, WO$_3$ stabilizes in a monoclinic ($P2_1/n$) phase at room temperature~\citep{wo3_stability}, and yet we have opted for a cubic $Pm\overline{3}m$ phase for calculations. This simplification is based on several factors, first, thin films of WO$_3$ can be epitaxially grown and stabilized in a high-symmetry tetragonal phase ~\citep{wo3_charge,wo3_tetragonal}, yielding a crucial four-fold rotational C$_{4v}$ symmetry of the (001) surface, which we will show plays a crucial role. Second, the cubic phase retains the corner-sharing WO$_{6}$ octahedral framework that governs the fundamental electronic structure, while allowing for the construction of a commensurate heterostructure with the cubic Cu layer. This approach effectively isolates the intrinsic electronic effects of the interface by minimizing complex reconstructions that would arise from a lower-symmetry monoclinic cell. Several earlier DFT-based works had also adopted this simple cubic $Pm\overline{3}m$ phase for investigating bulk properties ~\citep{wo3_dft, Djani2019,W-ter_stabilty}.

\section{Results and Discussion}
Figure ~\ref{fig:band_spin_w-ter} summarizes our key results, where in Fig. \ref{fig:band_spin_w-ter}(a) we show the full electronic structure along high symmetry path: $\overline{X}-\overline{\Gamma}-\overline{X}-\overline{M}-\overline{\Gamma}-\overline{M}$ of the surface Brillouin zone. Upon zooming in along the $\overline{M}-\overline{\Gamma}-\overline{M}$ path, we find three electron-like bands with minima located at approximately 0.25~eV, 0.65~eV, and 0.88~eV below the Fermi level, (Fig.~\ref{fig:band_spin_w-ter}(b)). We hereafter refer to these bands as $\alpha$, $\beta$, and $\gamma$. The $\beta$ and $\gamma$ bands exhibit a clear Rashba spin splitting, while the splitting is very weak for the $\alpha$ band. Spin-projected band analysis shows that only $\langle S_x \rangle$ and $\langle S_y \rangle$ contribute to the splitting, while the contribution from $\langle S_z \rangle$ is negligible (Figs.~\ref{fig:band_spin_w-ter}(c)-(d)). This observation suggests that the inversion symmetry-breaking electric-field-induced Rashba spin-orbit field locks the electron spins in the x-y plane parallel to the interface, yielding finite in-plane and negligible out-of-plane spin components.

The complexities of the in-plane spin-momentum locking, however, are evident from the spin texture plots shown in Figs.~\ref{fig:band_spin_w-ter}(e)-(f). Spin texture near the $\beta$-band bottom (Fig.~\ref{fig:band_spin_w-ter}(e)) exhibits a non-trivial precession, whereby the inner-outer spin contours do not follow the conventional clockwise–anticlockwise chiral pattern, and the projected in-plane spins consist of both tangential and radial components. In contrast, the spin-texture near the $\gamma$ band bottom (Fig.~\ref{fig:band_spin_w-ter}(f)) follows the typical clockwise–anticlockwise spin precession pattern, with a noticeably reduced radial component.

\begin{figure}[htbp]
    \centering
    \includegraphics[scale=0.85]{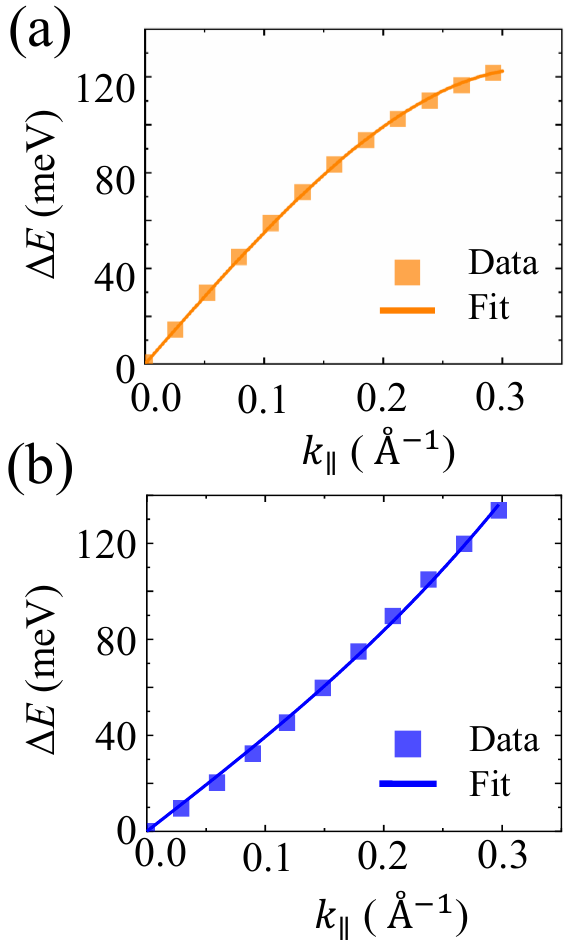}
    \caption{Momentum-resolved energy splitting along the $\bar{\Gamma}$–$\bar{M}$ path for the (a) the $\beta$ band and (orange squares) (b) the $\gamma$ band (blue squares). 
    The solid lines are fit to the model, $\Delta E(k) = 2\alpha_{R}k + 2\tilde{\alpha}_{R}k^{3}$.}
    \label{fig:rashba_parameter_fit}
\end{figure}

The non-trivial spin texture in Figs.~\ref{fig:band_spin_w-ter}(e)-(f) could arise due to a possible interplay between the linear Rashba effect and the cubic Rashba effect ~\citep{STO_cubic, STO_cubic_exp}. To test this hypothesis, we calculated the momentum-resolved energy splitting between the spin-up and spin-down branches: $\Delta E (k) = E_{\downarrow} - E_{\uparrow}$, where $E_{\downarrow}$ and $E_{\uparrow}$ are the energy of spin-down and spin-up branches, respectively. Fig.~\ref{fig:rashba_parameter_fit}(a) plots the extracted splitting, $\Delta E (k)$ for the $\beta$ band along the $\overline{\Gamma}-\overline{M}$ path; meanwhile, the splitting corresponding to the $\gamma$ band is shown in Fig.~\ref{fig:rashba_parameter_fit}(b). $\Delta E (k)$ is non-linear in $k_{||}$ for both bands and can be modeled with a two-parameter equation: $\Delta E (k) = 2\alpha_{R}k + 2\tilde{\alpha}_{R}k^{3}$, where $\alpha_{R}$ and $\tilde{\alpha}_{R}$ refer to the linear and cubic Rashba parameter, respectively. Fitting $\Delta E (k)$ in the $k$-range 0 to 0.3\AA$^{-1}$, we obtained ${\alpha}_{R}$ $\approx$ +0.3 eV$\cdot$\AA, and ${\tilde{\alpha}_{R}}$ $\approx$ -0.9 eV$\cdot$\AA$^{3}$~ for the $\beta$ band; while for the $\gamma$ band these values amount to $\approx$ +0.2 eV$\cdot$\AA~and +0.4 eV$\cdot$\AA$^{3}$, respectively. The linear Rashba parameters extracted from fitting are also consistent with values obtained using the standard formula $\alpha_{R}=2E_{R}/k_{R}$. 

The cubic Rashba contribution, to the best of our knowledge, has never been reported for Cu and Cu-based heterostructures, and it favors triple winding of spin \citep{STO_cubic_exp}. This triple winding, yields a finite radial component of spin. Meanwhile, the conventional linear Rashba effect yields purely tangential spin component. Thus the interplay between linear and cubic Rashba effects could explain the non-trivial spin texture as shown in Fig. \ref{fig:band_spin_w-ter}(e). By the same logic, a reduced cubic contribution in the $\gamma$ band (nearly halved in magnitude compared to the $\beta$ band) also explains why the characteristic spin-texture is more tangential in Fig. \ref{fig:band_spin_w-ter}(f).

\begin{figure*}[htbp]
    \centering
    \includegraphics[scale=0.95]{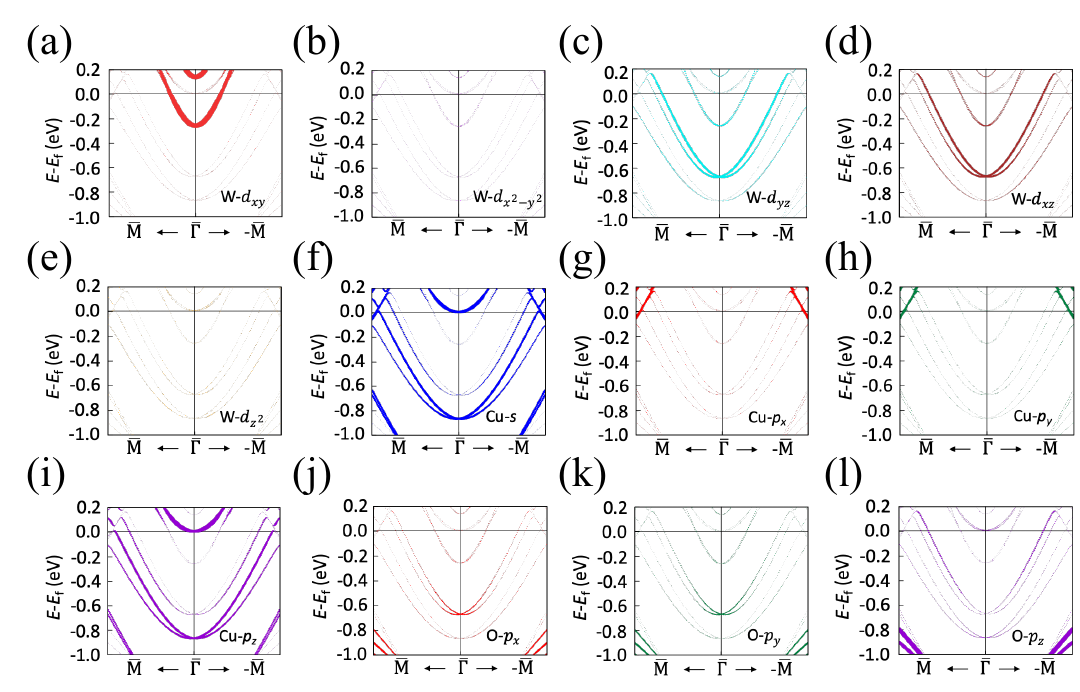}
    \caption{Orbital-resolved band structures. (a)–(e) Projections from the W ($d_{xy}$, $d_{x^{2}-y^{2}}$, $d_{yz}$, $d_{xz}$, $d_{z^{2}}$) orbitals; (f)–(i) projections from Cu ($s$, $p_{x}$, $p_{y}$, $p_{z}$) orbitals; and (j)–(l) projections from O ($p_{x}$, $p_{y}$, $p_{z}$) orbitals. The radii of the circles are  proportional to the weights of the corresponding atomic orbitals.}
    \label{fig:w-ter_orbital_projection}
\end{figure*}

To gain deeper insight into the hierarchical Rashba effect, we first identify the nature of the $\alpha$, $\beta$, and $\gamma$ bands by analyzing their real-space electron density distributions (see Supplementary Fig. S3 in~\citep{supplementary}). Our analysis reveals that, for the $\alpha$ band the electron cloud is confined to the terminating WO$_2$ layer and localized around W atoms. For the $\beta$-band, electrons are localized within the first few layers on either side of the interface, involving W, apical O, and Cu atoms, with the strongest localization around W atoms. The $\gamma$ band is delocalized across the entire Cu thickness, and gradually decays within the first few layers of WO$_3$, exhibiting a sizable localization around W and basal O atoms. Based on these observations, we identify the $\alpha$, $\beta$, and $\gamma$ bands as the 2DEG-like free electronic state, the interface state, and the Cu quantum well state (QWS), respectively.

In line with the above discussions, the orbital-resolved band structures, obtained by projecting the electronic states onto the localized atomic orbitals of all W, Cu, and O atoms in the slab model (Fig.~\ref{fig:w-ter_orbital_projection}), show that the $\alpha$ band has a predominantly W-$d_{xy}$ character. The $\beta$ band, on the other hand, exhibits a mixed character with contributions from W-($d_{xz}$,$d_{yz}$), Cu-($s$,$p_z$), and O-($p$) orbitals. The $\gamma$ band is mainly derived from the Cu-($s$,$p_z$) orbitals, with a minimal contribution from W-($d_{xz}$,$d_{yz}$) and O-($p$) orbitals. The orbital-resolved analysis thus clearly establishes a positive correlation between the W-orbital weight and the magnitudes of the characteristic linear and cubic Rashba parameters of the $\beta$ and $\gamma$ bands.  

In our model heterostructure, the Cu layer forms two interfaces, namely the Cu/WO$_{3}$ and vacuum/Cu interfaces. These two interfaces can interact and thereby influence the Rashba effect. To examine this possibility, we systematically varied the Cu thickness between 5 and 37 ML, while keeping the WO$_{3}$ thickness fixed. Increasing Cu thickness inevitably shifts the $\gamma$ band downward, which prevents a systematic assessment of how thickness affects its spin splitting, so, we restrict our analysis to the $\beta$ band. In addition, for some intermediate thicknesses new QWS emerges and interact with the $\beta$ band, strongly distorting it. Such thicknesses are excluded from our analysis. 

The thickness dependence of the linear ($\alpha_{R}$) and cubic ($\tilde{\alpha}_{R}$) Rashba parameters is summarized in Fig.~\ref{fig:cubic-linear} (see Supplementary Figs.~S4--S5 and Table~S2 in~\citep{supplementary}). The linear term varies only weakly with Cu thickness and saturates at $\approx 0.49$~eV$\cdot$\AA in the thick-film limit. In contrast, the magnitude of the cubic term increases by about a factor of five, from $0.48$ to a giant $1.93$~eV$\cdot$\AA$^{3}$, across the explored thickness range. For the 33‑ML case, a narrow 
$k$-interval where the $\beta$ band is distorted by interaction with a Cu QWS is excluded from the fit (see Fig. S5(e) in~\citep{supplementary}), which leads to the local anomaly in the plotted thickness dependence. The data in Fig.~\ref{fig:cubic-linear} are obtained by fitting $\Delta E(k)$ over $k = 0$--$0.3$~\AA$^{-1}$, but the overall trend remains robust for smaller fitting windows (see Table~S2 in~\citep{supplementary}). 

The pronounced thickness dependence of $\tilde{\alpha}_{R}$ thus signals cross‑coupling between the Cu/WO$_3$ and vacuum/Cu interfaces, which is most significant at smaller Cu thickness. As the Cu layer thickens, enhanced electronic screening progressively suppresses the influence of the vacuum/Cu interface and drives $\tilde{\alpha}_{R}$ toward its intrinsic giant value. We have cross‑checked and validated this trend for two Cu thicknesses, 5 ML and 15 ML, using the Quantum ESPRESSO code~\citep{QE-2009,QE-2017}, with details in~\citep{supplementary} and Supplementary Fig.~S6. Further evidence for interfacial cross‑coupling comes from the real‑space charge distribution of the $\beta$ band for heterostructures with 5‑ML and 15‑ML Cu (Supplementary Fig.~S7 in~\citep{supplementary}); for the 5‑ML film, the charge‑density tail extends from the Cu/WO$_3$ interface to the vacuum/Cu interface, indicating a finite contribution from the latter. In contrast to $\tilde{\alpha}_{R}$, the weak thickness dependence of $\alpha_{R}$ indicates that the linear Rashba term is largely insensitive to this interfacial cross‑coupling.

Overall, the thickness-dependent analysis suggests that for Cu thickness $\geq$ 33 ML, the Cu/WO$_{3}$ and vacuum/Cu interfaces effectively decouple, yielding $\alpha_{R}$ $\approx$ +0.49 eV$\cdot$\AA~and $\tilde{\alpha}_{R}$ $\approx$ -1.9 eV$\cdot$\AA$^{3}$. The $\alpha_{R}$ is significantly larger than those reported for the conducting interface of SrTiO$_{3}$, NdNiO$_{2}$/SrTiO$_{3}$, and NdNiO$_{2}$/KTaO$_{3}$ heterostructures; meanwhile, the giant magnitude of $\tilde{\alpha}_{R}$ is comparable to these systems \citep{STO_nickelate, LaO_STO}. 

\begin{figure}
    \centering
    \includegraphics[scale=0.85]{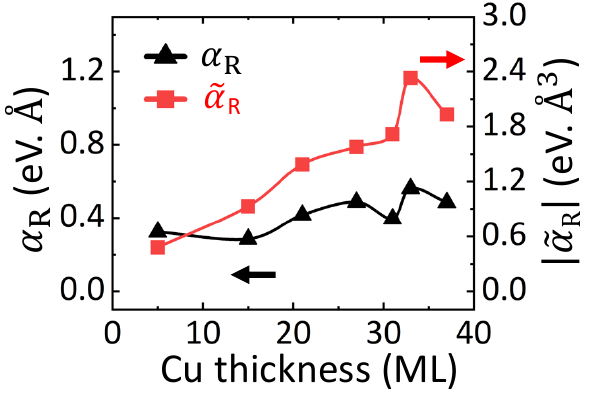}
    \caption{Thickness dependence of the linear (black triangles) and cubic (red squares) Rashba parameters. For clarity, we plotted the magnitude of the cubic Rashba parameter, $|\tilde{\alpha}_{R}|$. The solid lines are guides to the eye. }
    \label{fig:cubic-linear}
\end{figure}

To assess the robustness of the Rashba effect, we have also examined the Type-II and Type-III interface configurations (Supplementary Figs.~S8–S13 in~\citep{supplementary}). For all configurations, the characteristic $\alpha$, $\beta$, and $\gamma$ bands persist. In the decoupled-interface limit, the $\beta$ interface state in Type-II and Type-III still exhibits robust Rashba parameters, with $\alpha_{R} \approx +0.40$~eV$\cdot$\AA~and $\tilde{\alpha}_{R} \approx -1.9$~eV$\cdot$\AA$^{3}$ (Type-II) or $\approx -1.3$~eV$\cdot$\AA$^{3}$ (Type-III) (Tables~S3 and S4 in~\citep{supplementary}). In contrast, the Rashba splitting of the $\gamma$ quantum well state is strongly suppressed in the Type-II configuration, and the corresponding real-space charge density and orbital-resolved analysis reveal a reduced W contribution (Supplementary Figs.~S9–S10 in ~\citep{supplementary}). These trends confirm that robust Rashba effects are generic features of WO$_2$-terminated Cu/WO$_3$ interfaces, while their strengths are tuned by the local W–Cu orbital hybridization.  

Next, we shed light on the underlying mechanism leading to the robust Rashba effects. We first focus on the interfacial chemistry and the large tensile strain acting on the Cu layer. To this end, we performed an additional bandstructure calculation that features an alternative O-terminated interface (Figs.~S14 and S15 in~\citep{supplementary}). This configuration imposes the same  +4.3 \% tensile strain on the Cu layer, as the WO$_{2}$-terminated configuration but inhibits the direct W-Cu orbital hybridization. This system features only Cu-($s$,$p_z$)-derived quantum well states that hybridize with the O-$p_z$ orbitals. The Rashba splitting of these quantum well states are very weak and comparable to that of the bare Cu surface~\citep{Cu_111, Cu_111_exp}. These results unambiguously demonstrates that the W-Cu hybridization is the key enabler for the robust Rashba effects, and that the impact of oxygen atoms and strain are negligible. 

\begin{figure*}[htbp]
    \centering
    \includegraphics[scale=0.7]{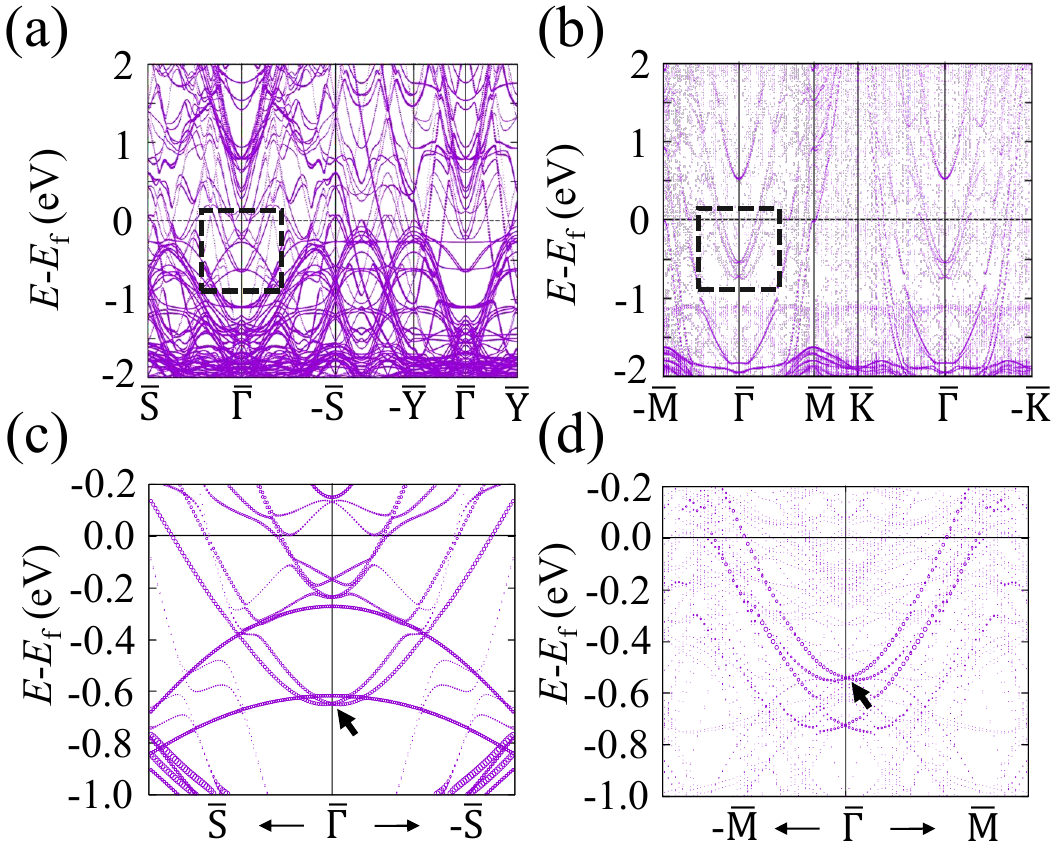}
    \caption{(a),(b) Unfolded band structures of Cu(110)/WO$_3$(001) and Cu(111)/WO$_3$(001) heterostructures, respectively. (c),(d) Magnified views around the $\overline{\Gamma}$ point, as marked by the dashed boxes in (a) and (b). Arrows indicate the interface states showing Rashba spin-splitting.
    }
    \label{fig:bandstructure_110_111}
\end{figure*}

Having established the negligible role of oxygen orbitals, we rationalize our observations in terms of interfacial W–Cu orbital hybridization. In bulk, centrosymmetric and insulating WO$_3$, the crystal field splits the W-$d$ orbitals into $e_g$ and $t_{2g}$ manifolds, with the latter primarily forming the conduction band minimum. In the WO$_2$-terminated heterostructure, the missing apical oxygen selectively redshifts the W-$d_{xz}$, $d_{yz}$, and $d_{z^{2}}$ levels. Consequently, for the Type-I and Type-III configurations, the low-lying W-$d_{xz}$ and $d_{yz}$ orbitals readily overlap with the Cu-$s$ and $p_{z}$ orbitals. This hybridization, acting together with the inversion-symmetry-breaking interfacial electric field, creates the Rashba-split $\beta$ interface state and also imprints a sizable spin splitting on the $\gamma$ quantum well state.

In the Type-II configuration, Cu and W atoms are laterally shifted by half a unit cell, which weakens the overlap between Cu-$s$,$p_{z}$ and W-$d_{xz}$,$d_{yz}$ orbitals and instead enhances hybridization with Cu-$p_{x}$,$p_{y}$ states (Supplementary Fig.~S10 in~\citep{supplementary}). As a result, the $\beta$ interface state retains a robust Rashba effect, whereas the spin splitting of the Cu-$s$,$p_{z}$-dominated $\gamma$ QWS is strongly reduced. For all configurations, the $\alpha$ band, which is primarily derived from W-$d_{xy}$ orbitals, remains only weakly affected by the local inversion-symmetry breaking, shows minimal hybridization with Cu states, and exhibits negligible Rashba splitting.

Regarding the $\beta$ interface state, it is important to note that a bare WO$_{2}$-terminated WO$_{3}$ surface inherently hosts a Rashba spin-split state (Supplementary Fig.~S16 in~\citep{supplementary}) with a robust linear contribution ($\alpha_{R} \approx +0.49$~eV$\cdot$\AA) but a negligible cubic Rashba contribution ($\tilde{\alpha}_{R} \approx -0.07$~eV$\cdot$\AA$^{3}$). This implies that the robust linear Rashba effect of the $\beta$ state is jointly governed by the inversion-symmetry-breaking interfacial electric field and the strong spin-orbit coupling of W atoms. In contrast, the giant cubic contribution requires an additional component: the  interfacial orbital hybridization between Cu and W atoms.     

The linear and cubic Rashba effects observed in the Cu(001)/WO$_3$(001) heterostructures can also be understood within a group-theoretic framework. In bulk, WO$_3$ adopts a non-polar O$_h$ point-group symmetry, which is reduced to a polar C$_{4v}$ symmetry at the WO$_2$-terminated surface. The Cu(001) surface is also C$_{4v}$-symmetric, so the Cu(001)/WO$_3$(001) interface can be described by the same C$_{4v}$ point group. Previous group-theoretical $k\cdot p$ model studies have shown that C$_{4v}$ symmetry explicitly permits the coexistence of both linear and cubic Rashba terms~\citep{spin-variation}. Thus, the simultaneous presence of linear and cubic Rashba effects in our calculations is fully consistent with these symmetry considerations.

To test the universality of the orbital hybridization-induced interface state and Rashba effect therein, we further studied Cu(110)/WO$_3$(001) and Cu(111)/WO$_3$(001) heterostructures. The computational details are given in Supplementary Figs. S17 and S18 and Table S5 ~\citep{supplementary}. To avoid unrealistically long computation times for supercell-based calculations, the thickness of  Cu(110) and Cu(111) is restricted to 5 ML and 4 ML, respectively, and one-unit-cell-thick WO$_3$ layer is used for the Cu(111)/WO$_3$(001) heterostructure. Consequently, we did not attempt to carry out a full thickness-dependent analysis, and restricted our investigation to hybridized interface state only. 

The unfolded band structure of the Cu(110)/WO$_3$(001) heterostructure along the $\overline{S}$–$\overline{\Gamma}$–$\overline{S}$ path exhibits electron-like bands (Fig.~\ref{fig:bandstructure_110_111}(a)). Upon magnifying around the $\overline{\Gamma}$ point, we find a spin-split interface state located at $\approx -0.65$~eV. The Cu(111)/WO$_3$(001) heterostructure likewise hosts a similar Rashba spin-split band at $\approx -0.58$~eV (Figs.~\ref{fig:bandstructure_110_111}(b) and (d)). The interface band for the Cu(110) orientation is primarily composed of W-$d_{xz}$,$d_{yz}$, Cu-$s$,$p_z$, and O-$p$ orbitals (Supplementary Fig.~S19 in~\citep{supplementary}), with the corresponding electron density predominantly localized around the interfacial atoms (Supplementary Fig.~S20 in~\citep{supplementary}). In contrast, for the Cu(111) orientation, contributions from W and O orbitals are comparatively suppressed, while the Cu contribution is noticeably enhanced (Supplementary Fig.~S21 in~\citep{supplementary}). The reduced W and enhanced Cu orbital weights can be attributed to the ultra-thin WO$_3$ layer and the higher planar packing fraction of the Cu(111) surface relative to the other orientations. Like for the other two orientations the real-space electron density analysis shows that the interface state preferably localizes around the interfacial W and Cu atoms (Supplementary Fig.~S22 in~\citep{supplementary}).

By analyzing the momentum-dependent energy splitting of these interface states (Supplementary Figs. S23 and S24 in~\citep{supplementary}), we obtained ${\alpha}_{R}$ $\approx$ +0.3 eV$\cdot$\AA~for both systems, while ${\tilde{\alpha}_{R}}$ amounts to $\approx$ -0.80 eV$\cdot$\AA$^{3}$~ and -0.53 eV$\cdot$\AA$^{3}$~for the Cu(110) and Cu(111) orientations, respectively. The spin texture at the $\overline{\Gamma}$ point for the Cu(110)/WO$_3$(001) heterostructure, reveals a non-trivial winding, confirming the interplay between the linear and cubic Rashba effects (Supplementary Fig. S25 in~\citep{supplementary}). For the Cu(111) orientation, the large supercell size and resulting dense set of folded bands hindered extracting the corresponding spin texture. Nevertheless, we expect a similar non‑trivial spin texture, consistent with the behavior observed for the other orientations. A comprehensive analysis of the Type‑II and Type‑III configurations for both Cu(110) and Cu(111) orientations also reveals interface states with sizable linear and cubic Rashba parameters (Supplementary Figs. S26–S29 and Tables S6–S7 in~\citep{supplementary}).
 
From symmetry considerations the  Cu(110) and Cu(111) surfaces have the C$_{2v}$ and C$_{3v}$ point group symmetries. Taking into consideration the anisotropic in-plane strain acting on these layers (see Table S5 in~\citep{supplementary}), hetero-interfaces with the WO$_3$(001) are expected to have C$_{s}$ or even lower C$_{1}$ symmetry. These low symmetry interfaces should support both linear and cubic Rashba effects like the C$_{4v}$-symmetric Cu(001)/WO$_3$(001) interface \citep{spin-variation}, which is consistent with our calculations. While, the symmetry arguments explain coexisting linear and cubic Rashba effects in these low symmetry interfaces, how these low symmetries influence the magnitude of Rashba parameters can only be understood by comparing the intrinsic Rashba parameters against the Cu(001)/WO$_3$(001) system. However, this requires systematic thickness dependent studies that are computationally intensive, and go beyond the scope of this present study. Overall, the above data unambiguously establish the universality of the linear and cubic Rashba effects in the WO$_2$-terminated Cu/WO$_3$ heterostructures.

Finally, in Table \ref{tab:cubic_rashba_abs} we compare Rashba parameters obtained for the Cu(001)/WO$_3$(001) heterostructures against several material system.

\vspace{-0.2 cm}
\begin{table}[htbp]
\centering
\caption{The comparison of cubic and linear Rashba parameters ($\tilde{\alpha}_R$ and $\alpha_R$) for various materials.}
\begin{tabular}{lccc}
\hline
System  &  $\tilde{\alpha}_R$ (eV$\cdot$\AA$^3$)  &  $\alpha_R$ (eV$\cdot$\AA)  &  Reference \\
\hline
SrTiO$_3$ surface    & 2.74   & 0.02    & \citep{STO_cubic} \\
LaAlO$_3$/SrTiO$_3$  & 4.0   & 0.008 & \citep{LaO_STO} \\
NdNiO$_2$/SrTiO$_3$  & -1.11  &  0.13   & \citep{STO_nickelate}\\
NdNiO$_2$/KTaO$_3$   & -1.09   &  0.12  & \citep{STO_nickelate} \\
Cu/Cu${_3}$N  & -- & 3.80 & \citep{cu/cu3n}\\
\textbf{Cu/WO$_3$} Type-I   & \textbf{-1.93} &  \textbf{ 0.49} & \textbf{This work} \\
\textbf{Cu/WO$_3$} Type-II   & \textbf{-1.86} &  \textbf{ 0.40} & \textbf{This work} \\
\textbf{Cu/WO$_3$} Type-III   & \textbf{-1.34} &  \textbf{ 0.40} & \textbf{This work} \\
\hline
\end{tabular}
\label{tab:cubic_rashba_abs}
\end{table}

\section{Summary and Conclusion}
In summary, we have comprehensively investigated the electronic structure of Cu/WO$_{3}$ interfaces using first-principles density functional theory, incorporating spin-orbit interaction. We show that the hybridization between W and Cu orbitals yields an interface state that exhibits a robust, yet previously unexplored, cubic Rashba effect alongside a large linear Rashba effect. This interface state with coexisting linear and cubic Rashba effects persists across Cu(001)/WO$_{3}$(001), Cu(110)/WO$_{3}$(001), and Cu(111)/WO$_{3}$(001) interfaces, suggesting that it is a generic feature of Cu/WO$_{3}$ interfaces. 

Our work establishes the Cu/WO$_3$ interface as a promising platform for exploring Rashba physics and suggests three design guidelines for optimizing spin-to-charge conversion. First, interface engineering via non-equilibrium growth techniques such as pulsed laser deposition or sputtering should be employed to kinetically stabilize the highly Rashba-active configuration, in which Cu atoms sit beneath W atoms, overcoming thermodynamic preferences for other configurations. Second, to suppress parasitic cross-coupling between the top and bottom Cu interfaces, the Cu layer thickness should be maintained at about 12 nm (thirty-three monolayers) or greater so that the system operates in the fully decoupled limit. Finally, within this decoupled and highly Rashba-active architecture, the Cu thickness can be tuned to exploit both the maximally Rashba-split interface state and the proximity-split Cu quantum-well states concurrently, thereby generating large spin currents from complex, non-trivial spin textures.

Moreover, our results qualitatively support the experimentally observed robust spin-current generation at W-rich interfaces in Bi$_2$WO$_6$-based heterostructures~\citep{das2023observation,itoh2024enhancement}. From an application standpoint, both WO$_3$ and Cu thin films can be readily grown using standard physical vapor deposition and atomic layer deposition techniques and are compatible with back-end-of-line (BEOL) CMOS processes, highlighting a realistic pathway toward technological integration. We hope that this work will stimulate experimental studies in which spin-orbit torque ferromagnetic resonance (ST-FMR) and spin-pumping measurements are used to harness the predicted robust Rashba spin splitting, while non-linear Hall measurements serve as a sensitive probe of the cubic Rashba effect.

\begin{acknowledgments}
Saikat Das acknowledges financial support from the ANRF, erstwhile SERB grant CRG/2022/008740. Md Aktar Hossain acknowledges the PMRF grant for providing his fellowship. Md Aktar Hossain acknowledges discussions with Dr. Tilak Das during the early stage of this project. We acknowledge National Supercomputing Mission (NSM) for providing computing resources of “PARAM Shakti” at IIT Kharagpur, which is implemented by C-DAC and supported by the Ministry of Electronics and Information Technology (MeitY) and Department of Science and Technology (DST), Government of India. 
\end{acknowledgments}

%\bibliographystyle{apsrev4-2}
%\bibliography{references}

\begin{thebibliography}{47}%
\makeatletter
\providecommand \@ifxundefined [1]{%
 \@ifx{#1\undefined}
}%
\providecommand \@ifnum [1]{%
 \ifnum #1\expandafter \@firstoftwo
 \else \expandafter \@secondoftwo
 \fi
}%
\providecommand \@ifx [1]{%
 \ifx #1\expandafter \@firstoftwo
 \else \expandafter \@secondoftwo
 \fi
}%
\providecommand \natexlab [1]{#1}%
\providecommand \enquote  [1]{``#1''}%
\providecommand \bibnamefont  [1]{#1}%
\providecommand \bibfnamefont [1]{#1}%
\providecommand \citenamefont [1]{#1}%
\providecommand \href@noop [0]{\@secondoftwo}%
\providecommand \href [0]{\begingroup \@sanitize@url \@href}%
\providecommand \@href[1]{\@@startlink{#1}\@@href}%
\providecommand \@@href[1]{\endgroup#1\@@endlink}%
\providecommand \@sanitize@url [0]{\catcode `\\12\catcode `\$12\catcode `\&12\catcode `\#12\catcode `\^12\catcode `\_12\catcode `\%12\relax}%
\providecommand \@@startlink[1]{}%
\providecommand \@@endlink[0]{}%
\providecommand \url  [0]{\begingroup\@sanitize@url \@url }%
\providecommand \@url [1]{\endgroup\@href {#1}{\urlprefix }}%
\providecommand \urlprefix  [0]{URL }%
\providecommand \Eprint [0]{\href }%
\providecommand \doibase [0]{https://doi.org/}%
\providecommand \selectlanguage [0]{\@gobble}%
\providecommand \bibinfo  [0]{\@secondoftwo}%
\providecommand \bibfield  [0]{\@secondoftwo}%
\providecommand \translation [1]{[#1]}%
\providecommand \BibitemOpen [0]{}%
\providecommand \bibitemStop [0]{}%
\providecommand \bibitemNoStop [0]{.\EOS\space}%
\providecommand \EOS [0]{\spacefactor3000\relax}%
\providecommand \BibitemShut  [1]{\csname bibitem#1\endcsname}%
\let\auto@bib@innerbib\@empty
%</preamble>
\bibitem [{\citenamefont {Manchon}\ \emph {et~al.}(2019)\citenamefont {Manchon}, \citenamefont {\ifmmode~\check{Z}\else \v{Z}\fi{}elezn\'y}, \citenamefont {Miron}, \citenamefont {Jungwirth}, \citenamefont {Sinova}, \citenamefont {Thiaville}, \citenamefont {Garello},\ and\ \citenamefont {Gambardella}}]{RevModPhys.91.035004}%
  \BibitemOpen
  \bibfield  {author} {\bibinfo {author} {\bibfnamefont {A.}~\bibnamefont {Manchon}}, \bibinfo {author} {\bibfnamefont {J.}~\bibnamefont {\ifmmode~\check{Z}\else \v{Z}\fi{}elezn\'y}}, \bibinfo {author} {\bibfnamefont {I.~M.}\ \bibnamefont {Miron}}, \bibinfo {author} {\bibfnamefont {T.}~\bibnamefont {Jungwirth}}, \bibinfo {author} {\bibfnamefont {J.}~\bibnamefont {Sinova}}, \bibinfo {author} {\bibfnamefont {A.}~\bibnamefont {Thiaville}}, \bibinfo {author} {\bibfnamefont {K.}~\bibnamefont {Garello}},\ and\ \bibinfo {author} {\bibfnamefont {P.}~\bibnamefont {Gambardella}},\ }\href {https://doi.org/10.1103/RevModPhys.91.035004} {\bibfield  {journal} {\bibinfo  {journal} {Rev. Mod. Phys.}\ }\textbf {\bibinfo {volume} {91}},\ \bibinfo {pages} {035004} (\bibinfo {year} {2019})}\BibitemShut {NoStop}%
\bibitem [{\citenamefont {Manipatruni}\ \emph {et~al.}(2019)\citenamefont {Manipatruni}, \citenamefont {Nikonov}, \citenamefont {Lin}, \citenamefont {Gosavi}, \citenamefont {Liu}, \citenamefont {Prasad}, \citenamefont {Huang}, \citenamefont {Bonturim}, \citenamefont {Ramesh},\ and\ \citenamefont {Young}}]{manipatruni2019scalable}%
  \BibitemOpen
  \bibfield  {author} {\bibinfo {author} {\bibfnamefont {S.}~\bibnamefont {Manipatruni}}, \bibinfo {author} {\bibfnamefont {D.~E.}\ \bibnamefont {Nikonov}}, \bibinfo {author} {\bibfnamefont {C.-C.}\ \bibnamefont {Lin}}, \bibinfo {author} {\bibfnamefont {T.~A.}\ \bibnamefont {Gosavi}}, \bibinfo {author} {\bibfnamefont {H.}~\bibnamefont {Liu}}, \bibinfo {author} {\bibfnamefont {B.}~\bibnamefont {Prasad}}, \bibinfo {author} {\bibfnamefont {Y.-L.}\ \bibnamefont {Huang}}, \bibinfo {author} {\bibfnamefont {E.}~\bibnamefont {Bonturim}}, \bibinfo {author} {\bibfnamefont {R.}~\bibnamefont {Ramesh}},\ and\ \bibinfo {author} {\bibfnamefont {I.~A.}\ \bibnamefont {Young}},\ }\href {https://doi.org/10.1038/s41586-018-0770-2} {\bibfield  {journal} {\bibinfo  {journal} {Nature}\ }\textbf {\bibinfo {volume} {565}},\ \bibinfo {pages} {35} (\bibinfo {year} {2019})}\BibitemShut {NoStop}%
\bibitem [{\citenamefont {Seifert}\ \emph {et~al.}(2016)\citenamefont {Seifert}, \citenamefont {Jaiswal}, \citenamefont {Martens}, \citenamefont {Hannegan}, \citenamefont {Braun}, \citenamefont {Maldonado}, \citenamefont {Freimuth}, \citenamefont {Kronenberg}, \citenamefont {Henrizi}, \citenamefont {Radu}, \citenamefont {Beaurepaire}, \citenamefont {Mokrousov}, \citenamefont {Oppeneer}, \citenamefont {Jourdan}, \citenamefont {Jakob}, \citenamefont {Turchinovich}, \citenamefont {Hayden}, \citenamefont {Wolf}, \citenamefont {M{\"u}nzenberg}, \citenamefont {Kl{\"a}ui},\ and\ \citenamefont {Kampfrath}}]{Seifert2016}%
  \BibitemOpen
  \bibfield  {author} {\bibinfo {author} {\bibfnamefont {T.}~\bibnamefont {Seifert}}, \bibinfo {author} {\bibfnamefont {S.}~\bibnamefont {Jaiswal}}, \bibinfo {author} {\bibfnamefont {U.}~\bibnamefont {Martens}}, \bibinfo {author} {\bibfnamefont {J.}~\bibnamefont {Hannegan}}, \bibinfo {author} {\bibfnamefont {L.}~\bibnamefont {Braun}}, \bibinfo {author} {\bibfnamefont {P.}~\bibnamefont {Maldonado}}, \bibinfo {author} {\bibfnamefont {F.}~\bibnamefont {Freimuth}}, \bibinfo {author} {\bibfnamefont {A.}~\bibnamefont {Kronenberg}}, \bibinfo {author} {\bibfnamefont {J.}~\bibnamefont {Henrizi}}, \bibinfo {author} {\bibfnamefont {I.}~\bibnamefont {Radu}}, \bibinfo {author} {\bibfnamefont {E.}~\bibnamefont {Beaurepaire}}, \bibinfo {author} {\bibfnamefont {Y.}~\bibnamefont {Mokrousov}}, \bibinfo {author} {\bibfnamefont {P.~M.}\ \bibnamefont {Oppeneer}}, \bibinfo {author} {\bibfnamefont {M.}~\bibnamefont {Jourdan}}, \bibinfo {author} {\bibfnamefont {G.}~\bibnamefont {Jakob}}, \bibinfo {author} {\bibfnamefont
  {D.}~\bibnamefont {Turchinovich}}, \bibinfo {author} {\bibfnamefont {L.~M.}\ \bibnamefont {Hayden}}, \bibinfo {author} {\bibfnamefont {M.}~\bibnamefont {Wolf}}, \bibinfo {author} {\bibfnamefont {M.}~\bibnamefont {M{\"u}nzenberg}}, \bibinfo {author} {\bibfnamefont {M.}~\bibnamefont {Kl{\"a}ui}},\ and\ \bibinfo {author} {\bibfnamefont {T.}~\bibnamefont {Kampfrath}},\ }\href {https://doi.org/10.1038/nphoton.2016.91} {\bibfield  {journal} {\bibinfo  {journal} {Nat. Photonics}\ }\textbf {\bibinfo {volume} {10}},\ \bibinfo {pages} {483} (\bibinfo {year} {2016})}\BibitemShut {NoStop}%
\bibitem [{\citenamefont {Ju}\ \emph {et~al.}(2022)\citenamefont {Ju}, \citenamefont {Xu}, \citenamefont {Li}, \citenamefont {Li}, \citenamefont {Tian}, \citenamefont {Chen},\ and\ \citenamefont {Li}}]{Rashba}%
  \BibitemOpen
  \bibfield  {author} {\bibinfo {author} {\bibfnamefont {W.}~\bibnamefont {Ju}}, \bibinfo {author} {\bibfnamefont {Y.}~\bibnamefont {Xu}}, \bibinfo {author} {\bibfnamefont {T.}~\bibnamefont {Li}}, \bibinfo {author} {\bibfnamefont {M.}~\bibnamefont {Li}}, \bibinfo {author} {\bibfnamefont {K.}~\bibnamefont {Tian}}, \bibinfo {author} {\bibfnamefont {J.}~\bibnamefont {Chen}},\ and\ \bibinfo {author} {\bibfnamefont {H.}~\bibnamefont {Li}},\ }\href {https://doi.org/https://doi.org/10.1016/j.apsusc.2022.153528} {\bibfield  {journal} {\bibinfo  {journal} {Appl. Surf. Sci.}\ }\textbf {\bibinfo {volume} {595}},\ \bibinfo {pages} {153528} (\bibinfo {year} {2022})}\BibitemShut {NoStop}%
\bibitem [{\citenamefont {Rashba}(1960)}]{rashba1960properties}%
  \BibitemOpen
  \bibfield  {author} {\bibinfo {author} {\bibfnamefont {E.~I.}\ \bibnamefont {Rashba}},\ }\href@noop {} {\bibfield  {journal} {\bibinfo  {journal} {Sov. Phys. Solid State}\ }\textbf {\bibinfo {volume} {2}},\ \bibinfo {pages} {1109} (\bibinfo {year} {1960})}\BibitemShut {NoStop}%
\bibitem [{\citenamefont {Wu}\ \emph {et~al.}(2021)\citenamefont {Wu}, \citenamefont {Garello}, \citenamefont {Kim}, \citenamefont {Gupta}, \citenamefont {Perumkunnil}, \citenamefont {Kateel}, \citenamefont {Couet}, \citenamefont {Carpenter}, \citenamefont {Rao}, \citenamefont {Van~Beek}, \citenamefont {Vudya~Sethu}, \citenamefont {Yasin}, \citenamefont {Crotti},\ and\ \citenamefont {Kar}}]{PhysRevApplied.15.064015}%
  \BibitemOpen
  \bibfield  {author} {\bibinfo {author} {\bibfnamefont {Y.}~\bibnamefont {Wu}}, \bibinfo {author} {\bibfnamefont {K.}~\bibnamefont {Garello}}, \bibinfo {author} {\bibfnamefont {W.}~\bibnamefont {Kim}}, \bibinfo {author} {\bibfnamefont {M.}~\bibnamefont {Gupta}}, \bibinfo {author} {\bibfnamefont {M.}~\bibnamefont {Perumkunnil}}, \bibinfo {author} {\bibfnamefont {V.}~\bibnamefont {Kateel}}, \bibinfo {author} {\bibfnamefont {S.}~\bibnamefont {Couet}}, \bibinfo {author} {\bibfnamefont {R.}~\bibnamefont {Carpenter}}, \bibinfo {author} {\bibfnamefont {S.}~\bibnamefont {Rao}}, \bibinfo {author} {\bibfnamefont {S.}~\bibnamefont {Van~Beek}}, \bibinfo {author} {\bibfnamefont {K.}~\bibnamefont {Vudya~Sethu}}, \bibinfo {author} {\bibfnamefont {F.}~\bibnamefont {Yasin}}, \bibinfo {author} {\bibfnamefont {D.}~\bibnamefont {Crotti}},\ and\ \bibinfo {author} {\bibfnamefont {G.}~\bibnamefont {Kar}},\ }\href {https://doi.org/10.1103/PhysRevApplied.15.064015} {\bibfield  {journal} {\bibinfo  {journal} {Phys. Rev. Appl.}\
  }\textbf {\bibinfo {volume} {15}},\ \bibinfo {pages} {064015} (\bibinfo {year} {2021})}\BibitemShut {NoStop}%
\bibitem [{\citenamefont {Bihlmayer}\ \emph {et~al.}(2022)\citenamefont {Bihlmayer}, \citenamefont {No{\"e}l}, \citenamefont {Vyalikh}, \citenamefont {Chulkov},\ and\ \citenamefont {Manchon}}]{Quantum_metric}%
  \BibitemOpen
  \bibfield  {author} {\bibinfo {author} {\bibfnamefont {G.}~\bibnamefont {Bihlmayer}}, \bibinfo {author} {\bibfnamefont {P.}~\bibnamefont {No{\"e}l}}, \bibinfo {author} {\bibfnamefont {D.~V.}\ \bibnamefont {Vyalikh}}, \bibinfo {author} {\bibfnamefont {E.~V.}\ \bibnamefont {Chulkov}},\ and\ \bibinfo {author} {\bibfnamefont {A.}~\bibnamefont {Manchon}},\ }\href {https://doi.org/10.1038/s42254-022-00490-y} {\bibfield  {journal} {\bibinfo  {journal} {Nat. Rev. Phys.}\ }\textbf {\bibinfo {volume} {4}},\ \bibinfo {pages} {642} (\bibinfo {year} {2022})}\BibitemShut {NoStop}%
\bibitem [{\citenamefont {Ohtomo}\ and\ \citenamefont {Hwang}(2004)}]{Ohtomo2004}%
  \BibitemOpen
  \bibfield  {author} {\bibinfo {author} {\bibfnamefont {A.}~\bibnamefont {Ohtomo}}\ and\ \bibinfo {author} {\bibfnamefont {H.~Y.}\ \bibnamefont {Hwang}},\ }\href {https://doi.org/10.1038/nature02308} {\bibfield  {journal} {\bibinfo  {journal} {Nature}\ }\textbf {\bibinfo {volume} {427}},\ \bibinfo {pages} {423} (\bibinfo {year} {2004})}\BibitemShut {NoStop}%
\bibitem [{\citenamefont {Lesne}\ \emph {et~al.}(2016)\citenamefont {Lesne}, \citenamefont {Fu}, \citenamefont {Oyarzun}, \citenamefont {Rojas-S{\'a}nchez}, \citenamefont {Vaz}, \citenamefont {Naganuma}, \citenamefont {Sicoli}, \citenamefont {Attan{\'e}}, \citenamefont {Jamet}, \citenamefont {Jacquet}, \citenamefont {George}, \citenamefont {Barth{\'e}l{\'e}my}, \citenamefont {Jaffr{\`e}s}, \citenamefont {Fert}, \citenamefont {Bibes},\ and\ \citenamefont {Vila}}]{LaAlO3_STO}%
  \BibitemOpen
  \bibfield  {author} {\bibinfo {author} {\bibfnamefont {E.}~\bibnamefont {Lesne}}, \bibinfo {author} {\bibfnamefont {Y.}~\bibnamefont {Fu}}, \bibinfo {author} {\bibfnamefont {S.}~\bibnamefont {Oyarzun}}, \bibinfo {author} {\bibfnamefont {J.~C.}\ \bibnamefont {Rojas-S{\'a}nchez}}, \bibinfo {author} {\bibfnamefont {D.~C.}\ \bibnamefont {Vaz}}, \bibinfo {author} {\bibfnamefont {H.}~\bibnamefont {Naganuma}}, \bibinfo {author} {\bibfnamefont {G.}~\bibnamefont {Sicoli}}, \bibinfo {author} {\bibfnamefont {J.-P.}\ \bibnamefont {Attan{\'e}}}, \bibinfo {author} {\bibfnamefont {M.}~\bibnamefont {Jamet}}, \bibinfo {author} {\bibfnamefont {E.}~\bibnamefont {Jacquet}}, \bibinfo {author} {\bibfnamefont {J.-M.}\ \bibnamefont {George}}, \bibinfo {author} {\bibfnamefont {A.}~\bibnamefont {Barth{\'e}l{\'e}my}}, \bibinfo {author} {\bibfnamefont {H.}~\bibnamefont {Jaffr{\`e}s}}, \bibinfo {author} {\bibfnamefont {A.}~\bibnamefont {Fert}}, \bibinfo {author} {\bibfnamefont {M.}~\bibnamefont {Bibes}},\ and\ \bibinfo {author}
  {\bibfnamefont {L.}~\bibnamefont {Vila}},\ }\href {https://doi.org/10.1038/nmat4726} {\bibfield  {journal} {\bibinfo  {journal} {Nat. Mater.}\ }\textbf {\bibinfo {volume} {15}},\ \bibinfo {pages} {1261} (\bibinfo {year} {2016})}\BibitemShut {NoStop}%
\bibitem [{\citenamefont {Al-Tawhid}\ \emph {et~al.}(2025)\citenamefont {Al-Tawhid}, \citenamefont {Sun}, \citenamefont {Comstock}, \citenamefont {Kumah}, \citenamefont {Sun},\ and\ \citenamefont {Ahadi}}]{KTaO3}%
  \BibitemOpen
  \bibfield  {author} {\bibinfo {author} {\bibfnamefont {A.~H.}\ \bibnamefont {Al-Tawhid}}, \bibinfo {author} {\bibfnamefont {R.}~\bibnamefont {Sun}}, \bibinfo {author} {\bibfnamefont {A.~H.}\ \bibnamefont {Comstock}}, \bibinfo {author} {\bibfnamefont {D.~P.}\ \bibnamefont {Kumah}}, \bibinfo {author} {\bibfnamefont {D.}~\bibnamefont {Sun}},\ and\ \bibinfo {author} {\bibfnamefont {K.}~\bibnamefont {Ahadi}},\ }\href {https://doi.org/10.1063/5.0247001} {\bibfield  {journal} {\bibinfo  {journal} {Appl. Phys. Lett.}\ }\textbf {\bibinfo {volume} {126}},\ \bibinfo {pages} {091601} (\bibinfo {year} {2025})}\BibitemShut {NoStop}%
\bibitem [{\citenamefont {Bordoloi}\ \emph {et~al.}(2024)\citenamefont {Bordoloi}, \citenamefont {Garcia-Castro}, \citenamefont {Romestan}, \citenamefont {Romero},\ and\ \citenamefont {Singh}}]{two-dimensional}%
  \BibitemOpen
  \bibfield  {author} {\bibinfo {author} {\bibfnamefont {A.}~\bibnamefont {Bordoloi}}, \bibinfo {author} {\bibfnamefont {A.~C.}\ \bibnamefont {Garcia-Castro}}, \bibinfo {author} {\bibfnamefont {Z.}~\bibnamefont {Romestan}}, \bibinfo {author} {\bibfnamefont {A.~H.}\ \bibnamefont {Romero}},\ and\ \bibinfo {author} {\bibfnamefont {S.}~\bibnamefont {Singh}},\ }\href {https://doi.org/10.1063/5.0212170} {\bibfield  {journal} {\bibinfo  {journal} {J. Appl. Phys.}\ }\textbf {\bibinfo {volume} {135}},\ \bibinfo {pages} {220901} (\bibinfo {year} {2024})}\BibitemShut {NoStop}%
\bibitem [{\citenamefont {Usachov}\ \emph {et~al.}(2020)\citenamefont {Usachov}, \citenamefont {Nechaev}, \citenamefont {Poelchen}, \citenamefont {G\"uttler}, \citenamefont {Krasovskii}, \citenamefont {Schulz}, \citenamefont {Generalov}, \citenamefont {Kliemt}, \citenamefont {Kraiker}, \citenamefont {Krellner}, \citenamefont {Kummer}, \citenamefont {Danzenb\"acher}, \citenamefont {Laubschat}, \citenamefont {Weber}, \citenamefont {S\'anchez-Barriga}, \citenamefont {Chulkov}, \citenamefont {Santander-Syro}, \citenamefont {Imai}, \citenamefont {Miyamoto}, \citenamefont {Okuda},\ and\ \citenamefont {Vyalikh}}]{si-terminated}%
  \BibitemOpen
  \bibfield  {author} {\bibinfo {author} {\bibfnamefont {D.~Y.}\ \bibnamefont {Usachov}}, \bibinfo {author} {\bibfnamefont {I.~A.}\ \bibnamefont {Nechaev}}, \bibinfo {author} {\bibfnamefont {G.}~\bibnamefont {Poelchen}}, \bibinfo {author} {\bibfnamefont {M.}~\bibnamefont {G\"uttler}}, \bibinfo {author} {\bibfnamefont {E.~E.}\ \bibnamefont {Krasovskii}}, \bibinfo {author} {\bibfnamefont {S.}~\bibnamefont {Schulz}}, \bibinfo {author} {\bibfnamefont {A.}~\bibnamefont {Generalov}}, \bibinfo {author} {\bibfnamefont {K.}~\bibnamefont {Kliemt}}, \bibinfo {author} {\bibfnamefont {A.}~\bibnamefont {Kraiker}}, \bibinfo {author} {\bibfnamefont {C.}~\bibnamefont {Krellner}}, \bibinfo {author} {\bibfnamefont {K.}~\bibnamefont {Kummer}}, \bibinfo {author} {\bibfnamefont {S.}~\bibnamefont {Danzenb\"acher}}, \bibinfo {author} {\bibfnamefont {C.}~\bibnamefont {Laubschat}}, \bibinfo {author} {\bibfnamefont {A.~P.}\ \bibnamefont {Weber}}, \bibinfo {author} {\bibfnamefont {J.}~\bibnamefont {S\'anchez-Barriga}}, \bibinfo {author}
  {\bibfnamefont {E.~V.}\ \bibnamefont {Chulkov}}, \bibinfo {author} {\bibfnamefont {A.~F.}\ \bibnamefont {Santander-Syro}}, \bibinfo {author} {\bibfnamefont {T.}~\bibnamefont {Imai}}, \bibinfo {author} {\bibfnamefont {K.}~\bibnamefont {Miyamoto}}, \bibinfo {author} {\bibfnamefont {T.}~\bibnamefont {Okuda}},\ and\ \bibinfo {author} {\bibfnamefont {D.~V.}\ \bibnamefont {Vyalikh}},\ }\href {https://doi.org/10.1103/PhysRevLett.124.237202} {\bibfield  {journal} {\bibinfo  {journal} {Phys. Rev. Lett.}\ }\textbf {\bibinfo {volume} {124}},\ \bibinfo {pages} {237202} (\bibinfo {year} {2020})}\BibitemShut {NoStop}%
\bibitem [{\citenamefont {Tsai}\ \emph {et~al.}(2018)\citenamefont {Tsai}, \citenamefont {Karube}, \citenamefont {Kondou}, \citenamefont {Yamaguchi}, \citenamefont {Ishii},\ and\ \citenamefont {Otani}}]{cu-bi2o3}%
  \BibitemOpen
  \bibfield  {author} {\bibinfo {author} {\bibfnamefont {H.}~\bibnamefont {Tsai}}, \bibinfo {author} {\bibfnamefont {S.}~\bibnamefont {Karube}}, \bibinfo {author} {\bibfnamefont {K.}~\bibnamefont {Kondou}}, \bibinfo {author} {\bibfnamefont {N.}~\bibnamefont {Yamaguchi}}, \bibinfo {author} {\bibfnamefont {F.}~\bibnamefont {Ishii}},\ and\ \bibinfo {author} {\bibfnamefont {Y.}~\bibnamefont {Otani}},\ }\href {https://doi.org/10.1038/s41598-018-23787-4} {\bibfield  {journal} {\bibinfo  {journal} {Sci. Rep.}\ }\textbf {\bibinfo {volume} {8}},\ \bibinfo {pages} {5564} (\bibinfo {year} {2018})}\BibitemShut {NoStop}%
\bibitem [{\citenamefont {Sugimoto}\ \emph {et~al.}(2019)\citenamefont {Sugimoto}, \citenamefont {Uzuhashi}, \citenamefont {Isogami}, \citenamefont {Ohkubo}, \citenamefont {Takahashi}, \citenamefont {Kasai},\ and\ \citenamefont {Hono}}]{metal_Bi}%
  \BibitemOpen
  \bibfield  {author} {\bibinfo {author} {\bibfnamefont {S.}~\bibnamefont {Sugimoto}}, \bibinfo {author} {\bibfnamefont {J.}~\bibnamefont {Uzuhashi}}, \bibinfo {author} {\bibfnamefont {S.}~\bibnamefont {Isogami}}, \bibinfo {author} {\bibfnamefont {T.}~\bibnamefont {Ohkubo}}, \bibinfo {author} {\bibfnamefont {Y.~K.}\ \bibnamefont {Takahashi}}, \bibinfo {author} {\bibfnamefont {S.}~\bibnamefont {Kasai}},\ and\ \bibinfo {author} {\bibfnamefont {K.}~\bibnamefont {Hono}},\ }\href {https://doi.org/10.1103/PhysRevMaterials.3.104410} {\bibfield  {journal} {\bibinfo  {journal} {Phys. Rev. Mater.}\ }\textbf {\bibinfo {volume} {3}},\ \bibinfo {pages} {104410} (\bibinfo {year} {2019})}\BibitemShut {NoStop}%
\bibitem [{\citenamefont {Das}\ \emph {et~al.}(2023)\citenamefont {Das}, \citenamefont {Sugimoto}, \citenamefont {Kushwaha}, \citenamefont {Kozuka},\ and\ \citenamefont {Kasai}}]{das2023observation}%
  \BibitemOpen
  \bibfield  {author} {\bibinfo {author} {\bibfnamefont {S.}~\bibnamefont {Das}}, \bibinfo {author} {\bibfnamefont {S.}~\bibnamefont {Sugimoto}}, \bibinfo {author} {\bibfnamefont {V.~K.}\ \bibnamefont {Kushwaha}}, \bibinfo {author} {\bibfnamefont {Y.}~\bibnamefont {Kozuka}},\ and\ \bibinfo {author} {\bibfnamefont {S.}~\bibnamefont {Kasai}},\ }\bibfield  {journal} {\bibinfo  {journal} {APL Mater.}\ }\textbf {\bibinfo {volume} {11}},\ \href {https://doi.org/10.1063/5.0142695} {10.1063/5.0142695} (\bibinfo {year} {2023})\BibitemShut {NoStop}%
\bibitem [{\citenamefont {Itoh}\ \emph {et~al.}(2024)\citenamefont {Itoh}, \citenamefont {Kozuka}, \citenamefont {Hirai},\ and\ \citenamefont {Uchida}}]{itoh2024enhancement}%
  \BibitemOpen
  \bibfield  {author} {\bibinfo {author} {\bibfnamefont {T.}~\bibnamefont {Itoh}}, \bibinfo {author} {\bibfnamefont {Y.}~\bibnamefont {Kozuka}}, \bibinfo {author} {\bibfnamefont {T.}~\bibnamefont {Hirai}},\ and\ \bibinfo {author} {\bibfnamefont {K.-i.}\ \bibnamefont {Uchida}},\ }\href {https://doi.org/https://doi.org/10.1002/adfm.202409557} {\bibfield  {journal} {\bibinfo  {journal} {Adv. Funct. Mater.}\ }\textbf {\bibinfo {volume} {34}},\ \bibinfo {pages} {2409557} (\bibinfo {year} {2024})}\BibitemShut {NoStop}%
\bibitem [{\citenamefont {Djani}\ \emph {et~al.}(2019)\citenamefont {Djani}, \citenamefont {Garcia-Castro}, \citenamefont {Tong}, \citenamefont {Barone}, \citenamefont {Bousquet}, \citenamefont {Picozzi},\ and\ \citenamefont {Ghosez}}]{Djani2019}%
  \BibitemOpen
  \bibfield  {author} {\bibinfo {author} {\bibfnamefont {H.}~\bibnamefont {Djani}}, \bibinfo {author} {\bibfnamefont {A.~C.}\ \bibnamefont {Garcia-Castro}}, \bibinfo {author} {\bibfnamefont {W.-Y.}\ \bibnamefont {Tong}}, \bibinfo {author} {\bibfnamefont {P.}~\bibnamefont {Barone}}, \bibinfo {author} {\bibfnamefont {E.}~\bibnamefont {Bousquet}}, \bibinfo {author} {\bibfnamefont {S.}~\bibnamefont {Picozzi}},\ and\ \bibinfo {author} {\bibfnamefont {P.}~\bibnamefont {Ghosez}},\ }\href {https://doi.org/10.1038/s41535-019-0190-z} {\bibfield  {journal} {\bibinfo  {journal} {npj Quantum Mater.}\ }\textbf {\bibinfo {volume} {4}},\ \bibinfo {pages} {51} (\bibinfo {year} {2019})}\BibitemShut {NoStop}%
\bibitem [{\citenamefont {Jeong}\ \emph {et~al.}(2021)\citenamefont {Jeong}, \citenamefont {Mun}, \citenamefont {Das}, \citenamefont {Kim}, \citenamefont {Kim}, \citenamefont {Peng}, \citenamefont {Kim},\ and\ \citenamefont {Noh}}]{acs_applied}%
  \BibitemOpen
  \bibfield  {author} {\bibinfo {author} {\bibfnamefont {J.}~\bibnamefont {Jeong}}, \bibinfo {author} {\bibfnamefont {J.}~\bibnamefont {Mun}}, \bibinfo {author} {\bibfnamefont {S.}~\bibnamefont {Das}}, \bibinfo {author} {\bibfnamefont {J.}~\bibnamefont {Kim}}, \bibinfo {author} {\bibfnamefont {J.~R.}\ \bibnamefont {Kim}}, \bibinfo {author} {\bibfnamefont {W.}~\bibnamefont {Peng}}, \bibinfo {author} {\bibfnamefont {M.}~\bibnamefont {Kim}},\ and\ \bibinfo {author} {\bibfnamefont {T.~W.}\ \bibnamefont {Noh}},\ }\href {https://doi.org/10.1021/acsaelm.1c00005} {\bibfield  {journal} {\bibinfo  {journal} {ACS Appl. Electron. Mater.}\ }\textbf {\bibinfo {volume} {3}},\ \bibinfo {pages} {1023} (\bibinfo {year} {2021})}\BibitemShut {NoStop}%
\bibitem [{\citenamefont {Ozaki}(2003)}]{ozaki2003variationally}%
  \BibitemOpen
  \bibfield  {author} {\bibinfo {author} {\bibfnamefont {T.}~\bibnamefont {Ozaki}},\ }\href {https://doi.org/10.1103/PhysRevB.67.155108} {\bibfield  {journal} {\bibinfo  {journal} {Phys. Rev. B}\ }\textbf {\bibinfo {volume} {67}},\ \bibinfo {pages} {155108} (\bibinfo {year} {2003})}\BibitemShut {NoStop}%
\bibitem [{\citenamefont {Ozaki}\ and\ \citenamefont {Kino}(2004)}]{ozaki2004numerical}%
  \BibitemOpen
  \bibfield  {author} {\bibinfo {author} {\bibfnamefont {T.}~\bibnamefont {Ozaki}}\ and\ \bibinfo {author} {\bibfnamefont {H.}~\bibnamefont {Kino}},\ }\href {https://doi.org/10.1103/PhysRevB.69.195113} {\bibfield  {journal} {\bibinfo  {journal} {Phys. Rev. B}\ }\textbf {\bibinfo {volume} {69}},\ \bibinfo {pages} {195113} (\bibinfo {year} {2004})}\BibitemShut {NoStop}%
\bibitem [{\citenamefont {Hamann}(2013)}]{Norm}%
  \BibitemOpen
  \bibfield  {author} {\bibinfo {author} {\bibfnamefont {D.~R.}\ \bibnamefont {Hamann}},\ }\href {https://doi.org/10.1103/PhysRevB.88.085117} {\bibfield  {journal} {\bibinfo  {journal} {Phys. Rev. B}\ }\textbf {\bibinfo {volume} {88}},\ \bibinfo {pages} {085117} (\bibinfo {year} {2013})}\BibitemShut {NoStop}%
\bibitem [{\citenamefont {Perdew}\ \emph {et~al.}(1996)\citenamefont {Perdew}, \citenamefont {Burke},\ and\ \citenamefont {Ernzerhof}}]{GGA}%
  \BibitemOpen
  \bibfield  {author} {\bibinfo {author} {\bibfnamefont {J.~P.}\ \bibnamefont {Perdew}}, \bibinfo {author} {\bibfnamefont {K.}~\bibnamefont {Burke}},\ and\ \bibinfo {author} {\bibfnamefont {M.}~\bibnamefont {Ernzerhof}},\ }\href {https://doi.org/10.1103/PhysRevLett.77.3865} {\bibfield  {journal} {\bibinfo  {journal} {Phys. Rev. Lett.}\ }\textbf {\bibinfo {volume} {77}},\ \bibinfo {pages} {3865} (\bibinfo {year} {1996})}\BibitemShut {NoStop}%
\bibitem [{\citenamefont {Han}\ \emph {et~al.}(2006)\citenamefont {Han}, \citenamefont {Ozaki},\ and\ \citenamefont {Yu}}]{DFT+U-1}%
  \BibitemOpen
  \bibfield  {author} {\bibinfo {author} {\bibfnamefont {M.~J.}\ \bibnamefont {Han}}, \bibinfo {author} {\bibfnamefont {T.}~\bibnamefont {Ozaki}},\ and\ \bibinfo {author} {\bibfnamefont {J.}~\bibnamefont {Yu}},\ }\href {https://doi.org/10.1103/PhysRevB.73.045110} {\bibfield  {journal} {\bibinfo  {journal} {Phys. Rev. B}\ }\textbf {\bibinfo {volume} {73}},\ \bibinfo {pages} {045110} (\bibinfo {year} {2006})}\BibitemShut {NoStop}%
\bibitem [{\citenamefont {Ryee}\ and\ \citenamefont {Han}(2018{\natexlab{a}})}]{DFT+U-2}%
  \BibitemOpen
  \bibfield  {author} {\bibinfo {author} {\bibfnamefont {S.}~\bibnamefont {Ryee}}\ and\ \bibinfo {author} {\bibfnamefont {M.~J.}\ \bibnamefont {Han}},\ }\href {https://doi.org/10.1088/1361-648X/aac79c} {\bibfield  {journal} {\bibinfo  {journal} {J. Phys.:Condens. Matter}\ }\textbf {\bibinfo {volume} {30}},\ \bibinfo {pages} {275802} (\bibinfo {year} {2018}{\natexlab{a}})}\BibitemShut {NoStop}%
\bibitem [{\citenamefont {Ryee}\ and\ \citenamefont {Han}(2018{\natexlab{b}})}]{DFT+U-3}%
  \BibitemOpen
  \bibfield  {author} {\bibinfo {author} {\bibfnamefont {S.}~\bibnamefont {Ryee}}\ and\ \bibinfo {author} {\bibfnamefont {M.~J.}\ \bibnamefont {Han}},\ }\href {https://doi.org/10.1038/s41598-018-27731-4} {\bibfield  {journal} {\bibinfo  {journal} {Sci. Rep.}\ }\textbf {\bibinfo {volume} {8}},\ \bibinfo {pages} {9559} (\bibinfo {year} {2018}{\natexlab{b}})}\BibitemShut {NoStop}%
\bibitem [{\citenamefont {Straumanis}\ and\ \citenamefont {Yu}(1969)}]{cu_exp}%
  \BibitemOpen
  \bibfield  {author} {\bibinfo {author} {\bibfnamefont {M.~E.}\ \bibnamefont {Straumanis}}\ and\ \bibinfo {author} {\bibfnamefont {L.~S.}\ \bibnamefont {Yu}},\ }\href {https://doi.org/10.1107/S0567739469001549} {\bibfield  {journal} {\bibinfo  {journal} {Acta Crystallogr. A}\ }\textbf {\bibinfo {volume} {25}},\ \bibinfo {pages} {676} (\bibinfo {year} {1969})}\BibitemShut {NoStop}%
\bibitem [{\citenamefont {Crichton}\ \emph {et~al.}(2003)\citenamefont {Crichton}, \citenamefont {Bouvier},\ and\ \citenamefont {Grzechnik}}]{wo3_exp}%
  \BibitemOpen
  \bibfield  {author} {\bibinfo {author} {\bibfnamefont {W.~A.}\ \bibnamefont {Crichton}}, \bibinfo {author} {\bibfnamefont {P.}~\bibnamefont {Bouvier}},\ and\ \bibinfo {author} {\bibfnamefont {A.}~\bibnamefont {Grzechnik}},\ }\href {https://doi.org/https://doi.org/10.1016/S0025-5408(02)01030-9} {\bibfield  {journal} {\bibinfo  {journal} {Mater. Res. Bull.}\ }\textbf {\bibinfo {volume} {38}},\ \bibinfo {pages} {289} (\bibinfo {year} {2003})}\BibitemShut {NoStop}%
\bibitem [{\citenamefont {Mazzola}\ \emph {et~al.}(2023)\citenamefont {Mazzola}, \citenamefont {Hassani}, \citenamefont {Amoroso}, \citenamefont {Chaluvadi}, \citenamefont {Fujii}, \citenamefont {Polewczyk}, \citenamefont {Rajak}, \citenamefont {Koegler}, \citenamefont {Ciancio}, \citenamefont {Partoens}, \citenamefont {Rossi}, \citenamefont {Vobornik}, \citenamefont {Ghosez},\ and\ \citenamefont {Orgiani}}]{wo3_tetragonal}%
  \BibitemOpen
  \bibfield  {author} {\bibinfo {author} {\bibfnamefont {F.}~\bibnamefont {Mazzola}}, \bibinfo {author} {\bibfnamefont {H.}~\bibnamefont {Hassani}}, \bibinfo {author} {\bibfnamefont {D.}~\bibnamefont {Amoroso}}, \bibinfo {author} {\bibfnamefont {S.~K.}\ \bibnamefont {Chaluvadi}}, \bibinfo {author} {\bibfnamefont {J.}~\bibnamefont {Fujii}}, \bibinfo {author} {\bibfnamefont {V.}~\bibnamefont {Polewczyk}}, \bibinfo {author} {\bibfnamefont {P.}~\bibnamefont {Rajak}}, \bibinfo {author} {\bibfnamefont {M.}~\bibnamefont {Koegler}}, \bibinfo {author} {\bibfnamefont {R.}~\bibnamefont {Ciancio}}, \bibinfo {author} {\bibfnamefont {B.}~\bibnamefont {Partoens}}, \bibinfo {author} {\bibfnamefont {G.}~\bibnamefont {Rossi}}, \bibinfo {author} {\bibfnamefont {I.}~\bibnamefont {Vobornik}}, \bibinfo {author} {\bibfnamefont {P.}~\bibnamefont {Ghosez}},\ and\ \bibinfo {author} {\bibfnamefont {P.}~\bibnamefont {Orgiani}},\ }\href {https://doi.org/10.1021/acs.jpclett.3c01546} {\bibfield  {journal} {\bibinfo  {journal} {J. Phys.
  Chem. Lett.}\ }\textbf {\bibinfo {volume} {14}},\ \bibinfo {pages} {7208} (\bibinfo {year} {2023})}\BibitemShut {NoStop}%
\bibitem [{sup()}]{supplementary}%
  \BibitemOpen
  \href@noop {} {\bibinfo {title} {{See the Supplementary Information for data on various interface configurations and orientations, real-space charge analysis, orbital-resolved analysis, and thickness-dependent Rashba parameters. The Supplemental Material also contains Refs.~\citep{suppl-1,cu/cu3n}.}}}\BibitemShut {Stop}%
\bibitem [{\citenamefont {Liu}\ \emph {et~al.}(2006)\citenamefont {Liu}, \citenamefont {Li}, \citenamefont {Zheng},\ and\ \citenamefont {Jiang}}]{suppl-1}%
  \BibitemOpen
  \bibfield  {author} {\bibinfo {author} {\bibfnamefont {W.}~\bibnamefont {Liu}}, \bibinfo {author} {\bibfnamefont {J.~C.}\ \bibnamefont {Li}}, \bibinfo {author} {\bibfnamefont {W.~T.}\ \bibnamefont {Zheng}},\ and\ \bibinfo {author} {\bibfnamefont {Q.}~\bibnamefont {Jiang}},\ }\href {https://doi.org/10.1103/PhysRevB.73.205421} {\bibfield  {journal} {\bibinfo  {journal} {Phys. Rev. B}\ }\textbf {\bibinfo {volume} {73}},\ \bibinfo {pages} {205421} (\bibinfo {year} {2006})}\BibitemShut {NoStop}%
\bibitem [{\citenamefont {MacDonald}\ and\ \citenamefont {Vosko}(1979)}]{relativistic}%
  \BibitemOpen
  \bibfield  {author} {\bibinfo {author} {\bibfnamefont {A.~H.}\ \bibnamefont {MacDonald}}\ and\ \bibinfo {author} {\bibfnamefont {S.~H.}\ \bibnamefont {Vosko}},\ }\href {https://doi.org/10.1088/0022-3719/12/15/007} {\bibfield  {journal} {\bibinfo  {journal} {J. Phys. C: Solid State Phys.}\ }\textbf {\bibinfo {volume} {12}},\ \bibinfo {pages} {2977} (\bibinfo {year} {1979})}\BibitemShut {NoStop}%
\bibitem [{\citenamefont {Theurich}\ and\ \citenamefont {Hill}(2001)}]{relativistic-1}%
  \BibitemOpen
  \bibfield  {author} {\bibinfo {author} {\bibfnamefont {G.}~\bibnamefont {Theurich}}\ and\ \bibinfo {author} {\bibfnamefont {N.~A.}\ \bibnamefont {Hill}},\ }\href {https://doi.org/10.1103/PhysRevB.64.073106} {\bibfield  {journal} {\bibinfo  {journal} {Phys. Rev. B}\ }\textbf {\bibinfo {volume} {64}},\ \bibinfo {pages} {073106} (\bibinfo {year} {2001})}\BibitemShut {NoStop}%
\bibitem [{\citenamefont {Ramana}\ \emph {et~al.}(2006)\citenamefont {Ramana}, \citenamefont {Utsunomiya}, \citenamefont {Ewing}, \citenamefont {Julien},\ and\ \citenamefont {Becker}}]{wo3_stability}%
  \BibitemOpen
  \bibfield  {author} {\bibinfo {author} {\bibfnamefont {C.~V.}\ \bibnamefont {Ramana}}, \bibinfo {author} {\bibfnamefont {S.}~\bibnamefont {Utsunomiya}}, \bibinfo {author} {\bibfnamefont {R.~C.}\ \bibnamefont {Ewing}}, \bibinfo {author} {\bibfnamefont {C.~M.}\ \bibnamefont {Julien}},\ and\ \bibinfo {author} {\bibfnamefont {U.}~\bibnamefont {Becker}},\ }\href {https://doi.org/10.1021/jp056664i} {\bibfield  {journal} {\bibinfo  {journal} {J. Phys. Chem. B}\ }\textbf {\bibinfo {volume} {110}},\ \bibinfo {pages} {10430} (\bibinfo {year} {2006})}\BibitemShut {NoStop}%
\bibitem [{\citenamefont {Mattoni}\ \emph {et~al.}(2018)\citenamefont {Mattoni}, \citenamefont {Filippetti}, \citenamefont {Manca}, \citenamefont {Zubko},\ and\ \citenamefont {Caviglia}}]{wo3_charge}%
  \BibitemOpen
  \bibfield  {author} {\bibinfo {author} {\bibfnamefont {G.}~\bibnamefont {Mattoni}}, \bibinfo {author} {\bibfnamefont {A.}~\bibnamefont {Filippetti}}, \bibinfo {author} {\bibfnamefont {N.}~\bibnamefont {Manca}}, \bibinfo {author} {\bibfnamefont {P.}~\bibnamefont {Zubko}},\ and\ \bibinfo {author} {\bibfnamefont {A.~D.}\ \bibnamefont {Caviglia}},\ }\href {https://doi.org/10.1103/PhysRevMaterials.2.053402} {\bibfield  {journal} {\bibinfo  {journal} {Phys. Rev. Mater.}\ }\textbf {\bibinfo {volume} {2}},\ \bibinfo {pages} {053402} (\bibinfo {year} {2018})}\BibitemShut {NoStop}%
\bibitem [{\citenamefont {Wang}\ \emph {et~al.}(2011)\citenamefont {Wang}, \citenamefont {Di~Valentin},\ and\ \citenamefont {Pacchioni}}]{wo3_dft}%
  \BibitemOpen
  \bibfield  {author} {\bibinfo {author} {\bibfnamefont {F.}~\bibnamefont {Wang}}, \bibinfo {author} {\bibfnamefont {C.}~\bibnamefont {Di~Valentin}},\ and\ \bibinfo {author} {\bibfnamefont {G.}~\bibnamefont {Pacchioni}},\ }\href {https://doi.org/10.1021/jp201057m} {\bibfield  {journal} {\bibinfo  {journal} {J. Phys. Chem. C}\ }\textbf {\bibinfo {volume} {115}},\ \bibinfo {pages} {8345} (\bibinfo {year} {2011})}\BibitemShut {NoStop}%
\bibitem [{\citenamefont {Zhang}\ \emph {et~al.}(2019)\citenamefont {Zhang}, \citenamefont {Ren}, \citenamefont {Xiao},\ and\ \citenamefont {Huang}}]{W-ter_stabilty}%
  \BibitemOpen
  \bibfield  {author} {\bibinfo {author} {\bibfnamefont {X.}~\bibnamefont {Zhang}}, \bibinfo {author} {\bibfnamefont {X.}~\bibnamefont {Ren}}, \bibinfo {author} {\bibfnamefont {Z.}~\bibnamefont {Xiao}},\ and\ \bibinfo {author} {\bibfnamefont {Y.}~\bibnamefont {Huang}},\ }\href {https://doi.org/https://doi.org/10.1016/j.rinp.2019.102670} {\bibfield  {journal} {\bibinfo  {journal} {Results Phys.}\ }\textbf {\bibinfo {volume} {15}},\ \bibinfo {pages} {102670} (\bibinfo {year} {2019})}\BibitemShut {NoStop}%
\bibitem [{\citenamefont {Shanavas}(2016)}]{STO_cubic}%
  \BibitemOpen
  \bibfield  {author} {\bibinfo {author} {\bibfnamefont {K.~V.}\ \bibnamefont {Shanavas}},\ }\href {https://doi.org/10.1103/PhysRevB.93.045108} {\bibfield  {journal} {\bibinfo  {journal} {Phys. Rev. B}\ }\textbf {\bibinfo {volume} {93}},\ \bibinfo {pages} {045108} (\bibinfo {year} {2016})}\BibitemShut {NoStop}%
\bibitem [{\citenamefont {Nakamura}\ \emph {et~al.}(2012)\citenamefont {Nakamura}, \citenamefont {Koga},\ and\ \citenamefont {Kimura}}]{STO_cubic_exp}%
  \BibitemOpen
  \bibfield  {author} {\bibinfo {author} {\bibfnamefont {H.}~\bibnamefont {Nakamura}}, \bibinfo {author} {\bibfnamefont {T.}~\bibnamefont {Koga}},\ and\ \bibinfo {author} {\bibfnamefont {T.}~\bibnamefont {Kimura}},\ }\href {https://doi.org/10.1103/PhysRevLett.108.206601} {\bibfield  {journal} {\bibinfo  {journal} {Phys. Rev. Lett.}\ }\textbf {\bibinfo {volume} {108}},\ \bibinfo {pages} {206601} (\bibinfo {year} {2012})}\BibitemShut {NoStop}%
\bibitem [{\citenamefont {Giannozzi}\ \emph {et~al.}(2009)\citenamefont {Giannozzi} \emph {et~al.}}]{QE-2009}%
  \BibitemOpen
  \bibfield  {author} {\bibinfo {author} {\bibfnamefont {P.}~\bibnamefont {Giannozzi}} \emph {et~al.},\ }\href {https://doi.org/10.1088/0953-8984/21/39/395502} {\bibfield  {journal} {\bibinfo  {journal} {J. Phys.:Condens. Matter}\ }\textbf {\bibinfo {volume} {21}},\ \bibinfo {pages} {395502} (\bibinfo {year} {2009})}\BibitemShut {NoStop}%
\bibitem [{\citenamefont {Giannozzi}\ \emph {et~al.}(2017)\citenamefont {Giannozzi} \emph {et~al.}}]{QE-2017}%
  \BibitemOpen
  \bibfield  {author} {\bibinfo {author} {\bibfnamefont {P.}~\bibnamefont {Giannozzi}} \emph {et~al.},\ }\href {https://doi.org/10.1088/1361-648X/aa8f79} {\bibfield  {journal} {\bibinfo  {journal} {J. Phys.:Condens. Matter}\ }\textbf {\bibinfo {volume} {29}},\ \bibinfo {pages} {465901} (\bibinfo {year} {2017})}\BibitemShut {NoStop}%
\bibitem [{\citenamefont {Geisler}(2023{\natexlab{a}})}]{STO_nickelate}%
  \BibitemOpen
  \bibfield  {author} {\bibinfo {author} {\bibfnamefont {B.}~\bibnamefont {Geisler}},\ }\href {https://doi.org/10.1103/PhysRevB.108.224502} {\bibfield  {journal} {\bibinfo  {journal} {Phys. Rev. B}\ }\textbf {\bibinfo {volume} {108}},\ \bibinfo {pages} {224502} (\bibinfo {year} {2023}{\natexlab{a}})}\BibitemShut {NoStop}%
\bibitem [{\citenamefont {Zhong}\ \emph {et~al.}(2013)\citenamefont {Zhong}, \citenamefont {T\'oth},\ and\ \citenamefont {Held}}]{LaO_STO}%
  \BibitemOpen
  \bibfield  {author} {\bibinfo {author} {\bibfnamefont {Z.}~\bibnamefont {Zhong}}, \bibinfo {author} {\bibfnamefont {A.}~\bibnamefont {T\'oth}},\ and\ \bibinfo {author} {\bibfnamefont {K.}~\bibnamefont {Held}},\ }\href {https://doi.org/10.1103/PhysRevB.87.161102} {\bibfield  {journal} {\bibinfo  {journal} {Phys. Rev. B}\ }\textbf {\bibinfo {volume} {87}},\ \bibinfo {pages} {161102} (\bibinfo {year} {2013})}\BibitemShut {NoStop}%
\bibitem [{\citenamefont {Ishida}(2014)}]{Cu_111}%
  \BibitemOpen
  \bibfield  {author} {\bibinfo {author} {\bibfnamefont {H.}~\bibnamefont {Ishida}},\ }\href {https://doi.org/10.1103/PhysRevB.90.235422} {\bibfield  {journal} {\bibinfo  {journal} {Phys. Rev. B}\ }\textbf {\bibinfo {volume} {90}},\ \bibinfo {pages} {235422} (\bibinfo {year} {2014})}\BibitemShut {NoStop}%
\bibitem [{\citenamefont {Yaji}\ \emph {et~al.}(2018)\citenamefont {Yaji}, \citenamefont {Harasawa}, \citenamefont {Kuroda}, \citenamefont {Li}, \citenamefont {Yan}, \citenamefont {Komori},\ and\ \citenamefont {Shin}}]{Cu_111_exp}%
  \BibitemOpen
  \bibfield  {author} {\bibinfo {author} {\bibfnamefont {K.}~\bibnamefont {Yaji}}, \bibinfo {author} {\bibfnamefont {A.}~\bibnamefont {Harasawa}}, \bibinfo {author} {\bibfnamefont {K.}~\bibnamefont {Kuroda}}, \bibinfo {author} {\bibfnamefont {R.}~\bibnamefont {Li}}, \bibinfo {author} {\bibfnamefont {B.}~\bibnamefont {Yan}}, \bibinfo {author} {\bibfnamefont {F.}~\bibnamefont {Komori}},\ and\ \bibinfo {author} {\bibfnamefont {S.}~\bibnamefont {Shin}},\ }\href {https://doi.org/10.1103/PhysRevB.98.041404} {\bibfield  {journal} {\bibinfo  {journal} {Phys. Rev. B}\ }\textbf {\bibinfo {volume} {98}},\ \bibinfo {pages} {041404} (\bibinfo {year} {2018})}\BibitemShut {NoStop}%
\bibitem [{\citenamefont {Mera~Acosta}\ \emph {et~al.}(2021)\citenamefont {Mera~Acosta}, \citenamefont {Yuan}, \citenamefont {Dalpian},\ and\ \citenamefont {Zunger}}]{spin-variation}%
  \BibitemOpen
  \bibfield  {author} {\bibinfo {author} {\bibfnamefont {C.}~\bibnamefont {Mera~Acosta}}, \bibinfo {author} {\bibfnamefont {L.}~\bibnamefont {Yuan}}, \bibinfo {author} {\bibfnamefont {G.~M.}\ \bibnamefont {Dalpian}},\ and\ \bibinfo {author} {\bibfnamefont {A.}~\bibnamefont {Zunger}},\ }\href {https://doi.org/10.1103/PhysRevB.104.104408} {\bibfield  {journal} {\bibinfo  {journal} {Phys. Rev. B}\ }\textbf {\bibinfo {volume} {104}},\ \bibinfo {pages} {104408} (\bibinfo {year} {2021})}\BibitemShut {NoStop}%
\bibitem [{\citenamefont {Wicaksono}\ \emph {et~al.}(2024)\citenamefont {Wicaksono}, \citenamefont {You}, \citenamefont {Gu}, \citenamefont {Evseev}, \citenamefont {Piyanzina}, \citenamefont {Kusakabe}, \citenamefont {Yunoki},\ and\ \citenamefont {Maekawa}}]{cu/cu3n}%
  \BibitemOpen
  \bibfield  {author} {\bibinfo {author} {\bibfnamefont {Y.}~\bibnamefont {Wicaksono}}, \bibinfo {author} {\bibfnamefont {J.-Y.}\ \bibnamefont {You}}, \bibinfo {author} {\bibfnamefont {B.}~\bibnamefont {Gu}}, \bibinfo {author} {\bibfnamefont {A.}~\bibnamefont {Evseev}}, \bibinfo {author} {\bibfnamefont {I.}~\bibnamefont {Piyanzina}}, \bibinfo {author} {\bibfnamefont {K.}~\bibnamefont {Kusakabe}}, \bibinfo {author} {\bibfnamefont {S.}~\bibnamefont {Yunoki}},\ and\ \bibinfo {author} {\bibfnamefont {S.}~\bibnamefont {Maekawa}},\ }\href {https://doi.org/10.1103/PhysRevB.110.L220408} {\bibfield  {journal} {\bibinfo  {journal} {Phys. Rev. B}\ }\textbf {\bibinfo {volume} {110}},\ \bibinfo {pages} {L220408} (\bibinfo {year} {2024})}\BibitemShut {NoStop}%
\end{thebibliography}
\providecommand{\noopsort}[1]{}\providecommand{\singleletter}[1]{#1}%
%

% --- YOUR MAIN TEXT ENDS HERE ---

% --- End of your main double-column paper text ---

% --- Your paper text ends here ---

% --- YOUR MAIN Paper TEXT ENDS HERE ---

% --- YOUR MAIN Paper TEXT ENDS HERE ---

% --- YOUR MAIN Paper TEXT ENDS HERE ---

% --- YOUR MAIN DOUBLE-COLUMN Paper TEXT ENDS HERE ---

% --- YOUR MAIN Paper TEXT ENDS HERE ---

\clearpage
\onecolumngrid
\clearpage

\makeatletter
\let\LS@rot\@undefined
\makeatother

\thispagestyle{empty}
% The magic array string prevents Page 1 from stacking on Page 2


\section*{Supplementary material}

# Supplementary Information

## Orbital Hybridization Induces Giant Cubic Rashba Effect at Cu/$WO_3$ Interface

Md Aktar Hossain[1] and Saikat Das[1]

[1]Department of Physics, Indian Institute of Technology, Kharagpur, 721302, India

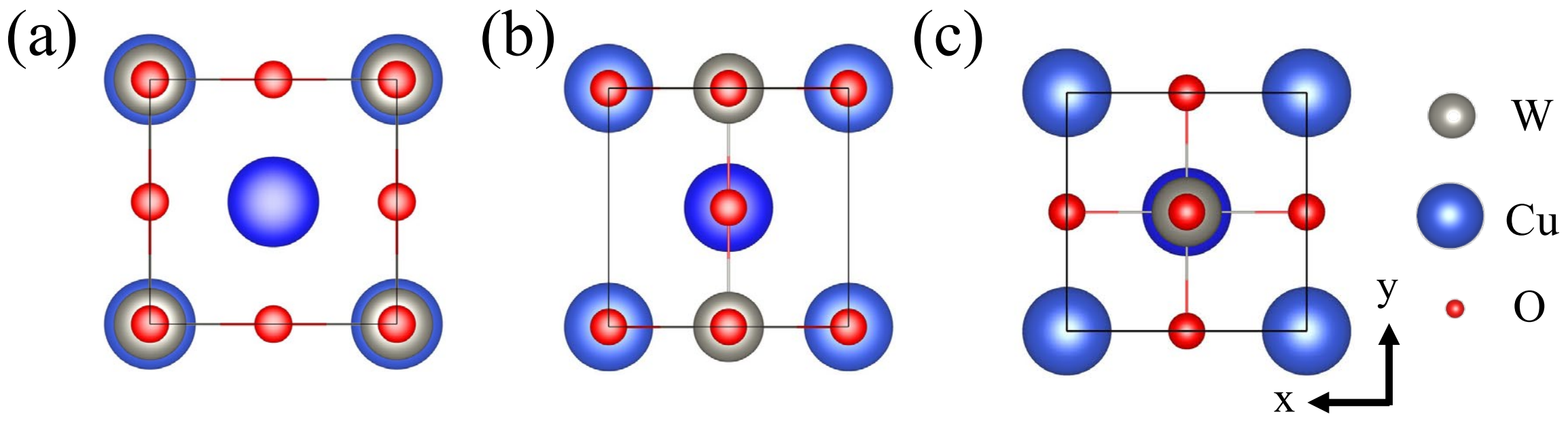


**FIG. S1.** Top view of the Cu(001)/$WO_3$(001) heterostructure slab model with different interface configurations: (a) Type-I, (b) Type-II, and (c) Type-III. For visual clarity, underlying Cu atoms are enlarged than the W atoms, and deep blue shade is used for face-centered Cu atoms.

### Interface and Formation Energies for Different Cu Orientations and Interface Stackings

The interface energy can be obtained using the following relation [1].

$$\gamma_{\mathrm{int}} = \frac{E_{\mathrm{Cu/WO_3}} - (E_{\mathrm{bulk}}^{\mathrm{Cu}} + E_{\mathrm{bulk}}^{\mathrm{WO_3}})}{A_{\mathrm{int}}} - (\gamma_{\mathrm{surf}}^{\mathrm{Cu}} + \gamma_{\mathrm{surf}}^{\mathrm{WO_3}}), \qquad (1)$$

where $E_{\mathrm{Cu/WO_3}}$ is the total energy of the Cu/$WO_3$ heterostructure consisting of the Cu and $WO_3$ slabs. $E_{\mathrm{bulk}}^{\mathrm{Cu}}$ and $E_{\mathrm{bulk}}^{\mathrm{WO_3}}$ are the bulk energies of Cu and $WO_3$, respectively, $A_{\mathrm{int}}$ is the interface area, and $\gamma_{\mathrm{surf}}^{\mathrm{Cu}}$ and $\gamma_{\mathrm{surf}}^{\mathrm{WO_3}}$ are the surface energies of the Cu and $WO_3$ slabs, respectively.

The formation energy can be obtained using the following relation [2].

$$E_{\mathrm{form}} = E_{\mathrm{total}} - \sum_i n_i \, E_i^{\mathrm{ref}}, \qquad (2)$$

where $E_{\mathrm{form}}$ is the formation energy, $E_{\mathrm{total}}$ is the total energy of the heterostructure, $n_i$ is the number of atoms of type $i$, and $E_i^{\mathrm{ref}}$ is the reference energy of the constituent elemental components in the heterostructure.

Using Eqs. (1) and (2), the interface and formation energies were calculated for different Cu orientations and interface configurations. Finally, in TABLE S1, we report the interface and formation energy per atom.

**TABLE S1.** Interface and Formation energy per atom for Type-I, Type-II, and Type-III interface configurations. The lowest energies are marked in bold.

| **Heterostructure** | | Type-I | Type-II | Type-III |
|---|---|---|---|---|
| Cu(001)/$WO_3$(001) | Interface Energy (eV/ Å$^2$) | 0.363 | **0.309** | 0.362 |
| | Formation Energy per atom (eV) | -4.399 | **-5.006** | -4.995 |
| Cu(110)/$WO_3$(001) | Interface Energy (eV/ Å$^2$) | **0.099** | 0.159 | 0.104 |
| | Formation Energy per atom (eV) | **-5.773** | -5.742 | -5.771 |
| Cu(111)/$WO_3$(001) | Interface Energy (eV/ Å$^2$) | **0.109** | 0.132 | 0.110 |
| | Formation Energy per atom (eV) | **-4.934** | -4.916 | -4.933 |

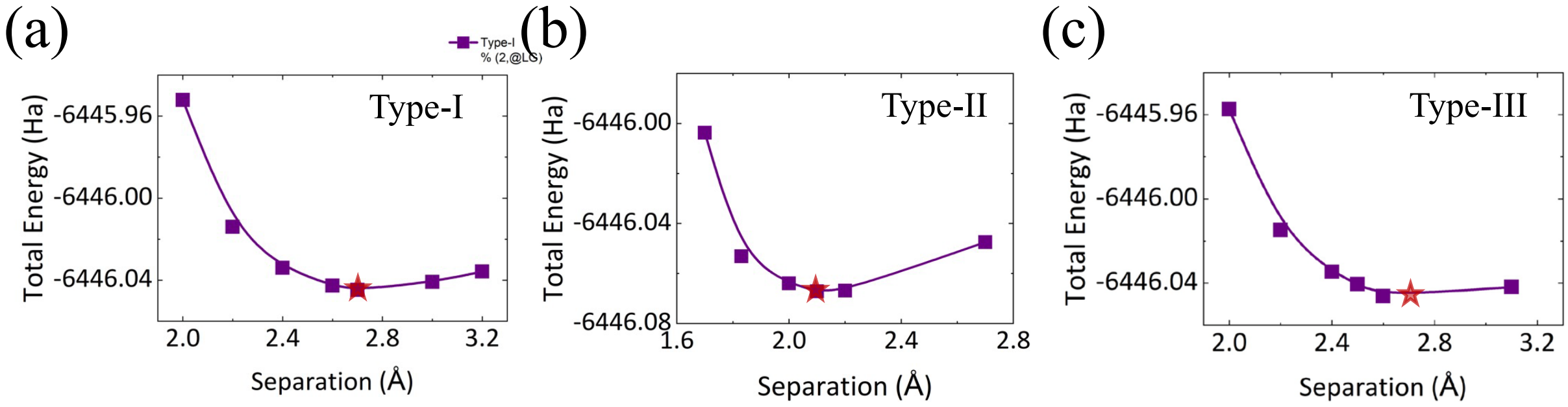


**FIG. S2.** Optimization of interfacial separation for the Cu(001)/$WO_3$(001) heterostructure with different interface configurations. (a) Type-I, (b) Type-II, and (c) Type-III. The Cu thickness is fixed at 15 ML for these calculations. The optimized separations are 2.7 Å, 2.1 Å, and 2.7 Å, respectively, and are marked by red stars. The lines serve as guides for the eye.

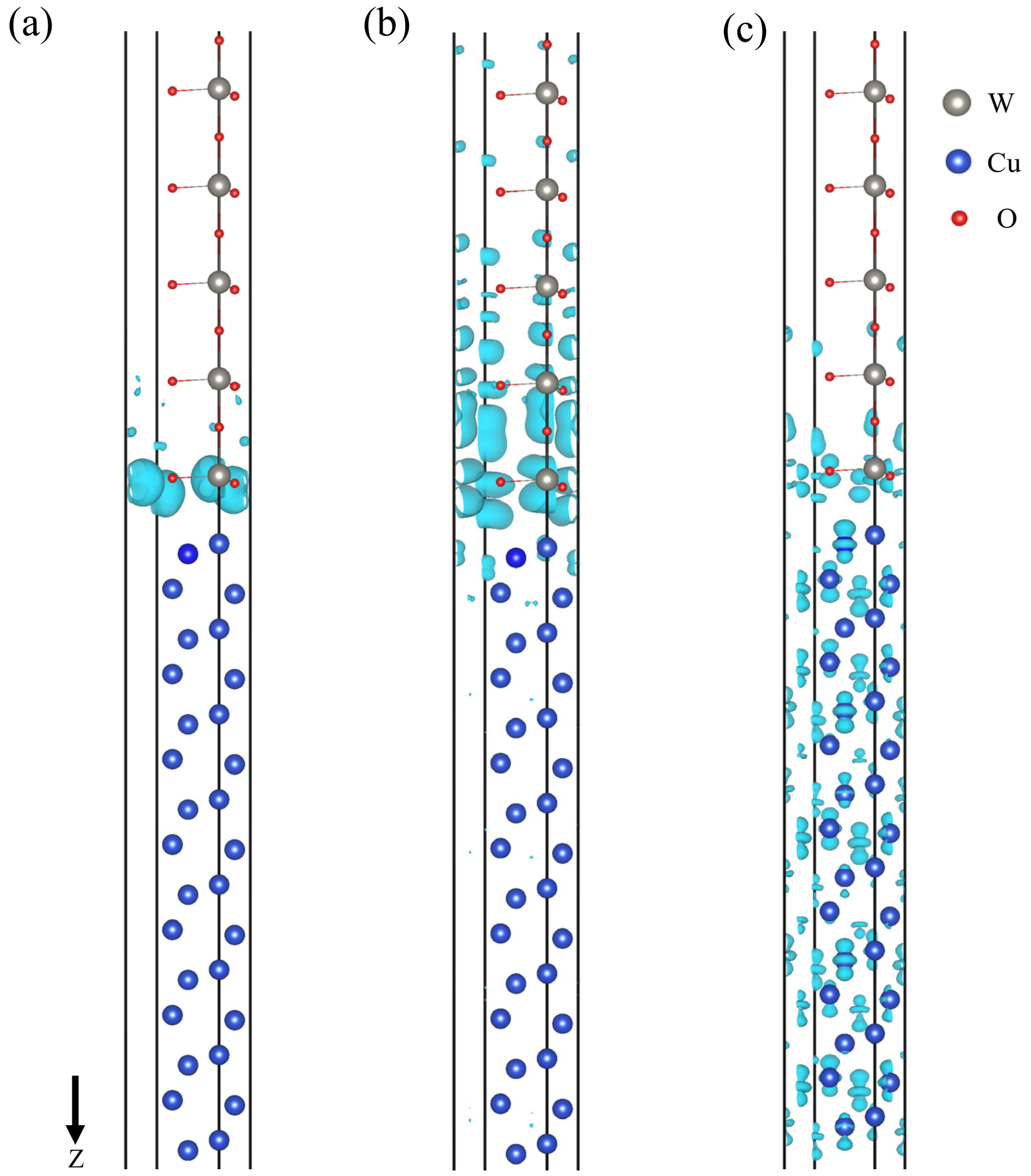


**FIG. S3.** Real-space electron density for the (a) α, (b) β, and (c) γ bands of the $Cu(001)/WO_3(001)$ heterostructure with Type-I interface configuration. The electron densities are evaluated at the band bottom at energies ≈−0.26 eV ( for α), −0.67 eV (for β), and −0.86 eV (for γ).

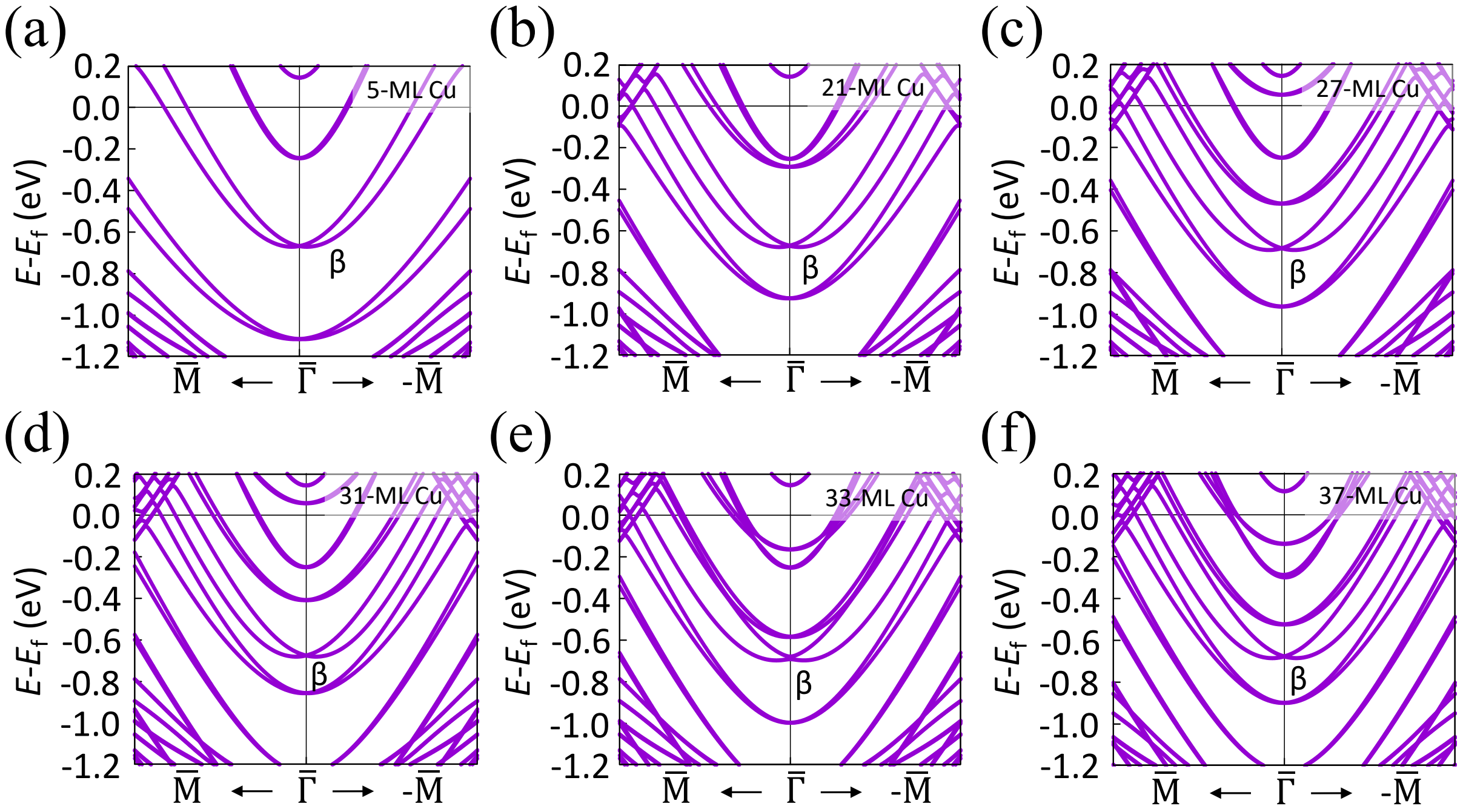


**FIG. S4.** Magnified view of band structure around the Γ point for the Cu(001)/$WO_3$(001) heterostructure (Type-I interface configuration) with Cu thickness: (a) 5-ML, (b) 21-ML, (c) 27-ML, (d) 31-ML, (e) 33-ML, and (f) 37-ML. For reference, the $\beta$ interface state is marked in each figure.

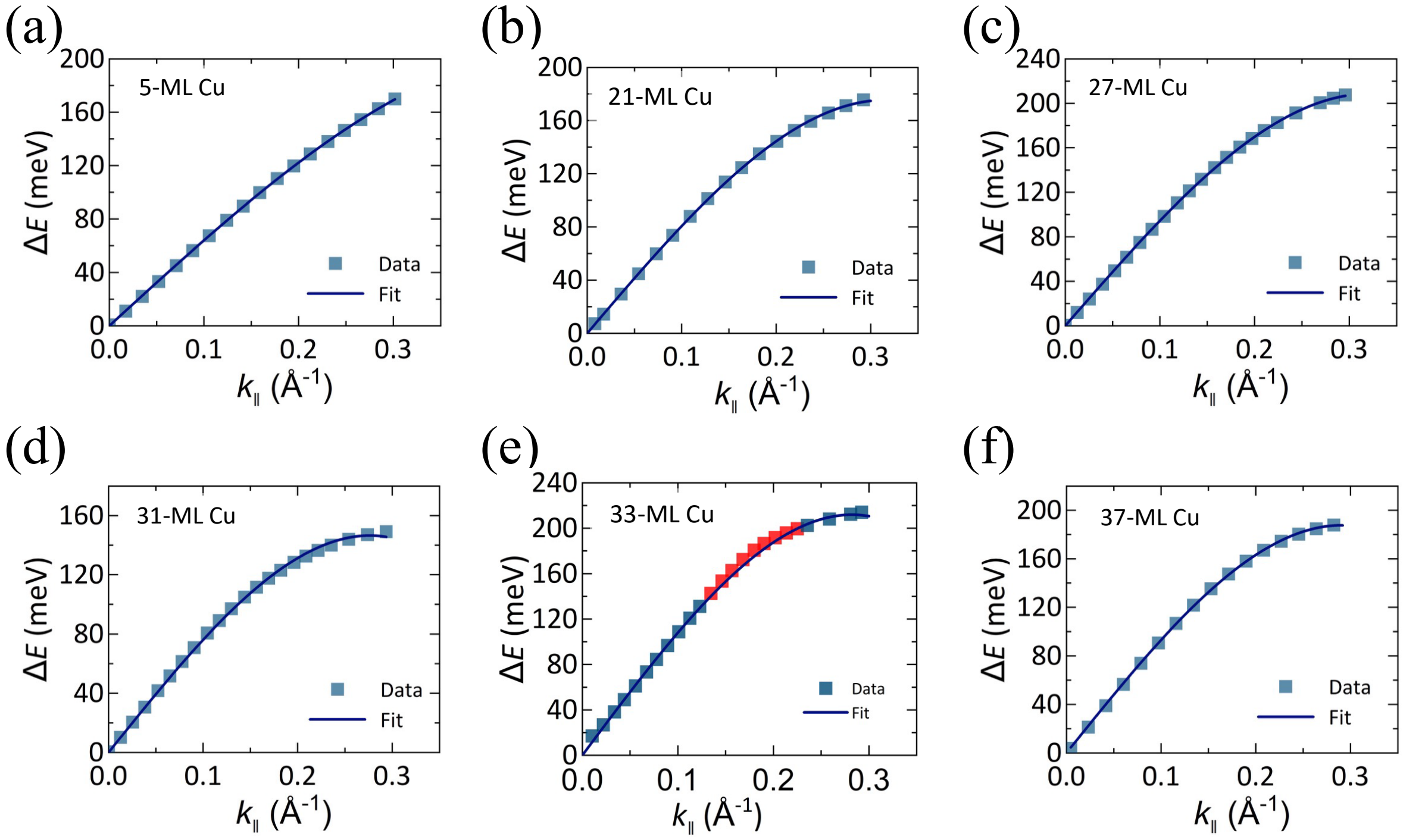


**FIG. S5.** Momentum-resolved energy splitting along the $\bar{\Gamma}-\bar{M}$ path of the surface Brillouin zone (SBZ) for the β band obtained from Fig. S4, with Cu thickness: (a) 5-ML, (b) 21-ML, (c) 27 ML, (d) 31 ML, (e) 33 ML, and (f) 37-ML. The solid lines represent fits to $\Delta E(k) = 2\alpha_R k +$

$2\,\tilde{\alpha}_R k^3$. In FIG. S5(e), data points marked in red are affected by the interaction between the QWS and the interface state, and hence are masked while fitting the data.

**TABLE S2.** Linear ($\alpha_R$) and cubic ($\tilde{\alpha}_R$) Rashba parameters obtained from fitting the momentum-resolved energy splitting in Fig. S5. A comparison of various fitting windows is summarised. In Fig. 5 of the main text, we have shown Rashba parameters that are obtained by using a broad fitting range between 0.0-0.3 Å$^{-1}$.

| **Thickness of the Cu layer** | **Fitting Range: 0.0-0.3 Å$^{-1}$** | |
|---|---|---|
| | $\tilde{\alpha}_R$ (eV. Å$^3$) | $\alpha_R$ (eV. Å) |
| 5-ML | -0.4793 ± 0.0023 | 0.3247 ± 0.0001 |
| 15-ML | -0.9276 ± 0.0049 | 0.2852 ± 0.0002 |
| 21-ML | -1.3870 ±0.0048 | 0.4165 ± 0.0003 |
| 27-ML | -1.5780 ±0.0041 | 0.4878 ± 0.0002 |
| 31-ML | -1.7163 ± 0.0106 | 0.3968 ± 0.0006 |
| 33-ML | -2.3194 ± 0.0129 | 0.5606 ± 0.0009 |
| 37-ML | -1.9317± 0.0051 | 0.4857 ± 0.0003 |
| | **Fitting Range: 0-0.2 Å$^{-1}$** | |
| 5-ML | -0.3988 ± 0.0092 | 0.3230 ± 0.0002 |
| 15-ML | -1.0988 ± 0.0063 | 0.2901 ± 0.0002 |
| 21-ML | -1.4565 ± 0.0102 | 0.4192 ± 0.0003 |
| 27-ML | -1.4643 ± 0.0150 | 0.4857 ± 0.0004 |
| 31-ML | -2.1083 ± 0.0055 | 0.4081 ± 0.0001 |
| 33-ML | - | - |
| 37-ML | -1.8754 ± 0.0155 | 0.4853 ± 0.0004 |
| | **Fitting Range : 0.1-0.3 Å$^{-1}$** | |
| 5-ML | -0.4925 ± 0.0014 | 0.3256 ± 0.0001 |
| 15-ML | -0.9179 ± 0.0059 | 0.2845 ± 0.0004 |
| 21-ML | -1.3899 ± 0.0061 | 0.4166 ± 0.0004 |
| 27-ML | -1.5997 ± 0.0037 | 0.4890 ± 0.0002 |
| 31-ML | -1.7339 ±0.0126 | 0.3967 ± 0.0007 |
| 33-ML | -2.3280 ± 0.0196 | 0.5613 ± 0.0014 |
| 37-ML | -1.9421 ± 0.0059 | 0.4864 ± 0.0003 |

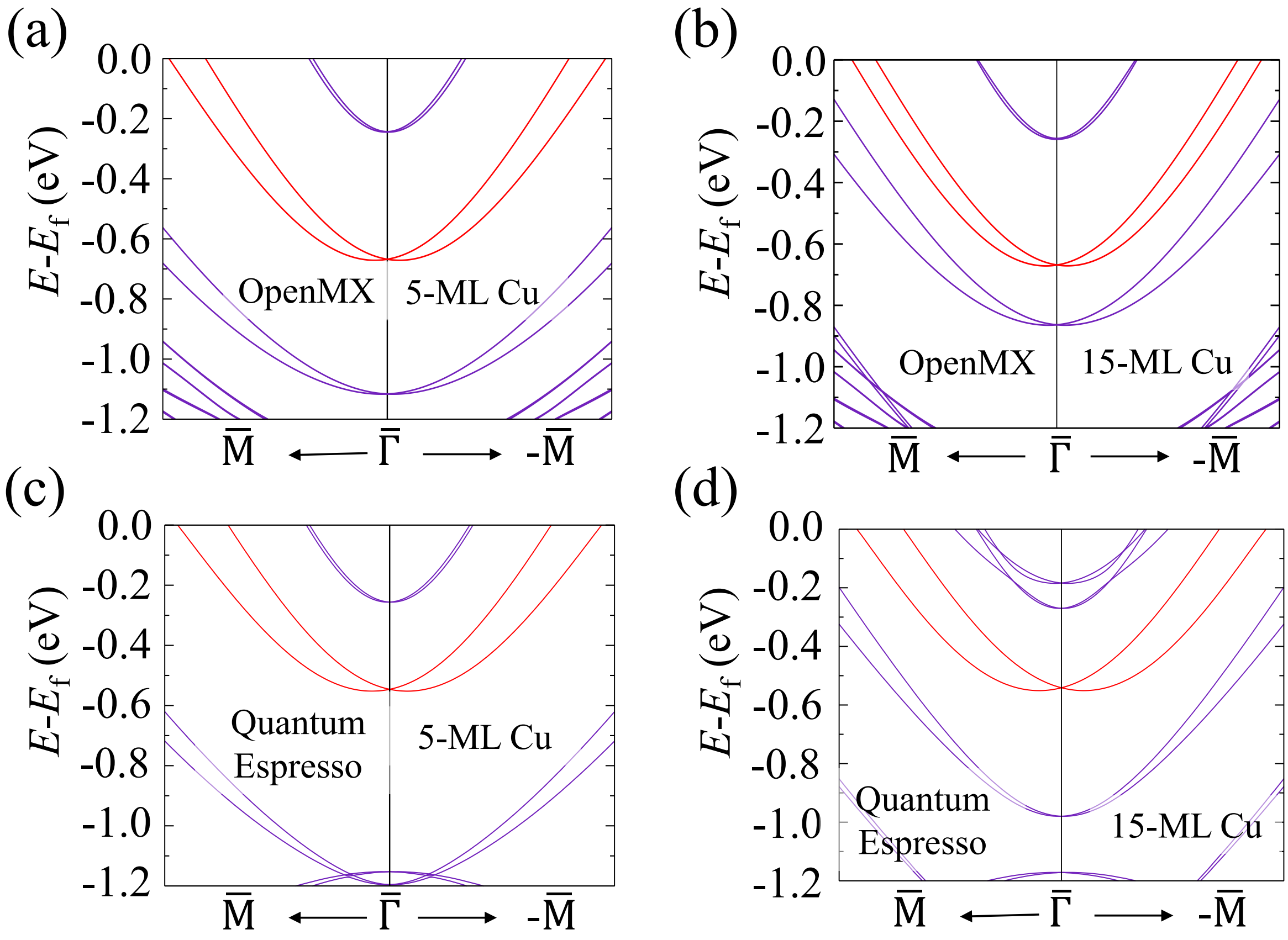


**FIG. S6.** Calculated band structures of the Cu(001)/$WO_3$(001) heterostructure with type-I interface configuration. (a) and (b) show results for 5 and 15-ML-thick Cu layer, respectively, that are obtained using the OpenMX codes. (c) and (d) show corresponding band structures obtained using the Quantum ESPRESSO. The β band is colored in red for clarity.

Density functional theory (DFT) calculations using the Quantum ESPRESSO package were performed with the generalized gradient approximation (GGA) of Perdew–Burke–Ernzerhof (PBE) for the exchange–correlation functional and norm-conserving pseudopotentials. Spin–orbit coupling (SOC) was included in all calculations. The DFT+U formalism was employed with U = 5.65 eV applied to the W–5d orbitals. For self-consistent calculations, a 16 × 16 × 1 Γ-centered k-point grid was used for both the 5-ML and 15-ML Cu(001)/$WO_3$(001) heterostructures with type-I interface configuration. A plane-wave cutoff energy of 70 Ry was adopted in all cases. For the β band, the linear Rashba parameter was determined to be $\alpha_R = 0.4053 \pm 0.0005$ eV.Å and the cubic Rashba parameter $\tilde{\alpha}_R = -0.341 \pm 0.009$ eV.Å$^3$ for the 5-ML Cu(001)/$WO_3$(001) heterostructure. For the 15-ML Cu(001)/$WO_3$(001) case, the corresponding parameters were $\alpha_R = 0.5000 \pm 0.0005$ eV.Å and $\tilde{\alpha}_R = -1.477 \pm 0.089$ eV.Å$^3$. These trends in the Rashba parameters are consistent with those obtained from OpenMX calculations. The quantitative differences may arise from the different basis-set implementations in OpenMX (linear combination of pseudoatomic orbitals) and Quantum Espresso (plane-wave basis).

(a)

(b)

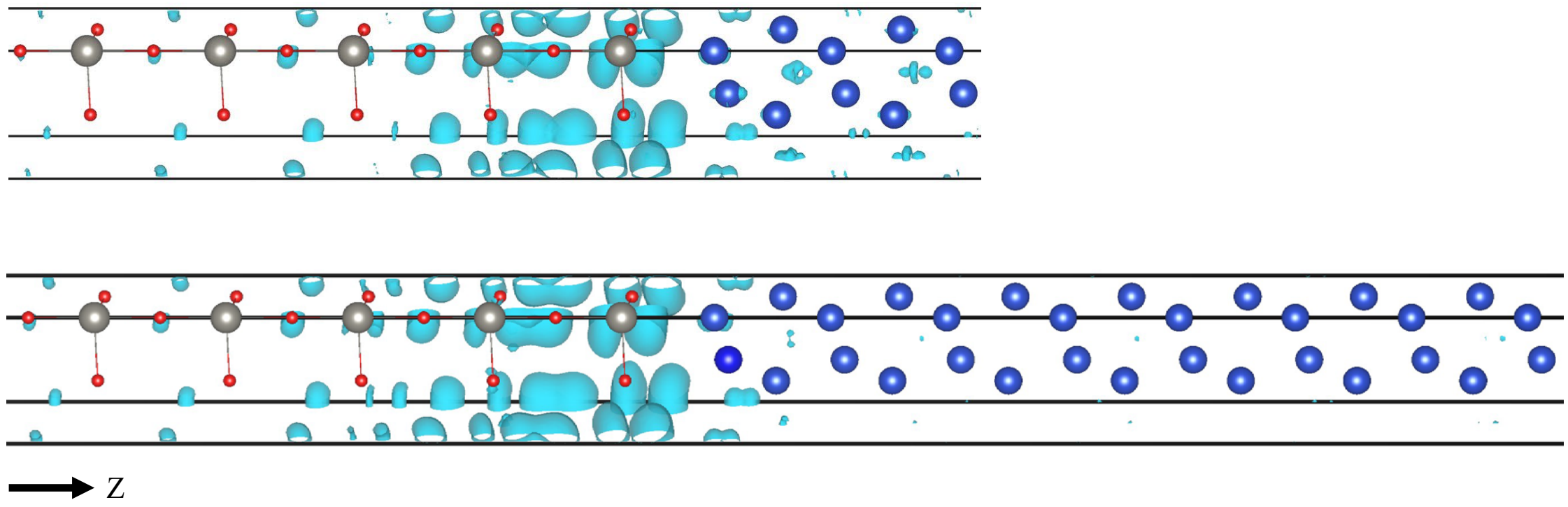


**FIG. S7.** Real-space electron density of the interface state for the Cu(001)/$WO_3$(001) heterostructure (Type-I interface configuration) with a Cu thickness of (a) 5-ML and (b) 15-ML. The electron density is obtained from the bottom of the $\beta$ interface state. Unlike the 15-ML thick Cu case, the tail of electron density extends up to the vacuum/Cu interface in the system with the 5-ML thick Cu layer.

**Discussion on Type II Configuration**

Similar to the Type-I interface configuration discussed in the main text; we have thoroughly analysed the Type-II interface configuration for the Cu(001)/$WO_3$(001) heterostructure. Bandstructures and analysis of Rashba parameters across various thicknesses ranging from 5 to 37 ML are show in Fig. S8. The band structure exhibits three electron-like bands (α, β, and γ) similar to the Type-I case, with minima at approximately –0.4 eV, –0.8 eV, and –0.6 eV, respectively. Notably, the β band shifts to a lower energy compared to the Type-I configuration. Meanwhile the γ band is shifted towards higher binding energy and splitting is noticeably suppressed compared to the Type-I case, with $\alpha_R \approx 0.06$ eV·Å and $\tilde{\alpha}_R \approx 0.06$ eV·Å$^3$. In the decoupled limit, the Rashba parameters from the β band are found to be $\alpha_R \approx 0.4$ eV·Å and $\tilde{\alpha}_R \approx -1.9$ eV·Å$^3$ (TABLE S3) , which are comparable to those of the Type-I interface.

To characterize the real-space electron density and the orbital make up, we selected focused on the 15 ML thick Cu case (band structure shown in Fig. S8 (c)). The real-space electron density (Fig. S9) reveals that the α-band is confined to the terminating $WO_2$ interfacial layer and primarily localizes around the W atoms, indicating a two-dimensional electron gas (2DEG)-like state. The β band localizes within the first few layers on both sides of the interface, involving W, apical O, and Cu atoms; maximum localization around the W atoms identifies it as a hybridized interface state. In contrast, the γ band delocalizes across the Cu slab—characteristic of a Cu quantum-well state (QWS)—and decays into the $WO_3$ layer. Unlike the Type-I interface, the QWS electron density on the $WO_3$ side localizes preferentially around the basal O atoms [Fig. S9(c)]. This, suggests a reduced Cu–W orbital overlap and explains the pronounced suppression of the Rashba splitting in the γ band [Fig. S8(c)].

Orbital-resolved analysis (Fig. S10) suggests that the β band originates from the hybridization of W-($d_{xz}$, $d_{yz}$), O-(p), and Cu orbitals. Different from the Type-I case, Cu (s,$p_z$)

orbital contributions are reduced, while the contribution from ($p_x$,$p_y$) orbitals are enhanced. This trend can be well understood noting the relative lateral shift between the Cu and W atoms by half-a lattice constant, which naturally favours hybridization between $p_x$-$d_{xz}$ and $p_y$-$d_{yz}$ orbitals over the s and $p_z$ counterparts. For the γ band, Cu (s,$p_z$) and O ($p_z$) orbitals mostly contribute, and contribution from the W orbitals are remarkably suppressed—consistent with the real-space charge analysis in Fig. S9.

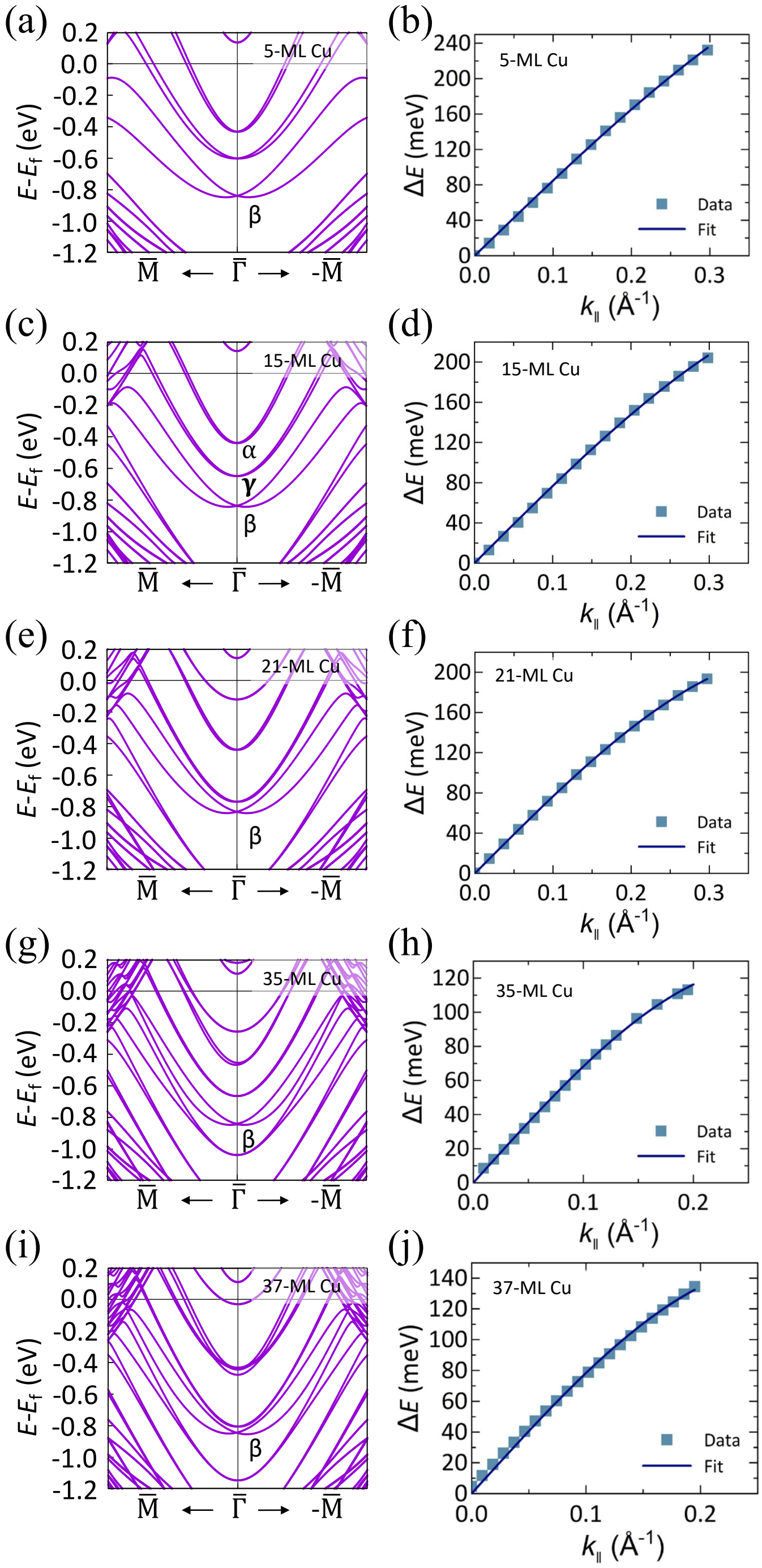


**FIG. S8.** Band structures and momentum-resolved energy splitting along the $\bar{\Gamma}-\bar{M}$ path for the $\beta$ band of the Cu(001)/$WO_3$(001) heterostructure (Type-II interface configuration) for Cu

thicknesses: (a, b) 5-ML, (c, d) 15-ML, (e, f) 21-ML, (g, h) 35-ML, and (i, j) 37-ML. Left panels show the band structures, while right panels show the corresponding momentum-resolved energy splitting (solid squares) and fitting (solid line) to the equation $\Delta E(k) = 2\alpha_R k + 2\,\tilde{\alpha}_R k^3$.

**TABLE S3.** Linear ($\alpha_R$) and cubic ($\tilde{\alpha}_R$) Rashba parameters obtained from fitting the momentum-resolved energy splitting in FIG. S8. A comparison of various fitting windows is summarised. For systems with the 35-ML and 37-ML-thick Cu layers, interaction between the QWS and the $\beta$ band hinders fitting for $k$ >0.2 Å$^{-1}$ .

| **Thickness of the Cu layer** | **Fitting Range: 0.0-0.3 Å$^{-1}$** | |
|---|---|---|
| | $\tilde{\alpha}_R$ (eV. Å$^3$) | $\alpha_R$ (eV. Å) |
| 5-ML | -0.3705 ± 0.0149 | 0.4281± 0.0009 |
| 15-ML | -0.4691 ± 0.0109 | 0.38803 ± 0.0006 |
| 21-ML | -0.7341 ± 0.0026 | 0.3899 ± 0.0002 |
| 35-ML | - | - |
| 37-ML | - | - |
| | **Fitting Range: 0.0-0.2 Å$^{-1}$** | |
| 5-ML | 0.3749 ± 0.0339 | 0.4096 ± 0.0009 |
| 15-ML | 0.0716 ± 0.0256 | 0.3746 ± 0.0007 |
| 21-ML | -0.8631 ± 0.0063 | 0.3931 ± 0.0002 |
| 35-ML | -1.6737 ± 0.0027 | 0.3578 ± 0.0007 |
| 37-ML | -1.8640 ± 0.0699 | 0.4104 ± 0.0018 |
| | **Fitting Range: 0.1-0.3 Å$^{-1}$** | |
| 5-ML | -0.4518 ± 0.0105 | 0.4334 ± 0.0006 |
| 15-ML | -0.5350 ± 0.0074 | 0.3922 ± 0.0004 |
| 21-ML | -0.7194 ± 0.0018 | 0.3889 ± 0.0001 |
| 35-ML | - | - |
| 37-ML | - | - |

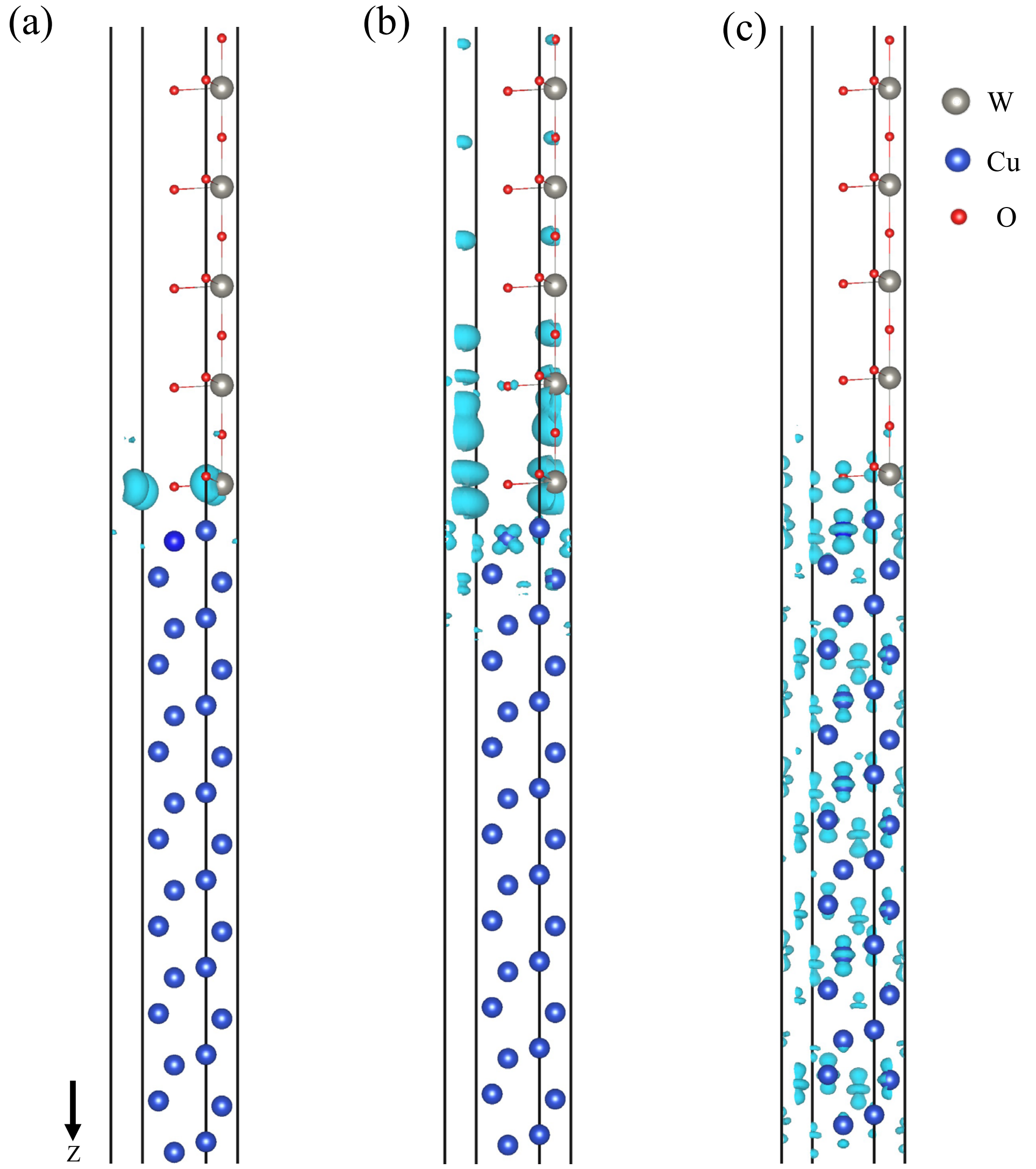


**FIG. S9.** Real-space electron density for the (a) α, (b) β, and (c) γ bands of the Cu(001)/$WO_3$(001) heterostructure with Type-II interface configuration. The electron densities are evaluated from the band bottom at energies ≈ −0.43 eV (for α), −0.84 eV (for β), and −0.64 eV (for γ).

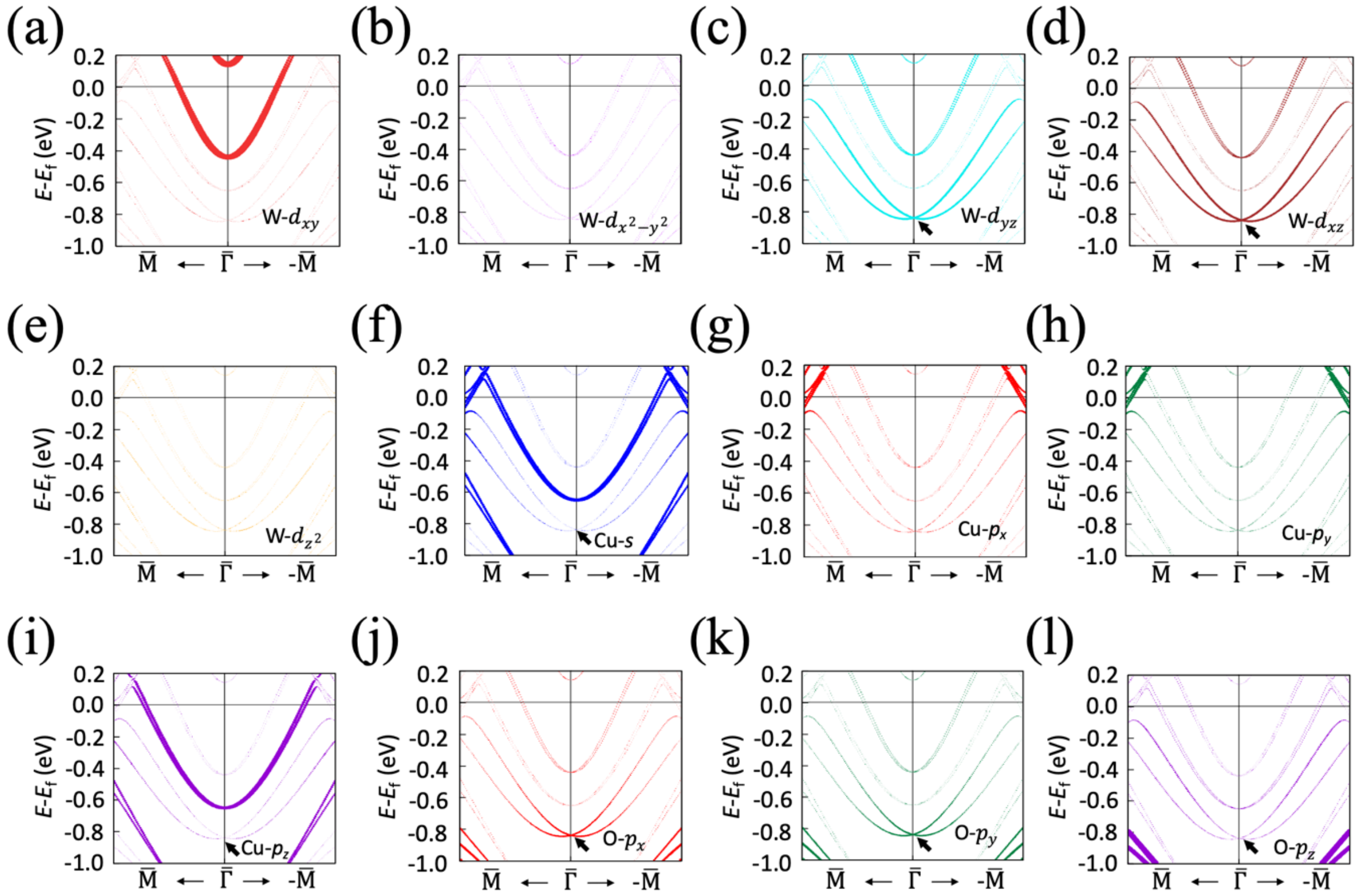


**FIG. S10.** Orbital-resolved band structure of the 15-ML Cu(001)/$WO_3$(001) heterostructure with the Type-II interface configuration. The radii of the circles are proportional to the weight of the corresponding atomic orbitals.

## Discussion on Type III Configuration

We have also analysed band structures for Type-III interface configuration of the Cu(001)/$WO_3$(001) heterostructure, maintaining the same reference thickness of 15-ML Cu(001). For brevity, we have only analysed bands for systems with 15 ML and 39 ML thick Cu layer (Fig. S11). Combining the band structures, real-space charge analysis (Fig. S12) , and orbital-resolved analysis (S13), we conclude that the Type-III interface configuration yields qualitatively similar features as in the Type-I configuration. With the interface β band yielding $\alpha_R \approx 0.4$ eV·Å and $\tilde{\alpha}_R \approx -1.3$ eV·Å$^3$ in the decoupled limit (TABLE S4). The Cu QWS (γ band) exhibits robust spin-splitting with the Rashba parameters $\alpha_R \approx 0.19$ eV·Å and $\tilde{\alpha}_R \approx 0.46$ eV·Å$^3$ is similar to the Type-I interface configuration.

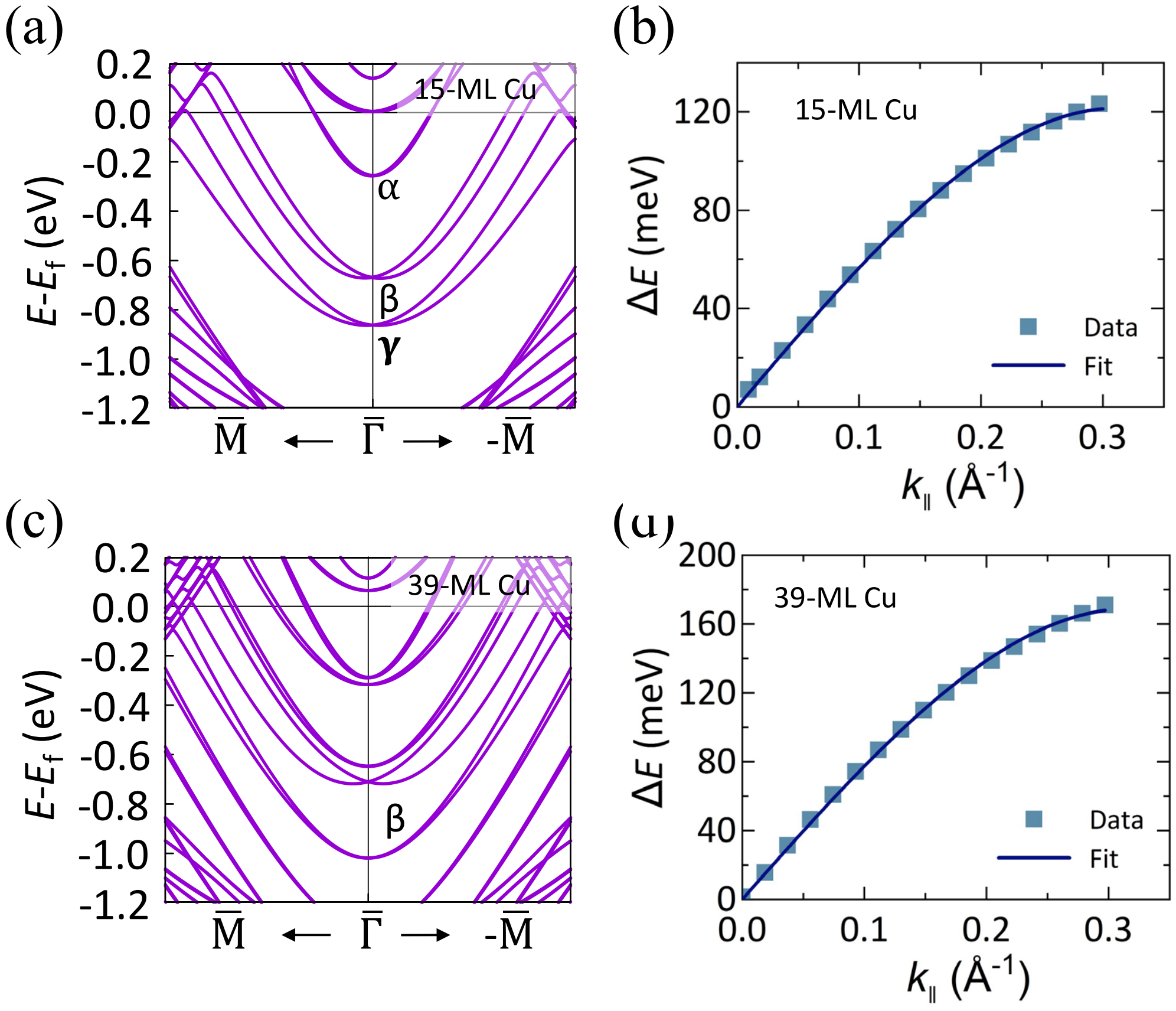


**FIG. S11.** Band structures and momentum-resolved energy splitting of β-band for the Cu(001)/$WO_3$(001) heterostructure (Type-III interface configuration) for Cu thicknesses: (a), (b) 15-ML and (c), (d) 39-ML. Solid line in (b) and (d) represents fit to the equation: $\Delta E(k) = 2\alpha_R k + 2\,\tilde{\alpha}_R k^3$

**TABLE S4.** Linear ($\alpha_R$) and cubic ($\tilde{\alpha}_R$) Rashba parameters for the β band of the Cu(001)/$WO_3$(001) Type-III heterostructure with varying Cu(001) film thickness.

| **Thickness of the Cu layer** | **Fitting Range: 0.0-0.3 Å$^{-1}$** | |
|---|---|---|
| | $\tilde{\alpha}_R$ (eV. Å$^3$) | $\alpha_R$ (eV. Å) |
| 15-ML | -1.0101± 0.0082 | 0.2927 ± 0.0005 |
| 39-ML | -1.3422 ± 0.0129 | 0.4004 ± 0.0008 |
| | **Fitting Range: 0.0-0.2 Å$^{-1}$** | |
| 15-ML | -1.3534 ± 0.0118 | 0.3019 ± 0.0003 |
| 39-ML | -1.9302 ± 0.0197 | 0.4157 ± 0.0005 |
| | **Fitting Range: 0.1-0.3 Å$^{-1}$** | |
| 15-ML | -0.9725 ± 0.0076 | 0.2904 ± 0.0005 |
| 39-ML | -1.2714 ± 0.0113 | 0.3961 ± 0.0007 |

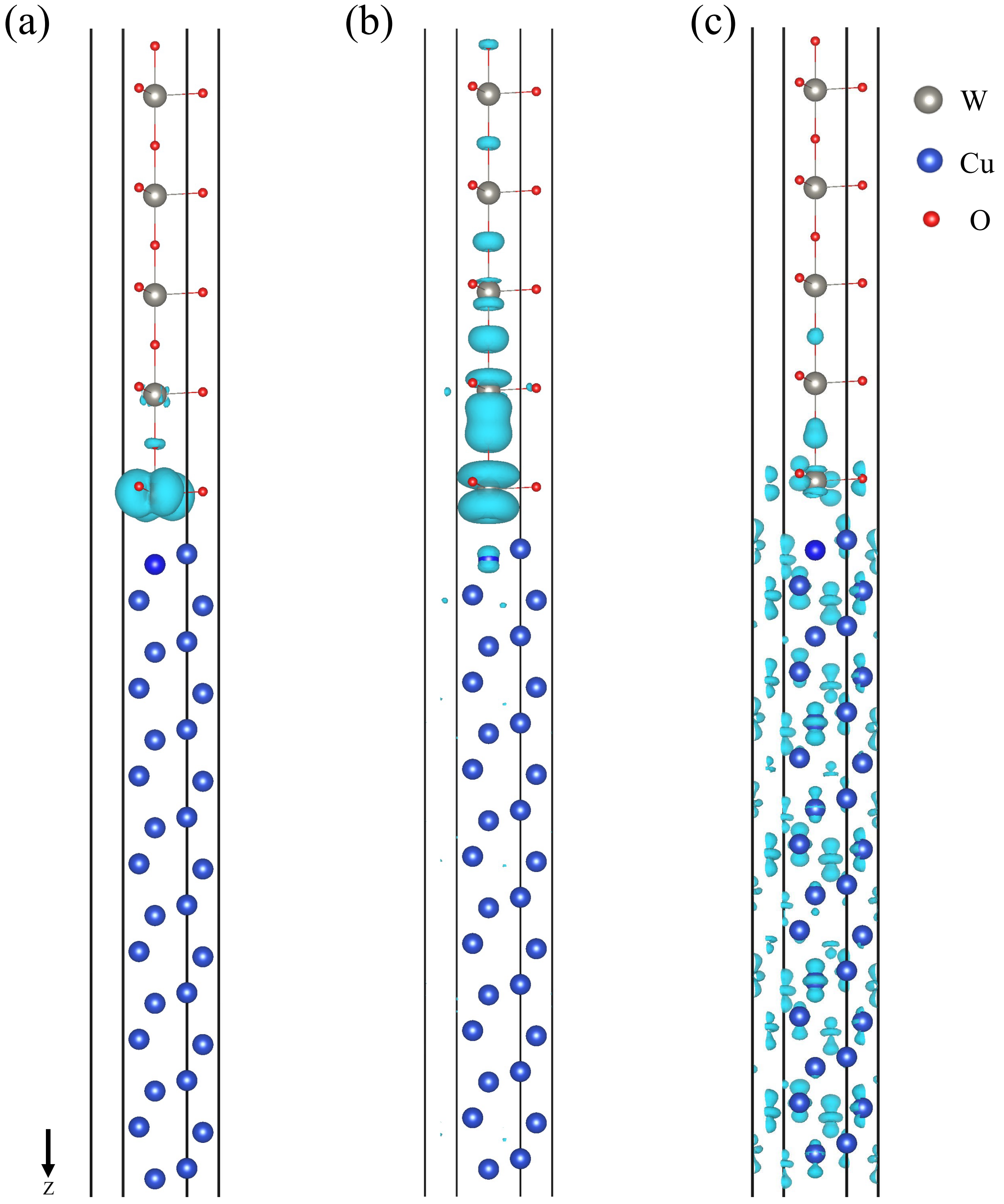


**FIG. S12.** Real-space electron density for the (a) α, (b) β, and (c) γ bands in the $Cu(001)/WO_3(001)$ heterostructure with Type-III interface stacking. The electron densities are evaluated around the band bottom at energies ≈ −0.25 eV (for α), −0.67 eV (for β), and −0.86 eV (for γ).

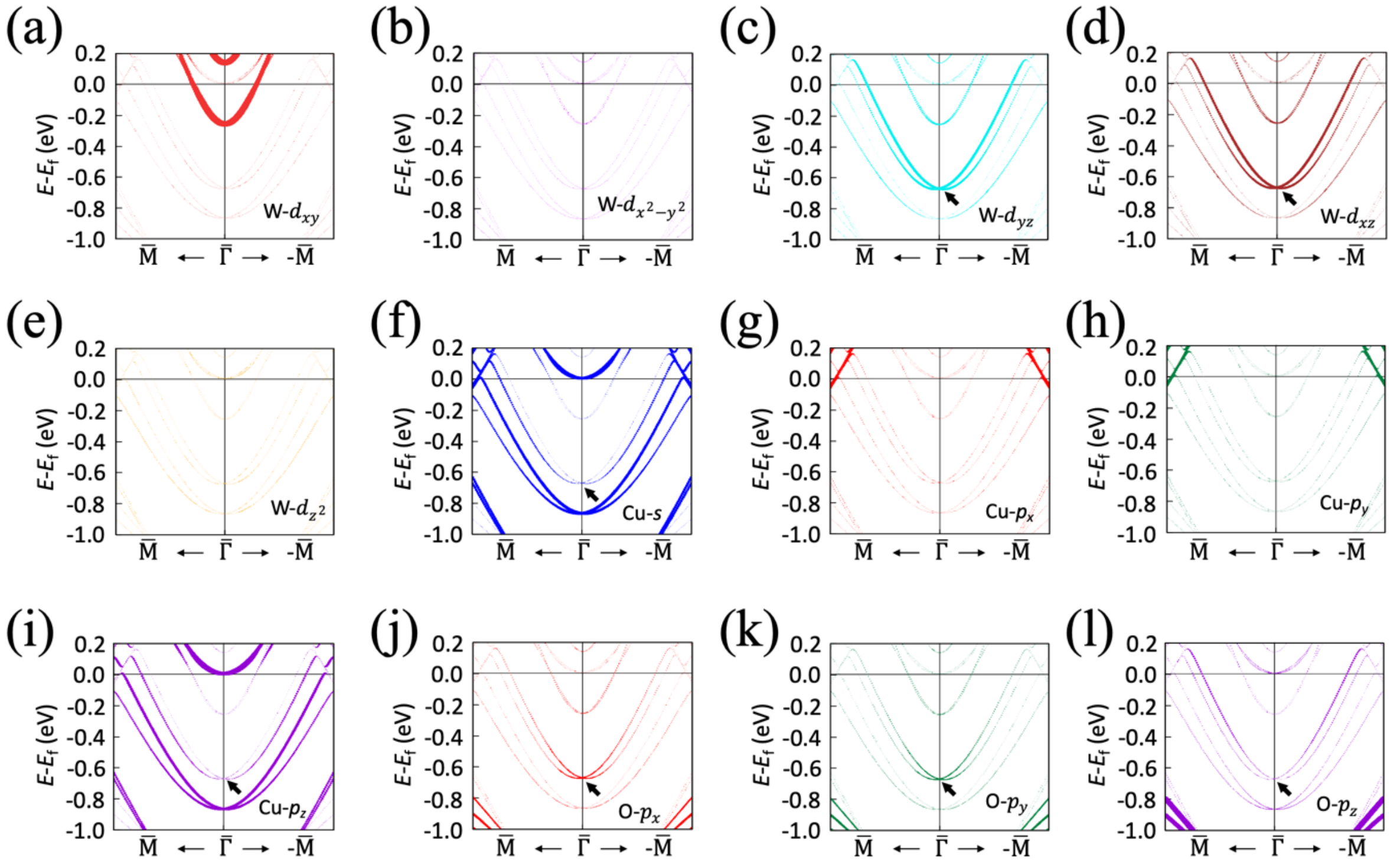


**FIG. S13.** Orbital-resolved band structure of the 15-ML Cu(001)/$WO_3$(001) heterostructure with the Type-III interface configuration. The radii of the circles are proportional to the weight of the corresponding atomic orbitals.

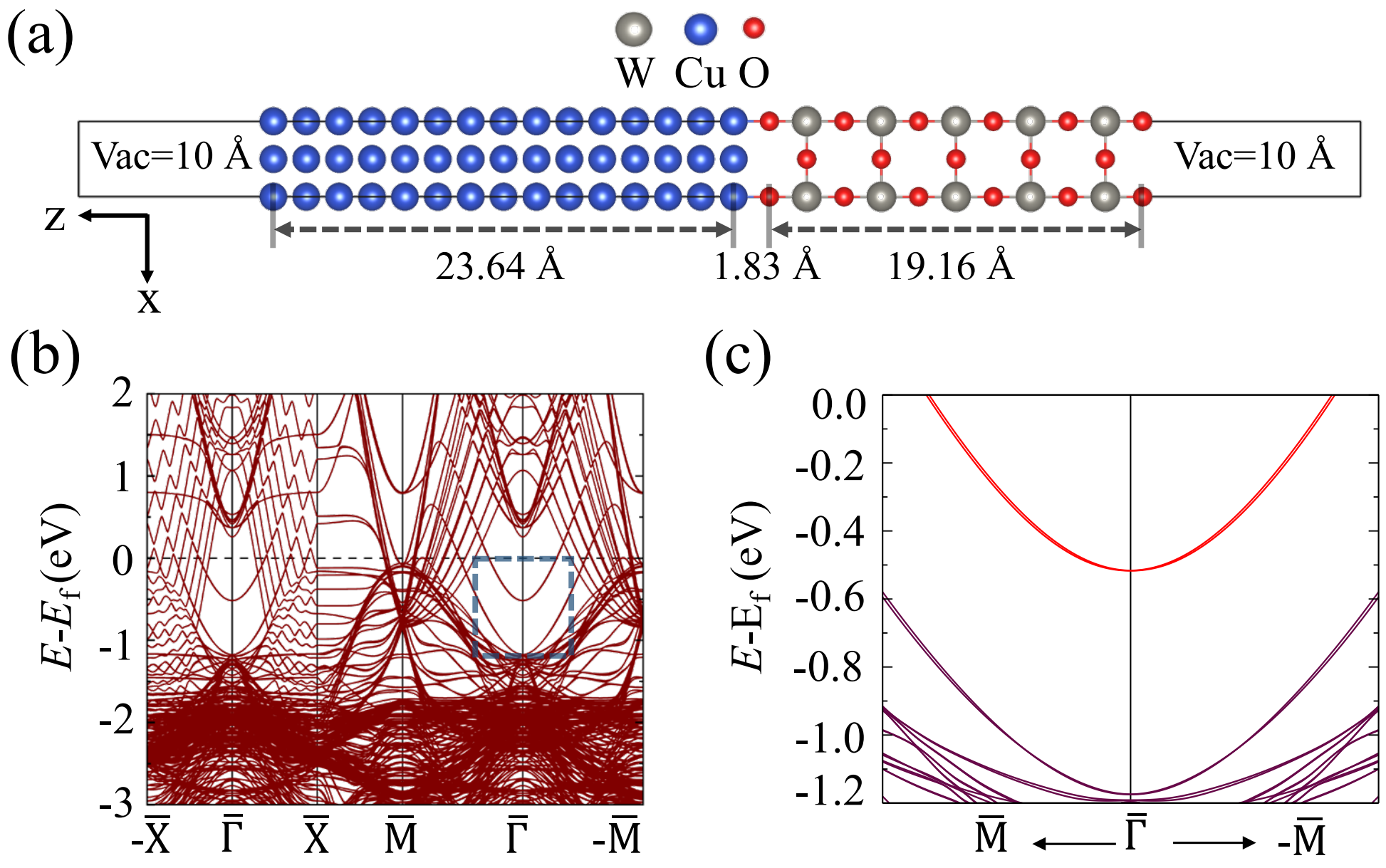


**FIG. S14.** (a) Slab model of the Cu(001)/$WO_3$(001) heterostructure with an O-terminated interface; (b) corresponding band structure; (c) magnified view of the region highlighted by the dark-blue square in (b). The band highlighted in red in (c) is a Cu QWS. The characteristic linear Rashba parameter extracted from the band bottom is $\alpha_R = 2E_R/k_R = 0.031$ eV.Å

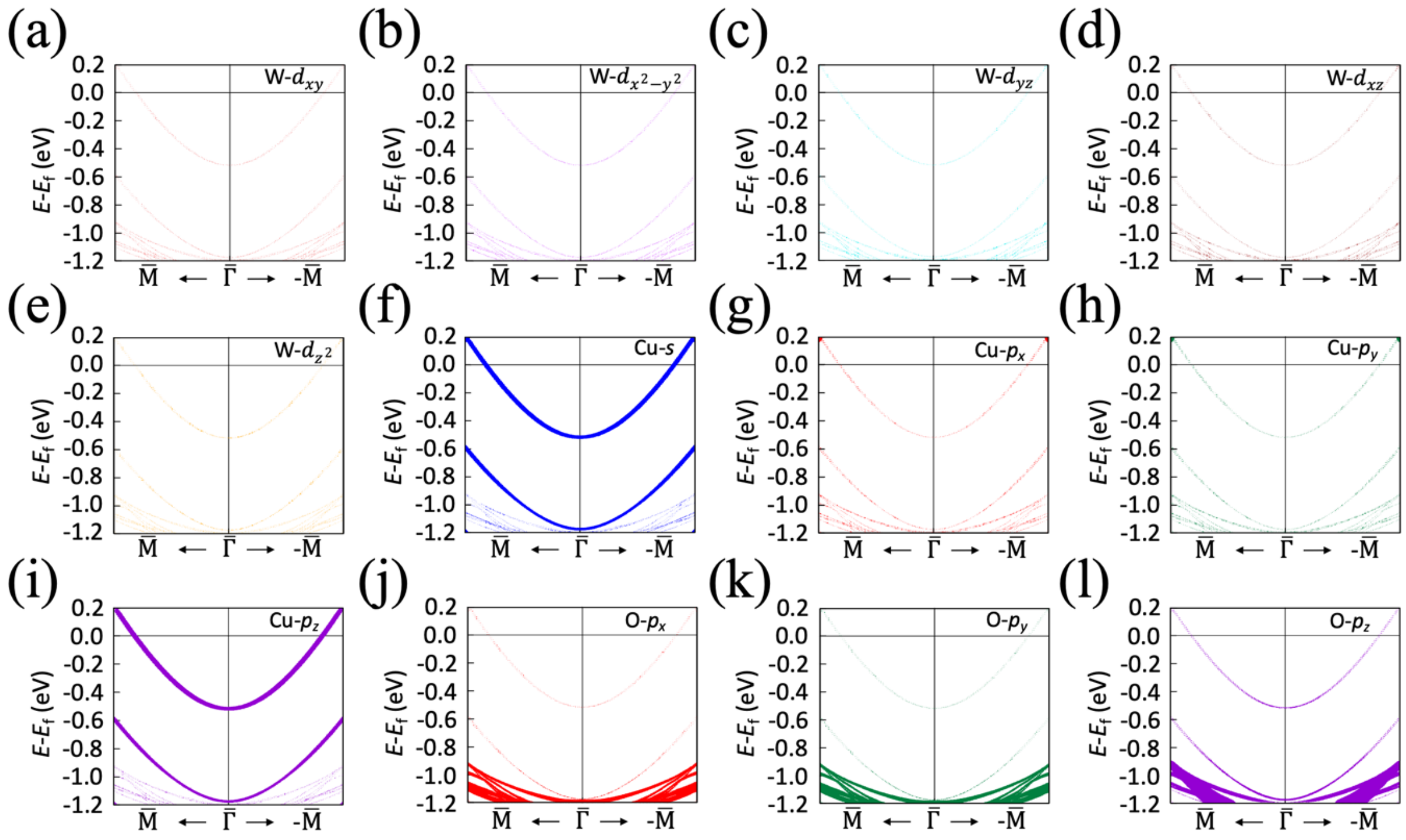


**FIG. S15.** Orbital-resolved band structure of the Cu(001)/$WO_3$(001) heterostructure with the O-terminated interface. The radii of the circles are proportional to the weight of the corresponding atomic orbitals.

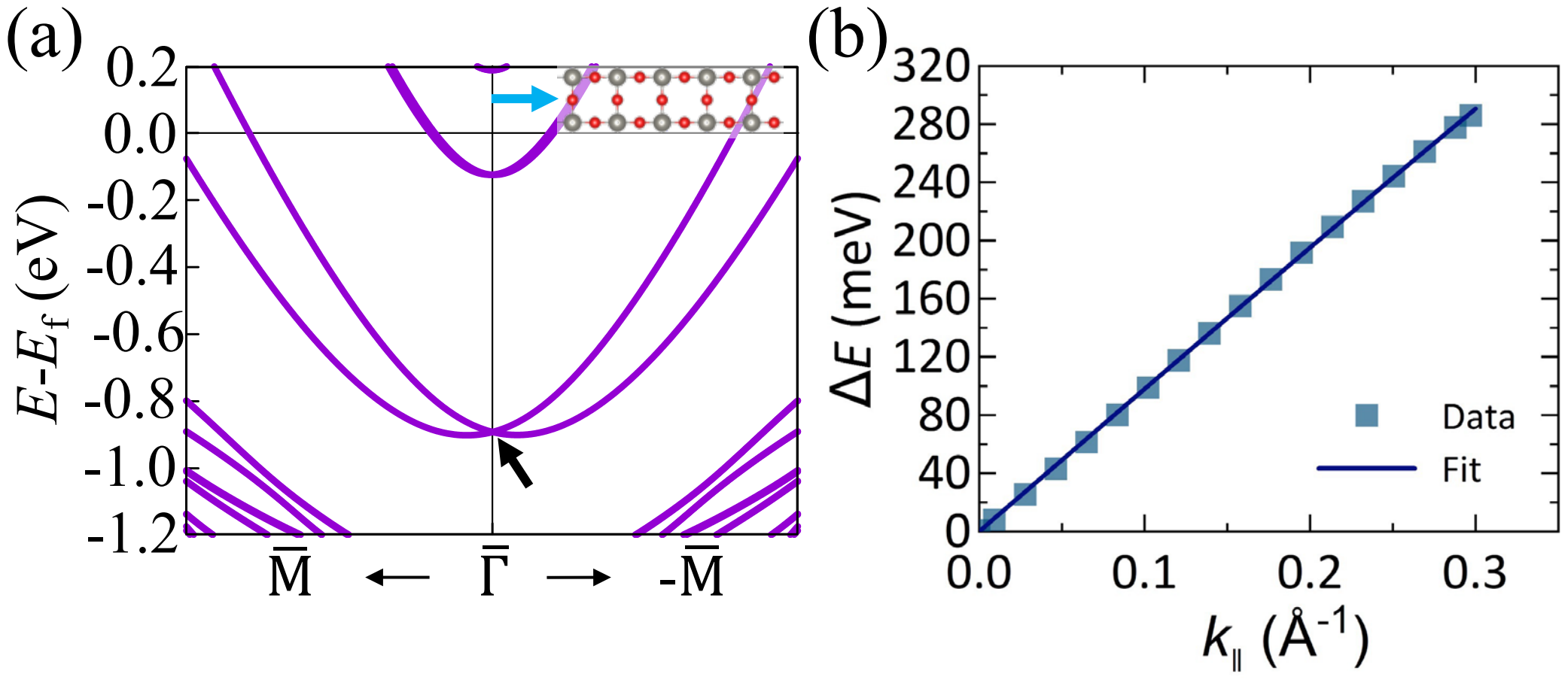


**FIG. S16.** Electronic structure of $WO_3$. (a) Band structure of the asymmetric, single-surface passivated paraelectric $WO_3$ (001) slab. The black arrow indicates the Rashba spin-split surface state localized at the unpassivated surface (marked by a light blue colored arrow in the inset slab model), which faces the Cu layer. (b) Momentum-resolved energy splitting, ΔE(k), of this surface state along the $\bar{\Gamma} - \bar{M}$ path (blue squares). The solid blue line represents the fit to the two-parameter model, $\Delta E(k) = 2\alpha_R k + 2\,\tilde{\alpha}_R k^3$. The fit yields a robust linear Rashba parameter of $\alpha_R \approx +0.49$ eV·Å and a negligible cubic Rashba parameter of $\tilde{\alpha}_R \approx -0.07$ eV·Å$^3$.

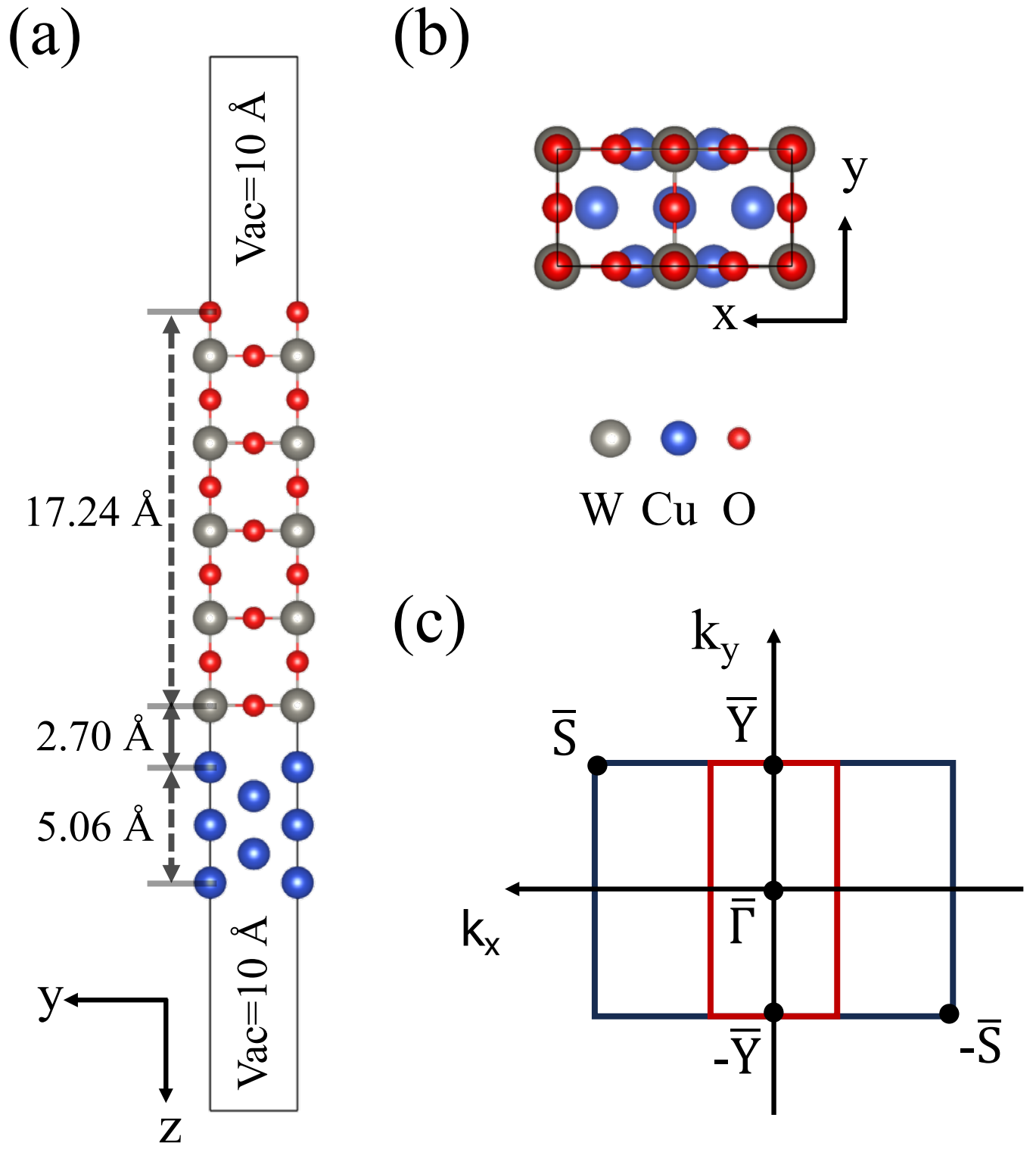


**FIG. S17.** Slab model of the Cu(110)/$WO_3$(001) heterostructure with Type-I interface configuration: (a) side view, (b) top view, and (c) corresponding SBZ. The first Brillouin Zone of the supercell is shown in red, while that of the primitive cell is shown in black. Band-unfolding was performed from the supercell Brillouin Zone to the SBZ of the primitive cell to obtain the electronic band structure shown in FIG. 6(a).

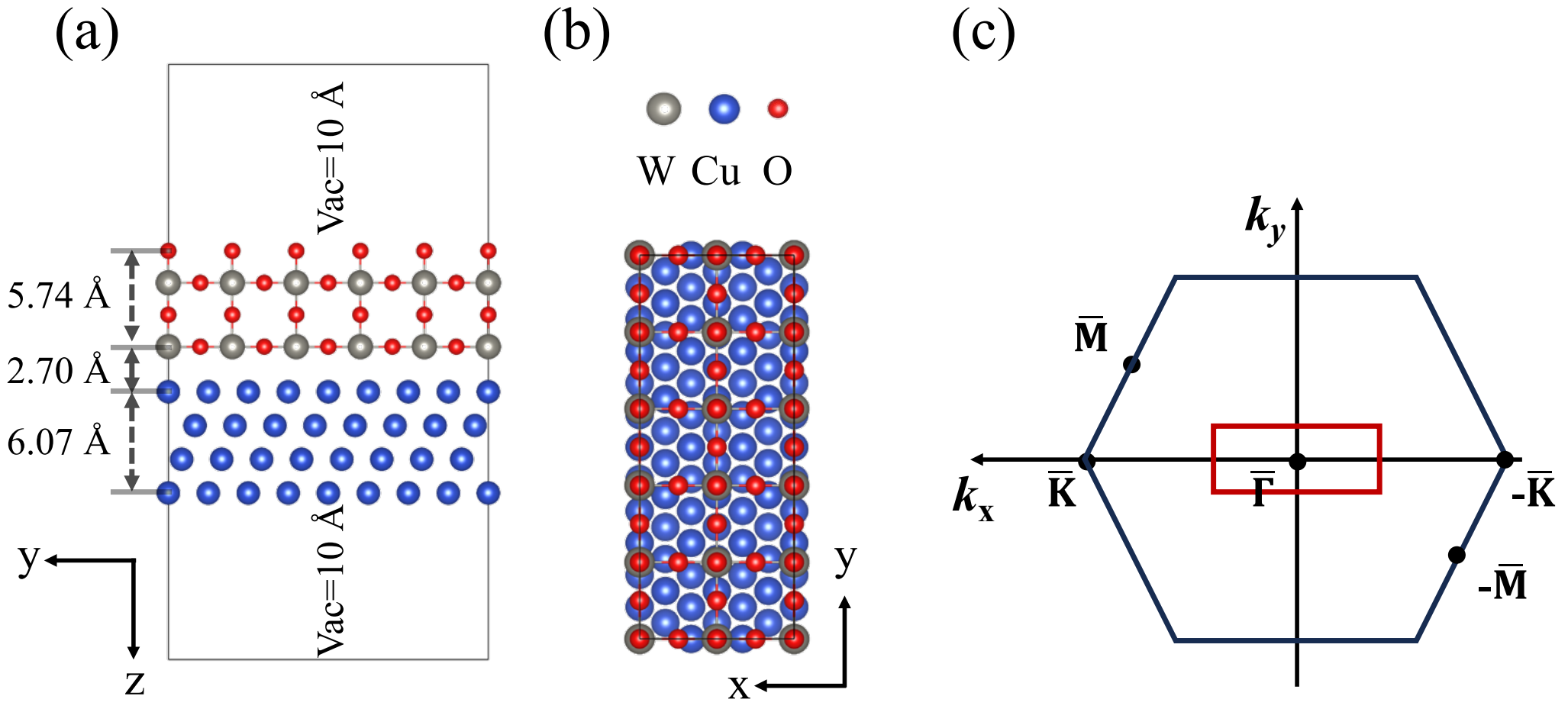


**FIG. S18.** Slab model of the Cu(111)/$WO_3$(001) heterostructure with a $WO_2$-terminated interface: (a) side view, (b) top view, and (c) corresponding SBZ. Shown here is the Type-I interface configuration. The first Brillouin Zone of the supercell is shown in red, while that of

the primitive cell is shown in black. Band-unfolding was performed from the supercell Brillouin Zone to the SBZ of the primitive cell to obtain the electronic band structure shown in FIG. 6(b)

**Structural and Computational Details for Cu(110)/$WO_3$(001) and Cu (111)/$WO_3$(001) Heterostructures:**

To model the Cu(110)/$WO_3$(001) heterostructure with a $WO_2$-terminated interface and reduce lattice mismatch, a 3 × 1-Cu(110) supercell and a 2 × 1-$WO_3$(001) supercell were employed. For the Cu(111)/$WO_3$(001) heterostructure with a $WO_2$-terminated interface, a 3 × 4-Cu(111) and a 2 × 5-$WO_3$(001) supercells were used. The Cu(110)/$WO_3$(001) slab consisted of 5-ML of Cu(110) and four-unit cells of $WO_3$(001) (Fig. S16), while the Cu(111)/$WO_3$(001) slab contained 4-ML of Cu(111) and one-unit-cell-thick $WO_3$(001) (Fig. S17).

The in-plane lattice parameters ($a$ and $b$) of Cu(110) and Cu(111) were constrained to match the $WO_3$, while the out-of-plane lattice parameter ($c$) was determined from volume conservation. The resulting in-plane strain values for both heterostructures are summarized in TABLE S5. Brillouin zone sampling was performed using an 8 × 16 × 1 $k$-point mesh for Cu(110)/$WO_3$(001) and an 8 × 4 × 1$k$-point mesh for Cu(111)/$WO_3$(001).

The Hubbard U correction with U = 5.65 eV was applied in all calculations, consistent with the Cu(001)/$WO_3$(001) system. Spin–orbit coupling (SOC) was included, and the same cut-off energy, SCF convergence criterion, and norm-conserving pseudopotentials as those used in the Cu(001)/$WO_3$(001) calculations were employed.

To extract spin-texture for the Type-I Cu(110)/$WO_3$(001) structure we employed OpenMX post-processing tools. We first identified the interface band in the folded supercell band structure, then computed the spin texture at the band bottom around the $\bar{\Gamma}$-point. For a shallow $k$-range around the $\bar{\Gamma}$-point of the folded band, the iso-energy contour serves as a qualitative proxy for the unfolded interface band, allowing us to indirectly probe its spin texture. For the Cu (111)/ $WO_3$(001) heterostructure, the very large supercell (193 atoms) generates an extremely dense folded band structure, which currently exceeds the practical limits of the OpenMX post-processing scripts. Hence, a similarly reliable extraction of the spin texture within the present computational setup was not possible.

**TABLE S5.** In-plane strain values (%) of Cu(110) and Cu(111) in the heterostructures. Positive values indicate tensile strain, while negative values indicate compressive strain.

| Heterostructures | Strain along a (%) | Strain along b (%) |
|---|---|---|
| Cu(110)/$WO_3$(001) | -1.6 | 4.3 |
| Cu(111)/$WO_3$(001) | -1.6 | 6.4 |

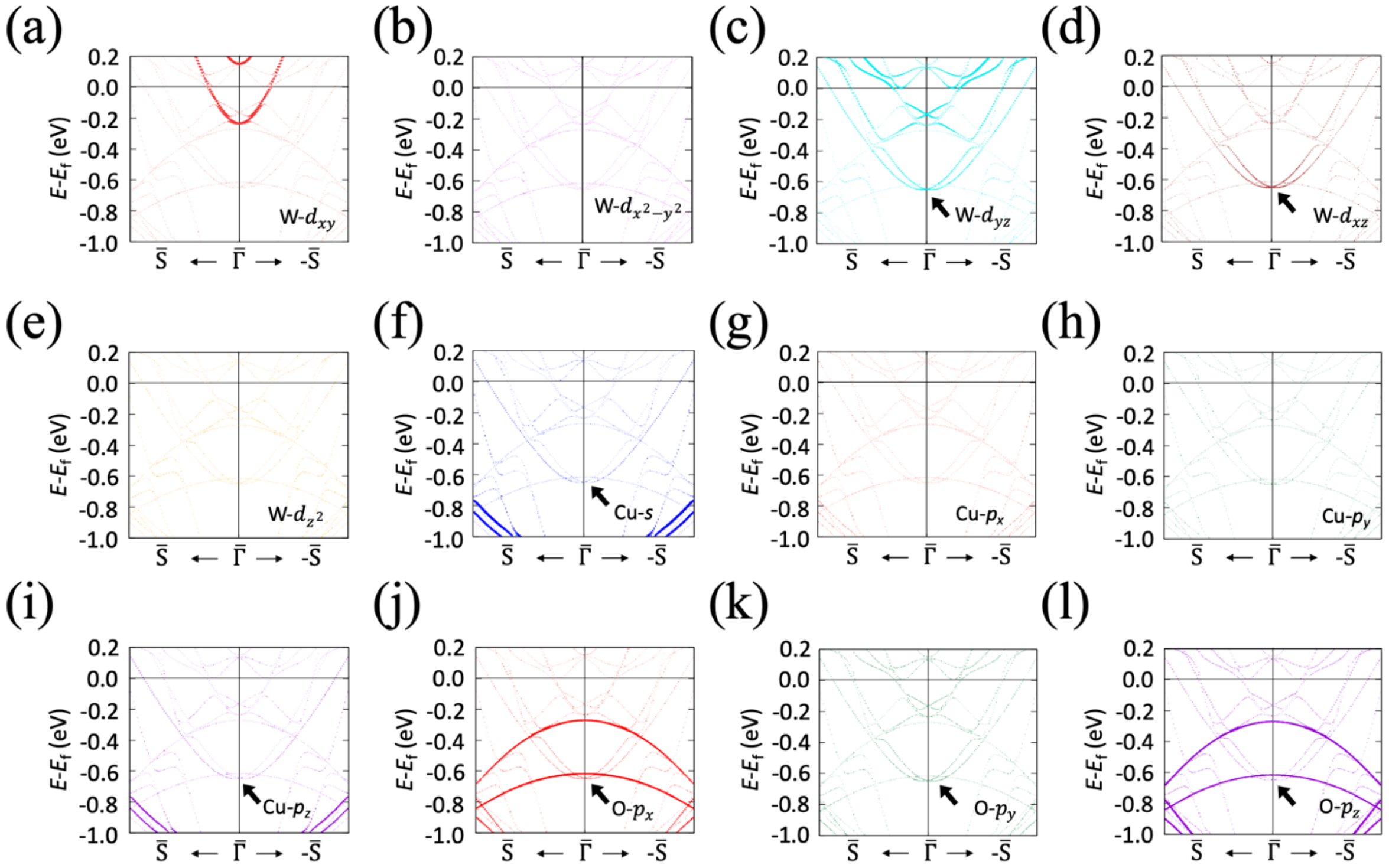

**FIG. S19.** Orbital-resolved spectral weight of the unfolded band structure for the Cu(110)/$WO_3$(001) heterostructure with the Type-I interface configuration that is discussed in the main text. The radii of the circles are proportional to the weight of the corresponding atomic orbitals. For clarity, the bottom of the interface band is marked by an arrow.

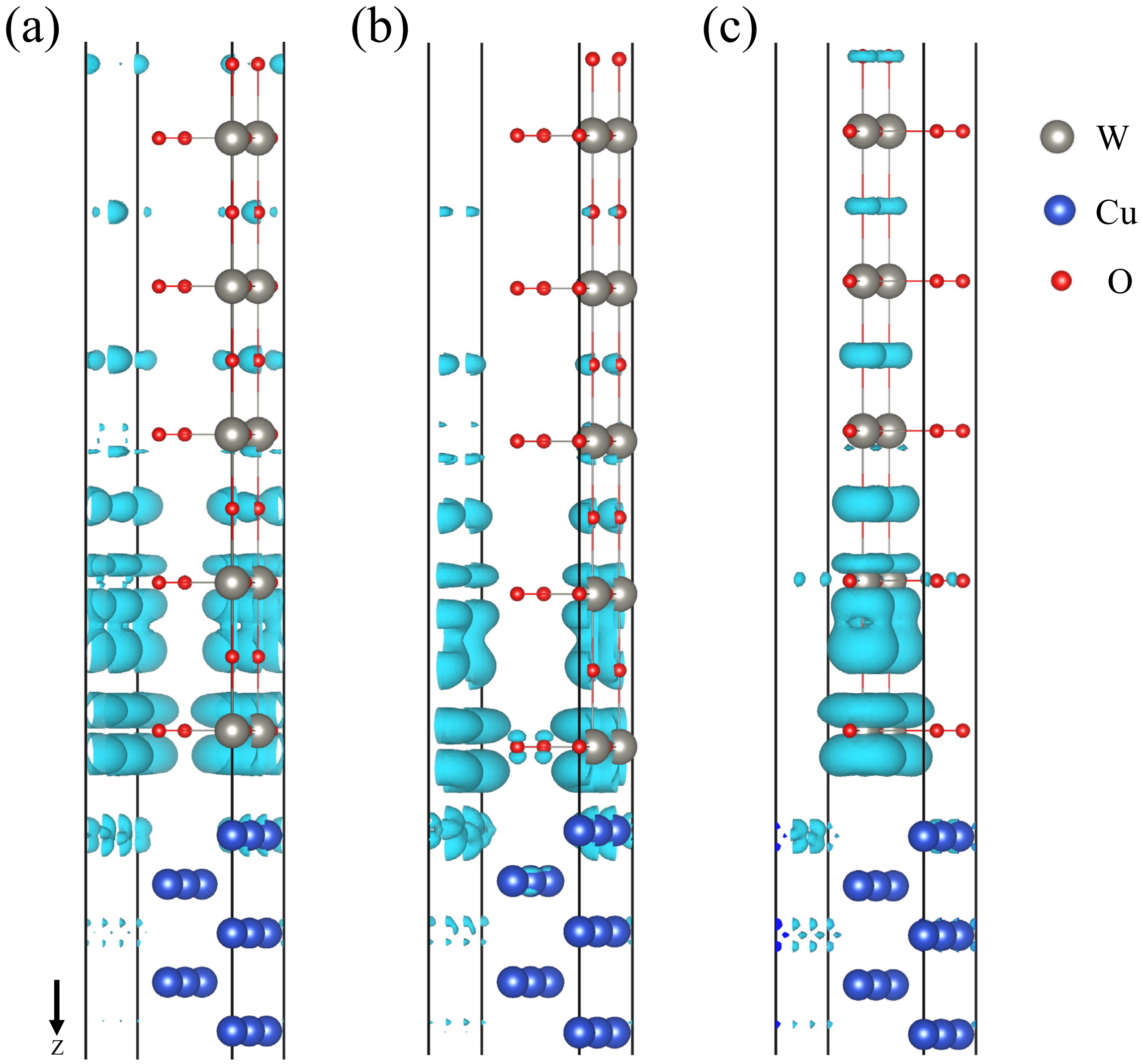


**FIG. S20**. Real-space electron density associated with the interface state of the $Cu(110)/WO_3(001)$ heterostructure with (a) Type-I, (b) Type-II, and (c) Type-III interface configuration. The electron densities are evaluated at the $\bar{\Gamma}$-point, from the band bottom at energies ≈ −0.65 eV, −0.59 eV, and −0.82 eV for Type-I, Type-II, and Type-III configurations, respectively.

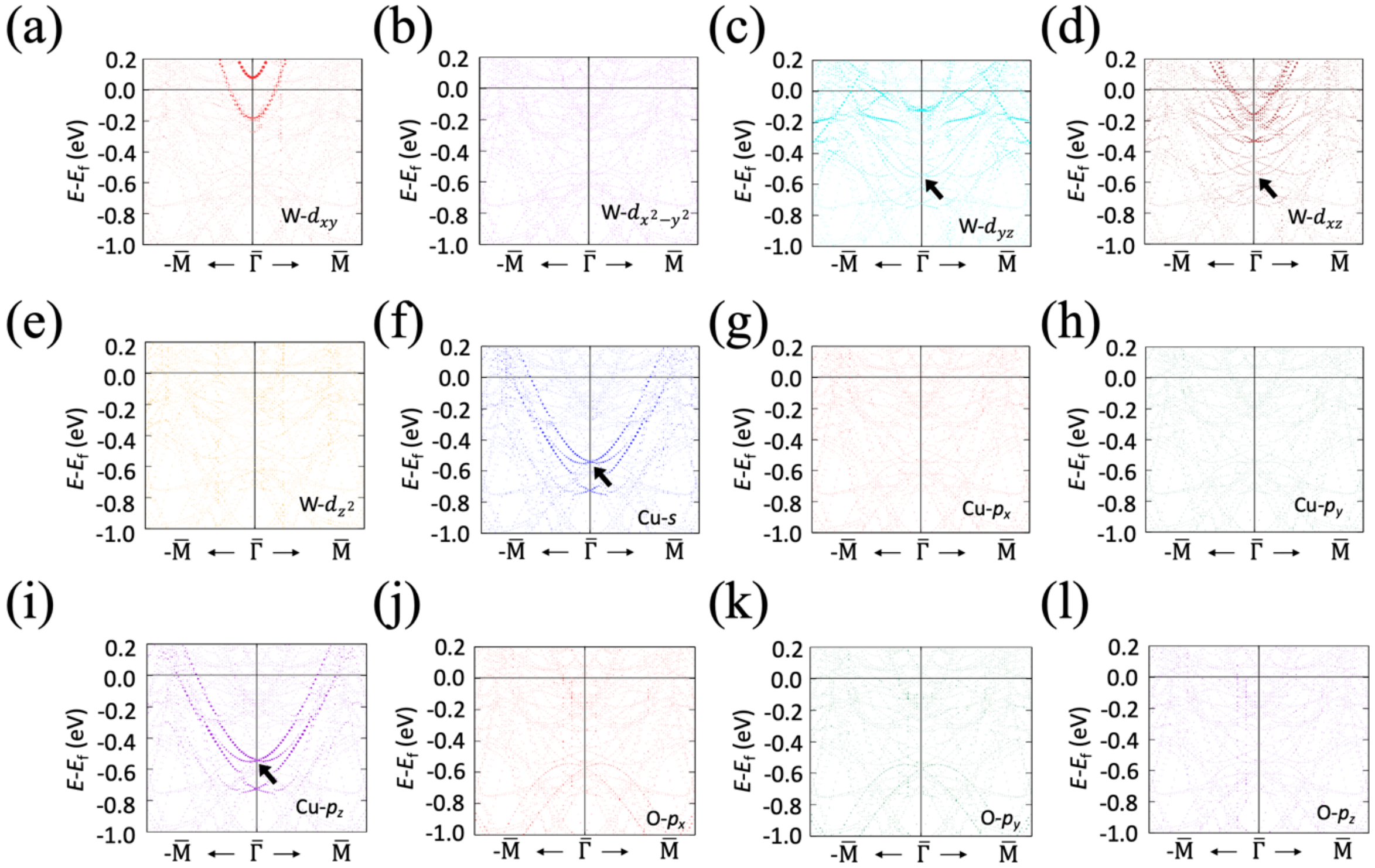


**FIG. S21**. Orbital-resolved spectral weight of the unfolded band structure for the Cu(111)/$WO_3$(001) heterostructure with the Type-I interface configuration that is discussed in the main text. The radii of the circles are proportional to the weight of the corresponding atomic orbitals. The bottom of the interface band is marked by an arrow. Notably, compared with other two Cu orientations, contribution from the Cu orbitals is noticeably enhanced, which could arise from the larger packing fraction of the Cu (111) surface than the Cu (001) and Cu (110) surfaces. Meanwhile, the one-unit-cell-thick $WO_3$ layer contains two $WO_2$ planes that are both strongly affected by the interfacial electric fields at the Cu/$WO_3$ and $WO_3$/vacuum interfaces. Consequently, the W and O orbitals that are more sensitive to the interfacial field yield a sizable incoherent, diffuse background in the unfolded spectra. Nonetheless, this analysis qualitatively demonstrates that the Cu-(s,$p_z$) and W-($d_{xy}$,$d_{yz}$) orbitals predominantly contribute to the interface state.

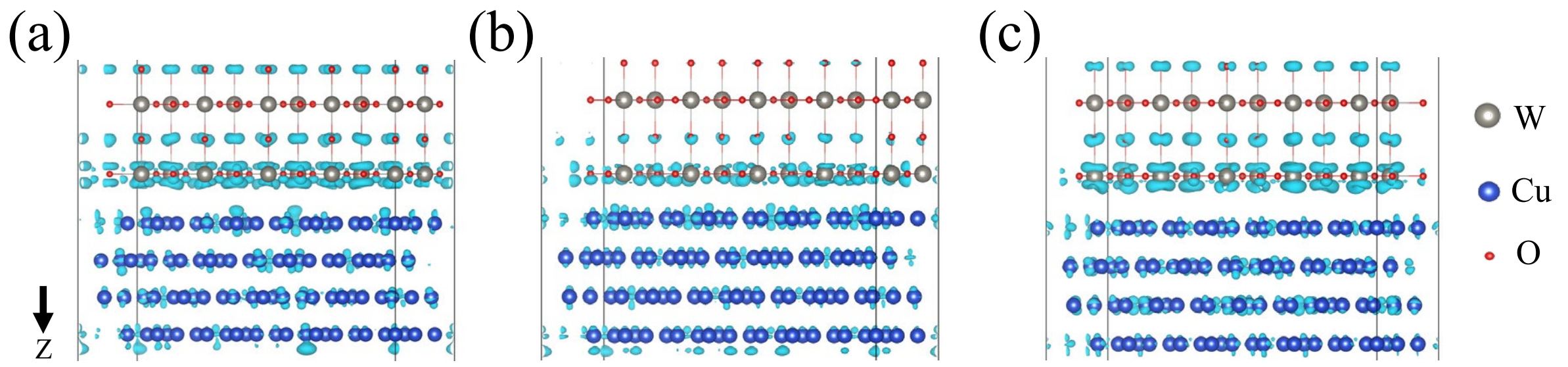


**FIG. S22.** Real-space electron density associated with the interface state of the Cu(111)/$WO_3$(001) heterostructure with (a) Type-I, (b) Type-II, and (c) Type-III interface stacking. The electron densities are evaluated at the $\bar{\Gamma}$-point, from the band bottom at energies $\approx -0.54$ eV, $-0.41$ eV, and $-0.53$ eV for Type-I, Type-II, and Type-III configuration, respectively.

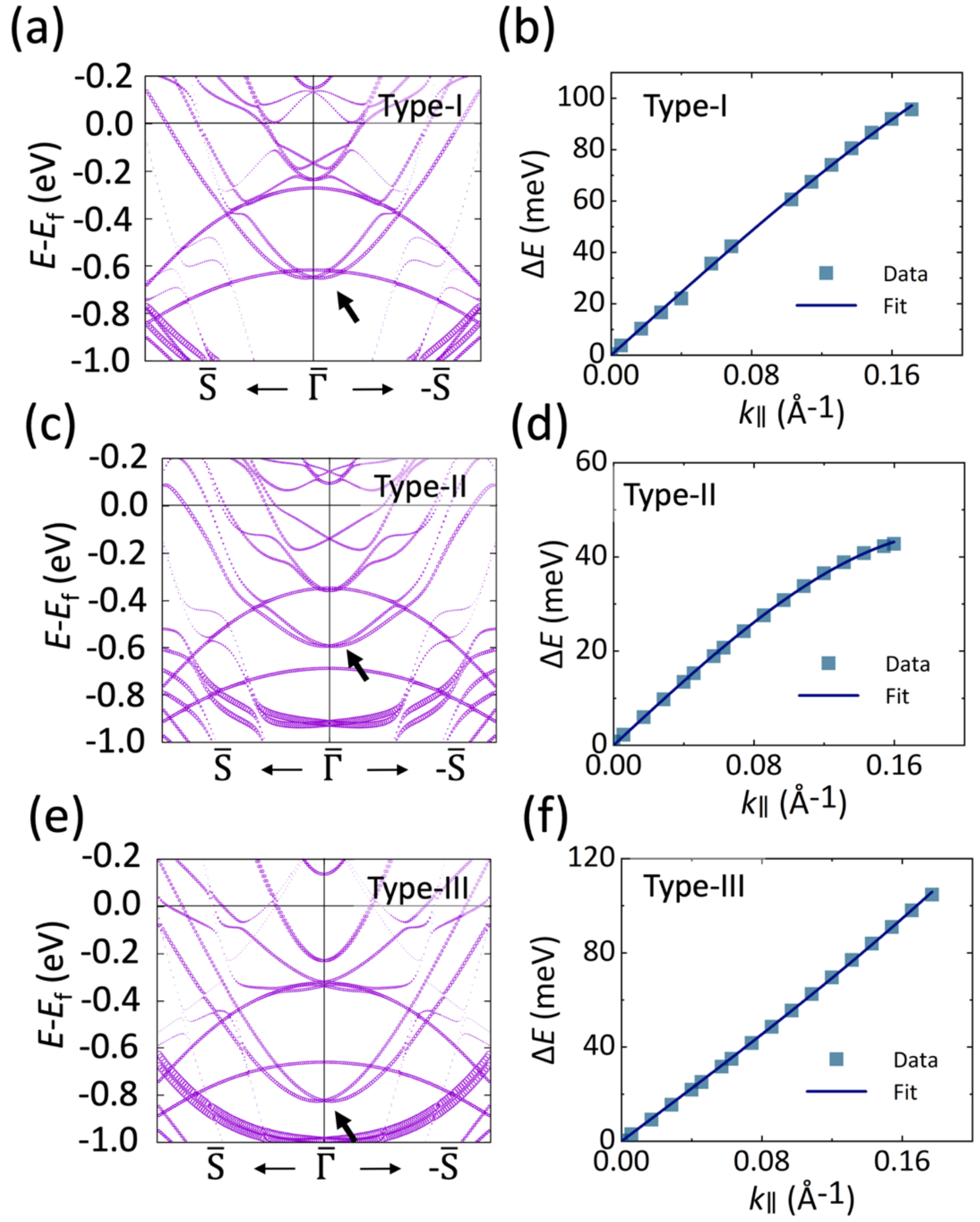


**FIG. S23**. Band structures and momentum-resolved energy splitting of the interface band for the Cu(110)/$WO_3$(001) heterostructure. (a), (c), and (e) Band structures for the Type-I, Type-II, and Type-III interface configurations, respectively. For clarity, the interface band is highlighted by an arrow. (b), (d), (f) Corresponding momentum-resolved energy splitting (solid squares) for the Type-I, Type-II, and Type-III interface configurations, respectively. The solid line represents fit to the equation $\Delta E(k) = 2\alpha_R k + 2\,\tilde{\alpha}_R k^3$.

**TABLE S6.** Characteristic linear ($\alpha_R$) and cubic ($\tilde{\alpha}_R$) Rashba parameters of the interface band of the Cu(110)/$WO_3$(001) heterostructure for different interface stacking.

| **Interface configuration** | $\tilde{\alpha}_R$ **(eV. Å$^3$)** | $\alpha_R$ **(eV. Å)** |
|---|---|---|
| Type-I | -0.7960±0.1150 | 0.3070±0.0020 |
| Type-II | -1.4621 ± 0.0339 | 0.1725 ± 0.0005 |
| Type-III | 0.6117 ± 0.0435 | 0.2800 ± 0.0009 |

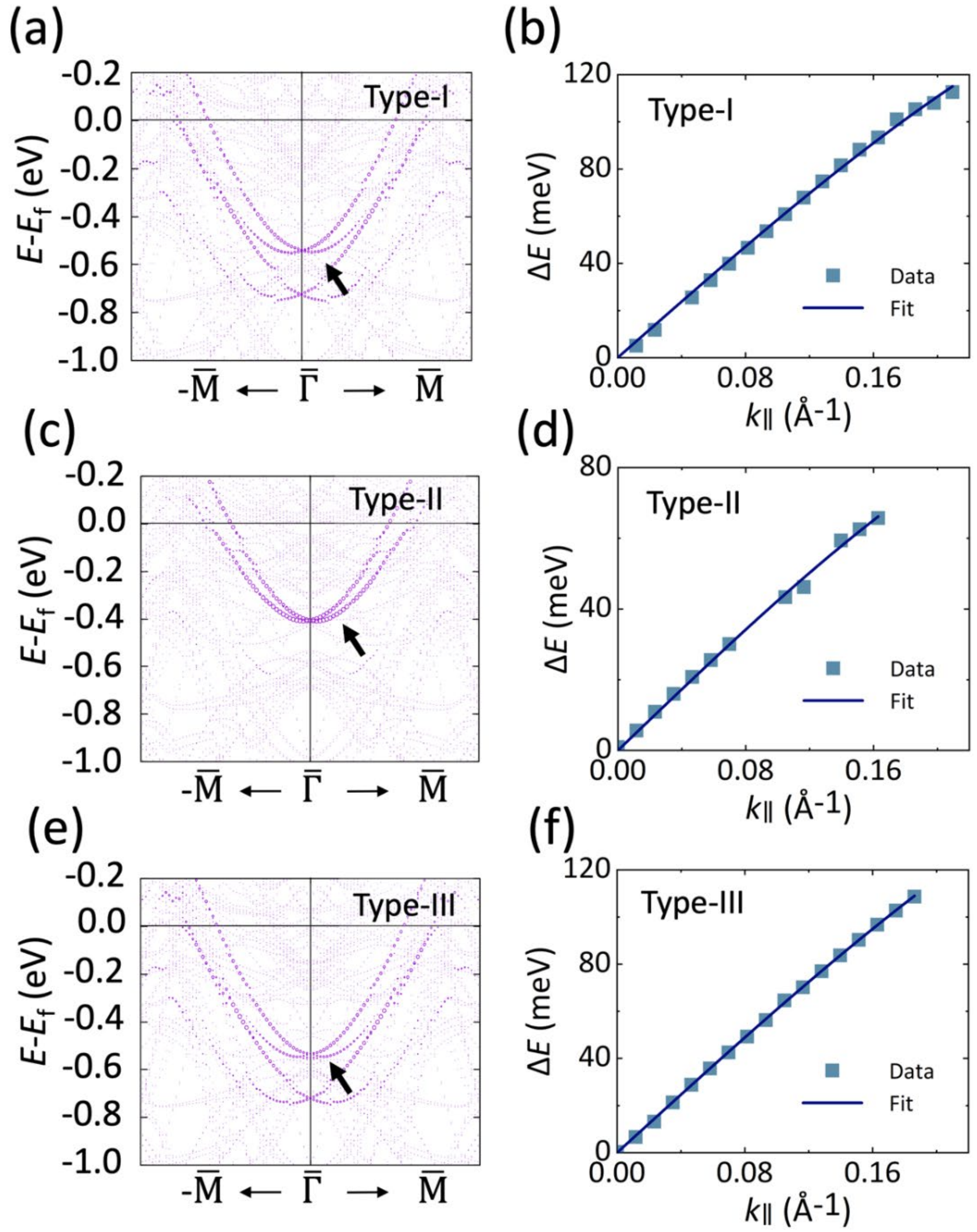


**FIG. S24**. Band structures and momentum-resolved energy splitting of the interface band for the Cu(111)/$WO_3$(001) heterostructure. (a), (c), and (e) Band structures for the Type-I, Type-II, and Type-III interface configurations, respectively. For clarity, the interface band is highlighted by an arrow. (b), (d), (f) Corresponding momentum-resolved energy splitting (solid squares) for the Type-I, Type-II, and Type-III interface configurations, respectively. The solid line represents a fit to the equation $\Delta E(k) = 2\alpha_R k + 2\,\tilde{\alpha}_R k^3$.

**TABLE S7**. Characteristic linear ($\alpha_R$) and cubic ($\tilde{\alpha}_R$) Rashba parameters of the interface band of the Cu(111)/$WO_3$(001) heterostructures with different interface configurations.

| **Interface configuration** | $\tilde{\alpha}_R$ **(eV. Å$^3$)** | $\alpha_R$ **(eV. Å)** |
|---|---|---|
| Type-I | -0.5330±0.1380 | 0.2980±0.0040 |
| Type-II | -0.5237 ± 0.2479 | 0.2169 ± 0.0049 |
| Type-III | -0.4589 ± 0.0549 | 0.3085 ± 0.0013 |

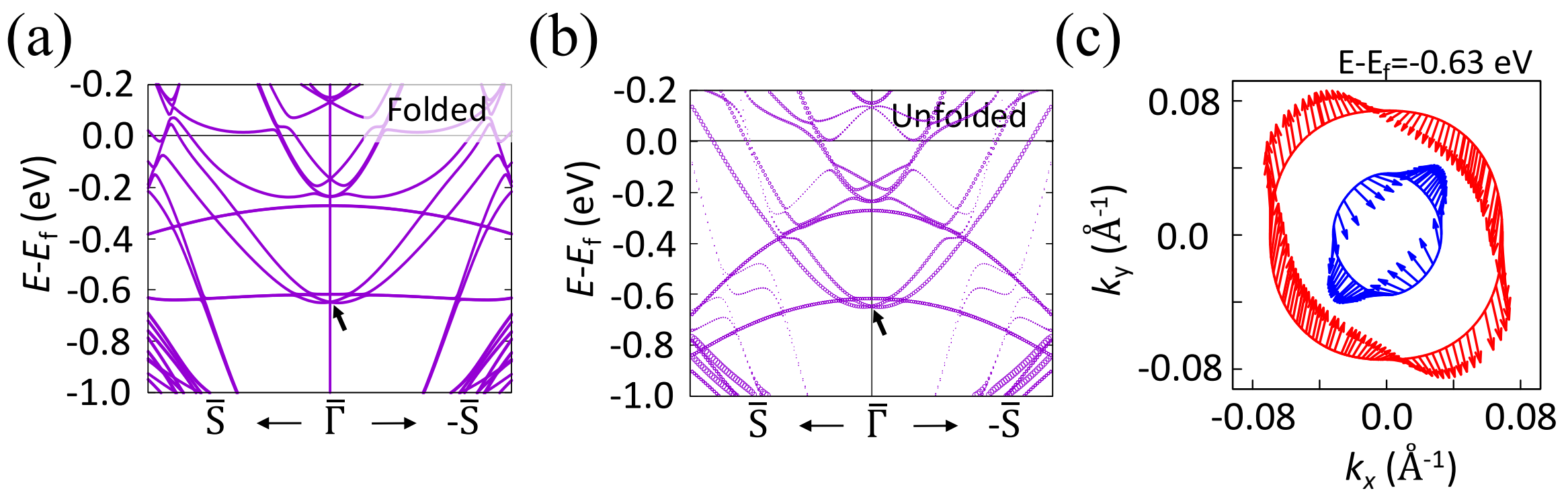


**FIG. S25**. (a) Folded band of the Cu(110)/$WO_3$(001) heterostructures with the Type-I interface configuration. (b) Corresponding unfolded band structure that is shown as Fig. 6(c) in the main text. (c) The spin texture obtained from the band bottom (-0.63 eV) of the folded band that is marked by an arrow. We first identified the interface band in the folded supercell band structure (Fig. S24 (a)), then computed the spin texture at the band bottom around the $\bar{\Gamma}$-point. For a shallow $k$-range around the $\bar{\Gamma}$-point of the folded band, the iso-energy contour serves as a qualitative proxy for the unfolded interface band in Fig. S24(b), allowing us to indirectly probe its spin texture. As shown in Fig. S24 (c), the resulting non-trivial spin winding clearly reflects the coexistence of linear and cubic Rashba effects in the interface state, similar to the Cu (001) case.

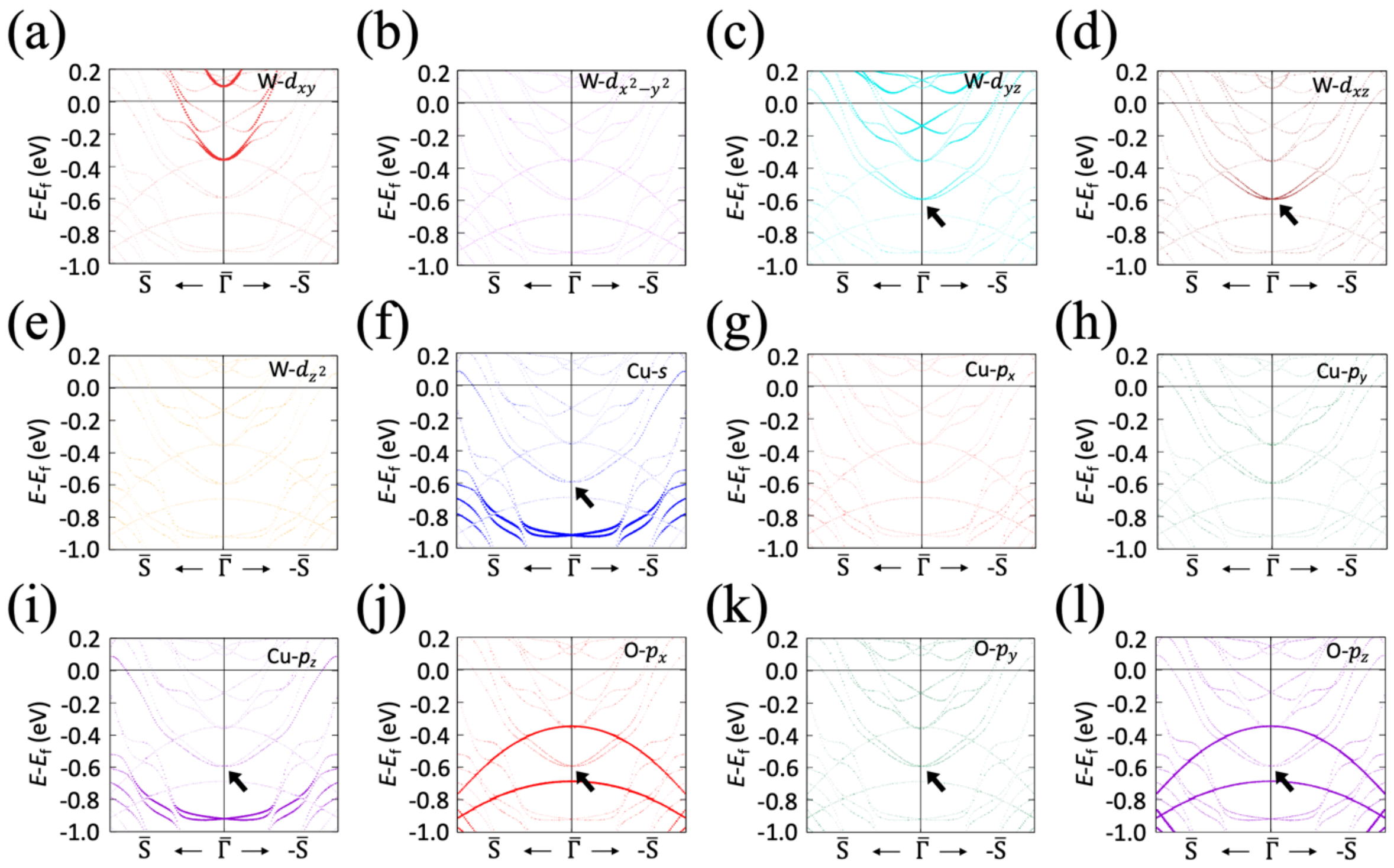


**FIG. S26.** Orbital-resolved spectral weight of the unfolded band structure for the Cu(110)/$WO_3$(001) heterostructure with the Type-II interface configuration. The radii of the circles are proportional to the weight of the corresponding atomic orbitals. For clarity, the bottom of the interface band is marked by an arrow.

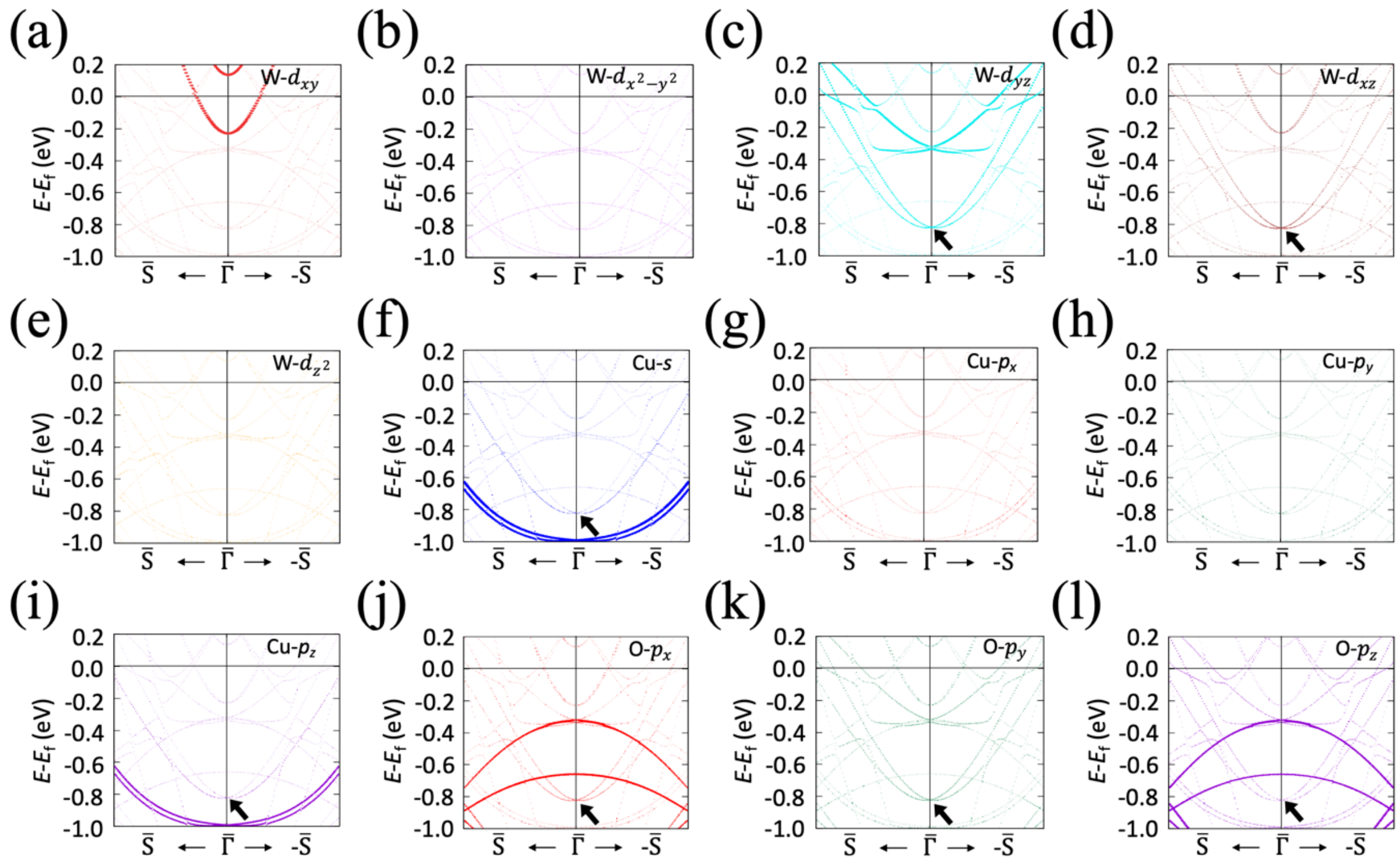


**FIG. S27.** Orbital-resolved spectral weight of the unfolded band structure for the Cu(110)/$WO_3$(001) heterostructure with the Type-III interface configuration The radii of the circles are proportional to the weight of the corresponding atomic orbitals. For clarity, the bottom of the interface band is marked by an arrow.

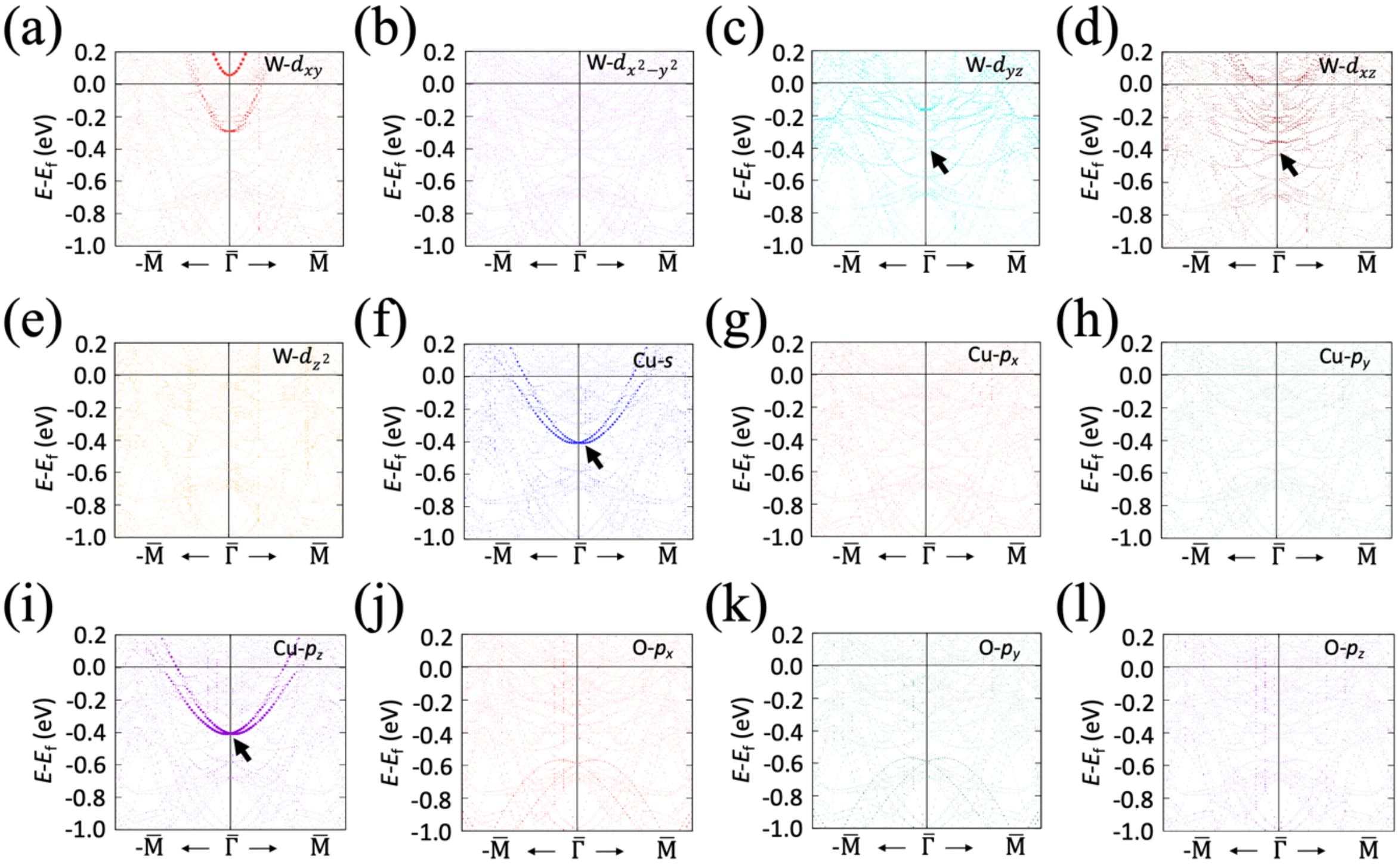


**FIG. S28.** Orbital-resolved spectral weight of the unfolded band structure for the Cu(111)/$WO_3$(001) heterostructure with the Type-II interface configuration. The radii of the circles are proportional to the weight of the corresponding atomic orbitals. The bottom of the interface band is marked by an arrow. Notably, compared with other two Cu orientations, contribution from the Cu orbitals is noticeably enhanced, which could arise from the larger packing fraction of the Cu (111) surface than the Cu (001) and Cu (110) surfaces. Meanwhile, the one-unit-cell-thick $WO_3$ layer contains two $WO_2$ planes that are both strongly affected by the interfacial electric fields at the Cu/$WO_3$ and $WO_3$/vacuum interfaces. Consequently, the W and O orbitals that are more sensitive to the interfacial field yield a sizable incoherent, diffuse background in the unfolded spectra. For this Type-II configuration, the interface state is pushed toward the Fermi level, with its band bottom shifted to −0.4 eV. Although it exhibits a prominent Cu-(s,$p_z$) orbital character, the incoherent diffused background makes it difficult to clearly resolve contribution from W-($d_{xy}$,$d_{yz}$) orbitals. Nonetheless, given the robust Rashba parameters obtained for this configuration (TABLE S7), we expect the interface state remain strongly hybridized in nature.

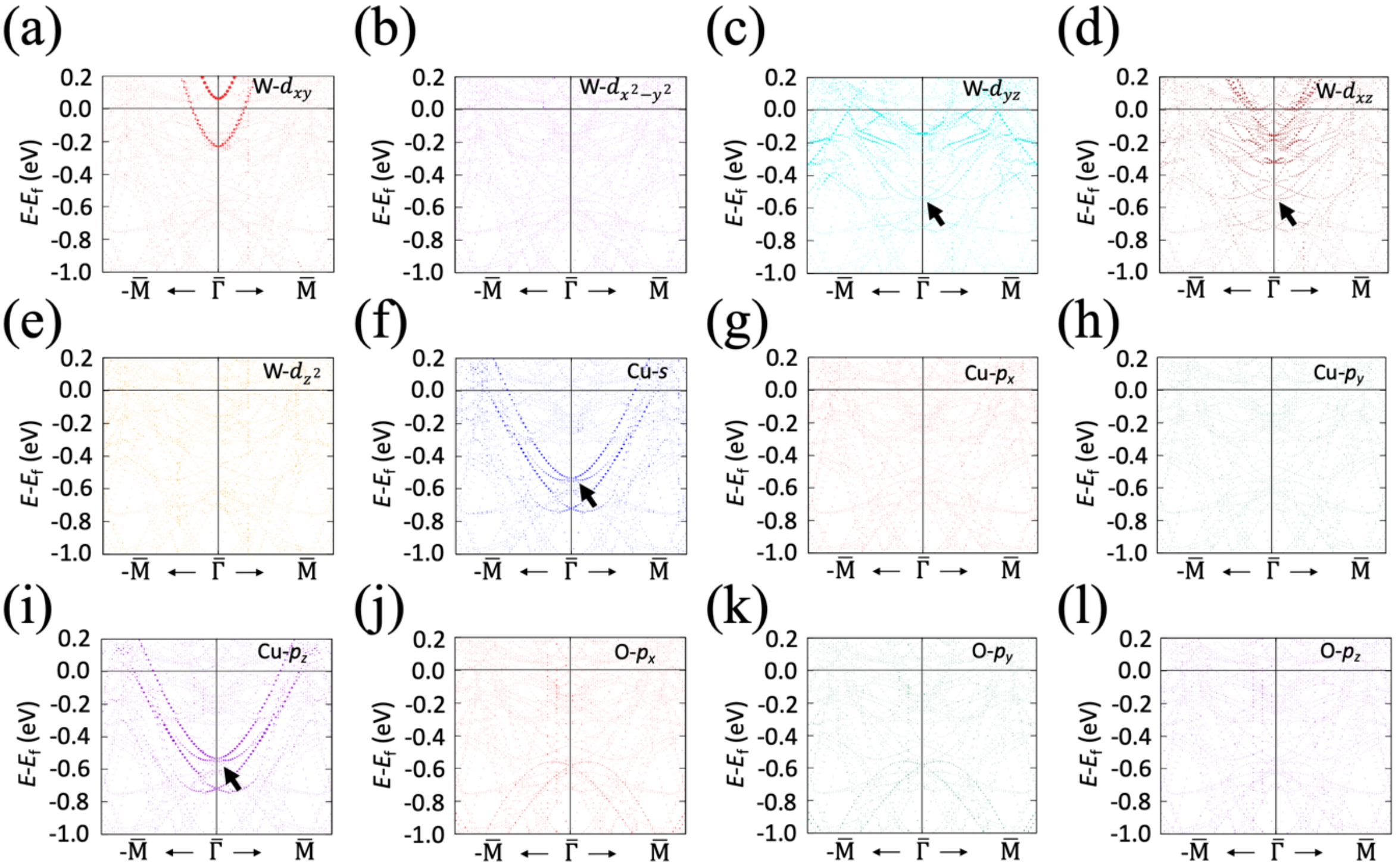


**FIG. S29.** Orbital-resolved spectral weight of the unfolded band structure for the Cu(111)/$WO_3$(001) heterostructure with the Type-III interface configuration. The radii of the circles are proportional to the weight of the corresponding atomic orbitals. The bottom of the interface band is marked by an arrow. Notably, compared with other two Cu orientations, contribution from the Cu orbitals is noticeably enhanced, which could arise from the larger packing fraction of the Cu (111) surface than the Cu (001) and Cu (110) surfaces. Meanwhile, the one-unit-cell-thick $WO_3$ layer contains two $WO_2$ planes that are both strongly affected by the interfacial electric fields at the Cu/$WO_3$ and $WO_3$/vacuum interfaces. Consequently, the W and O orbitals that are more sensitive to the interfacial field yield a sizable incoherent, diffuse background in the unfolded spectra. Nonetheless, this analysis qualitatively demonstrates that the Cu-($s$,$p_z$) and W-($d_{xy}$,$d_{yz}$) orbitals predominantly contribute to the interface state.

## References

[1] Liu, W., Li, J. C., Zheng, W. T. & Jiang, Q. NiAl(110)/Cr(110) interface: A density functional theory study. *Phys. Rev. B* **73**, 205421 (2006).

[2] Wicaksono, Y. *et al.* Giant Rashba spin-orbit coupling on light-element polar superlattices. *Phys. Rev. B* **110**, 220408 (2024).



\end{document}